\documentclass[aps,prl,twocolumn,superscriptaddress,nofootinbib]{revtex4-2}

\usepackage[T1]{fontenc}
\usepackage[utf8]{inputenc}
\usepackage{lmodern}
\usepackage{microtype}
\usepackage{amsmath,amssymb,amsthm,mathtools,mathrsfs,bm}
\usepackage{braket}
\usepackage{booktabs,enumitem}
\usepackage{graphicx}
\usepackage{float}
\usepackage[caption=false]{subfig}
\usepackage{adjustbox}
\usepackage{placeins}
\usepackage{needspace}
\usepackage{xcolor}
\usepackage[colorlinks=true,allcolors=blue!55!black]{hyperref}

\newcommand{\cP}{\mathcal P}
\newcommand{\cZ}{\mathcal Z}
\newcommand{\cM}{\mathcal M}
\newcommand{\cR}{\mathcal R}
\newcommand{\ee}{\mathrm e}
\newcommand{\ii}{\mathrm i}
\newcommand{\ff}{\mathrm{ff}}
\newcommand{\fplus}{\mathrm{f+}}
\newcommand{\pp}{\mathrm{++}}
\newcommand{\pmfix}{\mathrm{+-}}
\newcommand{\CT}{\operatorname{CT}}
\newcommand{\Sp}{\operatorname{Sp}}
\newcommand{\abs}[1]{\left|#1\right|}

\graphicspath{{./}}
\allowdisplaybreaks[2]
\newcommand{\dd}{\mathrm d}

\newcommand{\Pf}{\operatorname{Pf}}
\newcommand{\diag}{\operatorname{diag}}

\newcommand{\PBC}{\mathrm P}

\newcommand{\avg}[1]{\left\langle#1\right\rangle}

\newtheorem{theorem}{Theorem}

\begin{document}

\title{Pauli Spectra as a Boundary Spectroscope in Quantum Spin Chains}

\author{M. A. Rajabpour}
\affiliation{Instituto de F\'isica, Universidade Federal Fluminense, Av. Gal. Milton Tavares de Souza s/n, Gragoat\'a, 24210-346 Niter\'oi, RJ, Brazil}
\affiliation{Theoretical Physics III, Center for Electronic Correlations and Magnetism, Institute of Physics, University of Augsburg, D-86135 Augsburg, Germany}

\date{\today}

\begin{abstract}
We show that Pauli spectra provide a boundary spectroscope in critical quantum spin chains, resolving microscopic boundary fingerprints and finite responses to physical boundary changes. Universal logarithmic corrections to stabilizer R\'enyi entropy can remain identical across distinct boundary configurations, while their finite contributions retain information about how the physical boundaries are related. Using critical Ising chains, we demonstrate that suitable combinations of Pauli moments isolate normalization-independent responses to boundary reversal and reveal nonadditive contributions associated with the joint boundary configuration. These responses become analytically accessible through an exact term-by-term correspondence that organizes individual Pauli magnitudes into classical-group statistical structures, with their moments represented by discrete Selberg sums associated with unitary, orthogonal, and symplectic ensembles. In particular, fixed-boundary reversal admits an exact symplectic-character representation, while the same framework provides access to boundary-sensitive behavior at the R\'enyi transition. Our results establish a new connection between nonstabilizerness and boundary criticality through the exact statistical structures of classical-group ensembles.
\end{abstract}

\maketitle

At a quantum critical point, boundaries and defects possess universal physics of their own, beyond what is determined by bulk critical exponents~\cite{AffleckLudwig1991,AffleckReview2009,Cardy1984,Cardy1989}. Boundary conformal field theory (BCFT) characterizes this physics through boundary states, boundary entropies, and boundary-condition-changing operators, while entanglement studies provide access to boundary renormalization-group flows~\cite{Holzhey1994,Vidal2003,CalabreseCardy2004,CalabreseCardy2009,Laflorencie2006,CornfeldSela2017,XavierRajabpour2020}, mixed-boundary configurations~\cite{Estienne2025}, and conformal interfaces and junctions~\cite{SakaiSatoh2008,BrehmBrunner2015,GutperleMiller2017}. Wave-function statistics offer a complementary perspective, as demonstrated by universal contributions to participation and Shannon--R\'enyi entropies and their dependence on measurement basis and replica geometry~\cite{StephanEtAl2009,StephanMisguichPasquier2010,LuitzAletLaflorencie2014,AlcarazRajabpour2013,Stephan2014,AlcarazRajabpour2014,TarighiEtAl2022}. These developments suggest that the statistical structure of quantum many-body states may reveal boundary information not exhausted by their leading scaling signatures. This raises a further question: can the statistics of quantum observables resolve microscopic distinctions within the same boundary universality class and, more importantly, uncover finite responses arising from the interplay between the two ends of a critical system rather than from independent endpoint contributions?

Pauli expectation values provide a natural setting for addressing this question. Their moments define stabilizer R\'enyi entropies (SREs), which quantify nonstabilizerness, a computational resource distinct from entanglement~\cite{OlivieroLeoneHamma2022,Gottesman1997,BravyiKitaev2005,Veitch2012,Howard2014,Veitch2014,HaugPiroliMonotones2023,LeoneBittel2024,LamiCollura2024,ColluraEtAl2026}. Studies of critical spin chains, interacting many-body systems, and quantum spin liquids have revealed distinctive critical behavior and transitions of nonstabilizerness~\cite{Sarkar2020,White2021,LeoneIsing2022,TarabungaCritical2024,ZhangZhouSun2026,LammaBauerDalmonteCollura2026}, while Clifford disentangling has clarified its relationship with entanglement~\cite{FanEtAl2025,FrauEtAl2025,FuxEtAl2024} and recent studies have explored spectral diagnostics~\cite{Korbany2025,Turkeshi2025,Hallam2026}, thermal and off-critical crossovers~\cite{KhassehRajabpour2026,KhassehRamirezTrinoRajabpour2026}, and topological boundary responses~\cite{RajabpourBoundary2026}. Quantum measurement protocols~\cite{OlivieroProcessor2022,HaugLeeKim2024}, tensor-network formulations~\cite{HaugPiroli2023,TarabungaMPS2024}, and Pauli-sampling techniques~\cite{LamiCollura2023,DingWangYan2025} have further made these observables accessible in increasingly complex many-body settings. Yet each stabilizer R\'enyi moment compresses an exponentially large distribution of Pauli amplitudes into a single number. The complete Pauli spectrum contains richer statistical information~\cite{Turkeshi2025,TurkeshiSierantTarabunga2026,Hallam2026,LiChang2026}, raising the possibility of understanding how its underlying distribution changes under physical boundary operations.

Recent developments have connected SRE with boundary conformal field theory (BCFT) through the effective boundary condition imposed by Pauli/Bell measurements in a replicated theory~\cite{HoshinoOshikawaAshida2026,MatsudaHoshinoAshida2026}. For a critical chain with conformal boundaries $A$ and $B$, the boundary-resolved SRE has, within a given replica-boundary regime, the scaling structure
\begin{equation}
\cM_{\alpha,L}^{AB}=m_\alpha L
+\frac{\gamma_\alpha^{AB}}{\alpha-1}\ln L
+C_\alpha^{AB}+o(1),
\label{eq:BCFT-finite-C-scaling}
\end{equation}
where the logarithmic term is governed by replicated-BCFT corners and boundary-condition-changing operators. Remarkably, all nine elementary Ising boundary combinations are predicted to share the same logarithmic SRE correction within the corresponding replica-boundary regime, while their finite $O(1)$ contributions were left undetermined in that analysis~\cite{HoshinoAshida2026}. These finite terms can retain information about the relative physical boundary configuration, but may also contain nonuniversal endpoint contributions. Can suitable combinations isolate finite boundary-sensitive responses while eliminating independent local endpoint normalizations? The effective Pauli-measurement boundary can itself undergo a R\'enyi-index transition, modifying the logarithmic scaling and the associated boundary data~\cite{RamirezTrinoRajabpour2026}.

Here we show that the finite part of stabilizer R\'enyi entropy responds to how critical boundaries are combined, even when their logarithmic corrections coincide. Critical transverse-field Ising chains provide an exact setting for this boundary spectroscopy, while three microscopic terminations sharing the same free--free Ising boundary condition in BCFT retain distinct fingerprints in their complete Pauli-magnitude distributions. The striking feature is the exact collective organization of these fingerprints: individual nonzero Pauli magnitudes map term by term to weighted Weyl alternants, whose moments generate discrete Selberg ensembles associated with the root systems $A$, $B$, $C$, and $D$ of the classical unitary, orthogonal, and symplectic Lie groups. The corresponding interacting-particle ensembles organize the full Pauli spectrum through direct and reflected endpoint collisions and distinct trigonometric wall structures, with the three free--free terminations realizing the $B$, $C$, and $D$ families. Extending the earlier periodic $A$-type construction~\cite{Companion}, symmetry-breaking boundary fields produce further structures within the $C$ family, including rectangular, pinned, and character-inserted ensembles.

The same framework gives exact access to finite boundary responses. From four physical open-boundary sectors, we construct two independent ratios of Pauli moments whose associated SRE responses isolate normalization-independent combinations of the finite coefficients $C_\alpha^{AB}$. One compares free and fixed boundary configurations; the other resolves the response to reversing the relative sign of fixed boundary fields. Remarkably, the latter operation inserts the fundamental character of a symplectic group, so that the opposite- to equal-fixed boundary-moment ratio becomes an exact classical-group character expectation at arbitrary finite size and R\'enyi index. Exact special-index results and conditional thermodynamic formulas indicate a distinctive R\'enyi-index dependence: the free/fixed SRE combination is predicted to be independent of $\alpha$ throughout the unlocked regime, while the opposite-fixed response is predicted to retain nontrivial replica dependence; the candidate marginal limits differ from the unlocked continuations. Together, these quantities provide normalization-independent microscopic targets for finite boundary and defect--measurement-boundary junction data in replicated BCFT. Pauli spectra thus provide two complementary forms of boundary spectroscopy: they resolve exact boundary fingerprints within a common infrared boundary universality class and give analytical access to finite responses associated with physical boundary changes.

To establish this boundary spectroscopy, we begin with boundary-resolved Pauli moments. For a pure state on $L$ spins, the unnormalized Pauli moment and SRE are
\begin{equation}
\begin{aligned}
\cZ_{\alpha,L}^{\mathsf b}
&=\sum_{P\in\cP_L}\abs{\langle P\rangle_{\mathsf b}}^{2\alpha},
&&\alpha>0,\\
\cM_{\alpha,L}^{\mathsf b}
&=\frac{\log\!\left(2^{-L}\cZ_{\alpha,L}^{\mathsf b}\right)}{1-\alpha},
&&\alpha\ne1.
\end{aligned}
\label{eq:def}
\end{equation}
Here $\cP_L=\{I,X,Y,Z\}^{\otimes L}$ denotes the phase-free Pauli strings and $\mathsf b$ labels the physical boundary condition; at $\alpha=1$ the continuous extension is understood. The stabilizer-resource foundations~\cite{Gottesman1997,BravyiKitaev2005,Veitch2012,Howard2014,Veitch2014} and subsequent monotonicity and structural results~\cite{HaugPiroliMonotones2023,LeoneBittel2024,LamiCollura2024,ColluraEtAl2026} provide the quantum-information setting. We extend the periodic checkerboard Selberg construction~\cite{Companion} to resolve physical boundary fingerprints and boundary-changing responses.

\begin{figure*}[t]
\centering
\includegraphics[width=\textwidth]{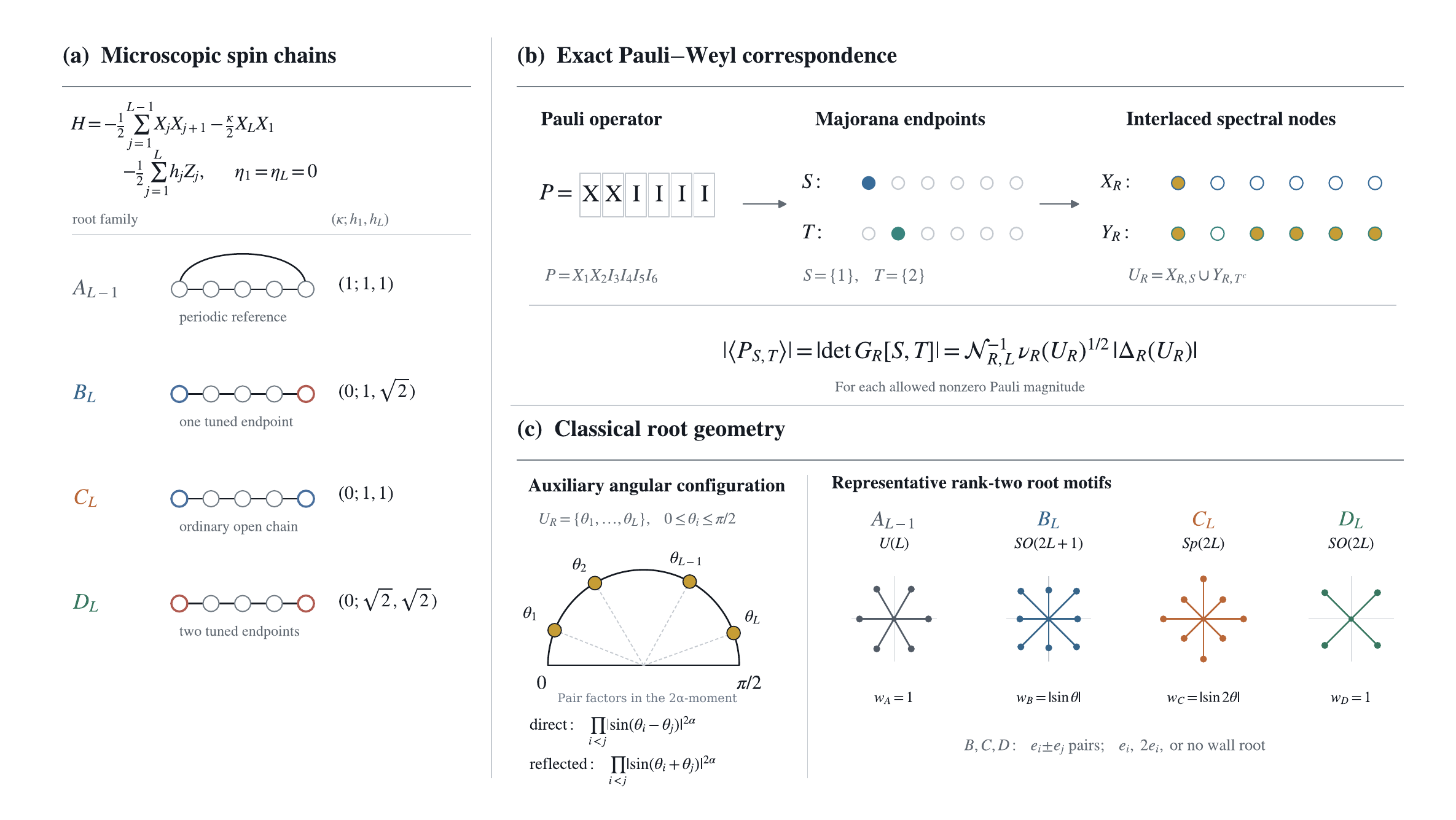}
\caption{Microscopic spin-chain terminations and the exact Pauli--Weyl correspondence.
(a) Critical Ising chains with vanishing longitudinal endpoint fields: the periodic
$A_{L-1}$ reference ($\kappa=1$) and open $B_L$, $C_L$, and $D_L$ terminations
($\kappa=0$), with transverse endpoint fields $(h_1,h_L)=(1,\sqrt2)$,
$(1,1)$, and $(\sqrt2,\sqrt2)$, respectively. The three open realizations share
the infrared free--free Ising boundary condition.
(b) An illustrative Pauli operator $P=X_1X_2I_3I_4I_5I_6$ determines balanced
Majorana index sets $S=\{1\}$ and $T=\{2\}$. Filled discs indicate selected
indices; the union $U_R(S,T)=X_{R,S}\cup Y_{R,T^c}$ selects $L$ angular nodes
(gold discs) from two interlaced grids. Each checkerboard-allowed nonzero Pauli
magnitude equals a weighted Weyl alternant, as shown by the termwise identity.
The factors $\nu_R$ are discrete grid weights, distinct from the wall factors below.
(c) The auxiliary angular nodes, not physical particles of the chain, carry
direct and reflected pair factors; $w_B(\theta)=|\sin\theta|$,
$w_C(\theta)=|\sin2\theta|$, and $w_D(\theta)=1$ distinguish the interval wall
geometries. Radial motifs schematically depict representative rank-two roots.
The group labels classify
the alternant geometry, not physical internal symmetries.}
\label{fig:pauli_weyl_geometry}
\end{figure*}

\textit{Boundary fingerprints and Pauli root geometry.--}
Both forms of boundary spectroscopy arise within the local Hamiltonian family
\begin{equation}
\begin{aligned}
H(\kappa;\bm h,\bm\eta)={}&-\frac12\sum_{j=1}^{L-1}X_jX_{j+1}
-\frac{\kappa}{2}X_LX_1
-\frac12\sum_{j=1}^{L}h_jZ_j\\
&-\frac12\left(\eta_1X_1+\eta_LX_L\right),
\end{aligned}
\label{eq:H}
\end{equation}
with $h_j=1$ away from the endpoints. Two physically distinct endpoint operations underlie the spectroscopy. Tuning the transverse fields $h_1,h_L$ preserves Ising parity and changes the microscopic Pauli reflection geometry within the infrared free--free boundary class; longitudinal fields $\eta_1,\eta_L$ instead break Ising parity and generate physical free-to-fixed boundary flows~\cite{Pfeuty1970,Campostrini2015,XavierRajabpour2020}. Figure~\ref{fig:pauli_weyl_geometry} connects the endpoint terminations to classical root families and follows individual Pauli strings through Majorana minors and interlaced spectral nodes to weighted Weyl alternants. Hamiltonian conventions, parity sectors, and the treatment of zero modes are given in Sec.~\ref{sec:S1} of the Supplemental Material (SM) below.

The mechanism can be stated without diagonalizing every row separately. Using  Jordan--Wigner transformation, Eq.~\eqref{eq:H} becomes $H=-(\ii/2)\bm b^{\mathsf T}B\bm a$~\cite{JordanWigner1928,LiebSchultzMattis1961}; the critical bulk is an alternating self-dual Majorana path. Standard correlation-matrix methods for quadratic ground states~\cite{Peschel2003,PeschelEisler2009} give the polar matrix $G=B(B^{\mathsf T}B)^{-1/2}$; with the Hamiltonian convention in Eq.~\eqref{eq:H}, the physical block $-\ii\langle b_j a_k\rangle$ is $-G$. Since only absolute minors enter, every checkerboard-allowed Pauli magnitude can be labelled by equally sized endpoint sets $S,T$ and obeys $\abs{\langle P_{S,T}\rangle}
 =\abs{\det G[S,T]}$.
Thus free-fermion solvability evaluates a single Pauli string, whereas SRE demands a power sum over all balanced minors. The root-system result is stronger: it identifies each minor with one many-particle alternant on a reflected lattice, so the correspondence holds before the sum over Pauli strings is taken.

Under Jordan--Wigner transformation, Pauli strings yield balanced fermionic order/disorder endpoints; coincident endpoints force antisymmetric minors to vanish. Periodicity supplies direct factors associated with $e_i-e_j$ and the $A_{L-1}$ denominator. Unfolding an interval adds reflected factors $e_i+e_j$ and, depending on the termination, wall roots $e_i$, $2e_i$, or none. Up to node-dependent measure factors, the interval alternants are
\begin{equation}
\begin{aligned}
\Delta_R(\bm\theta)={}&
\prod_{i<j}\sin(\theta_i-\theta_j)\sin(\theta_i+\theta_j)\\
&\times\prod_i w_R(\theta_i),\qquad R=B,C,D.
\end{aligned}
\label{eq:weyl}
\end{equation}
Here $w_B(\theta)=\sin\theta$, $w_C(\theta)=\sin2\theta$, and
$w_D(\theta)=1$ denote the one-body Weyl wall factors, distinct from
the discrete grid weights $\nu_R$ below.
\Needspace{4\baselineskip}
Thus closing the chain gives $A_{L-1}$; ordinary open termination gives $C_L$; tuning one endpoint transverse field from $1$ to $\sqrt2$ gives $B_L$; tuning both gives $D_L$. These labels classify the trigonometric Weyl-wall factors of the Pauli amplitudes, not internal symmetries of $H$; in particular, $B_L$ and $C_L$ share the same finite linear reflection hyperplanes but differ in root-length and wall-zero structure.

Transverse endpoint tuning selects the $B/C/D$ wall factors of square-minor ensembles, whereas longitudinal boundary fields produce rectangular, pinned, and character-inserted structures within the $C$ family. Their complete microscopic parameter assignments and minor structures are listed in Table~\ref{tab:classification} of \hyperlink{end:A}{End Matter A}. To test the microscopic robustness of boundary responses, we also consider longitudinal boundary fields in the endpoint-tuned $B/C/D$ chains. We denote the corresponding moments by $\cZ_{\alpha,L}^{\mathsf b|R}$, $R=B,C,D$; their finite-rank construction is given in \hyperref[sec:S5-folded-fixed]{Sec.~S5.E} of the SM.

The exact singular-value decompositions produce weighted Cauchy kernels. For the configuration $U_R(S,T)$ obtained by replacing selected nodes of one interlaced grid by nodes of the other, Cauchy determinants and complementary-root identities give the termwise relation
\begin{equation}
 \abs{\det G_R[S,T]}
 =\mathcal N_{R,L}^{-1}
 \nu_R\!\left(U_R(S,T)\right)^{1/2}
 \abs{\Delta_R\!\left(U_R(S,T)\right)}.
\label{eq:termwise}
\end{equation}
Here $\mathcal N_{R,L}$ is the finite-grid normalization factor specified in the SM.
It follows, for every real $\alpha>0$ and checkerboard fugacity $u$, that
\begin{equation}
\begin{aligned}
\cZ_{\alpha,L}^{R}(u)
={}&\mathcal N_{R,L}^{-2\alpha}
\sum_{U\in\mathfrak C_{R,L}}
\nu_R(U)^{\alpha}\,u^{k_R(U)}\\
&\times\abs{\Delta_R(U)}^{2\alpha},
\qquad R=A,B,C,D.
\end{aligned}
\label{eq:Selberg}
\end{equation}
Here $\mathfrak C_{R,L}$ is the appropriate half-filled reflected grid, $\nu_R$ contains fixed-node quadrature weights, and $k_R$ resolves the Pauli endpoint degree. Together with the exact selection-rule zeros, Eq.~\eqref{eq:termwise} therefore classifies the complete Pauli-magnitude distribution, not only its total moment. These discrete Coulomb-gas/Selberg sums connect to inverse-square and counting structures~\cite{Sutherland1971,Haldane1988,Shastry1988,Gaudin1973,Mehta1975,Tsukerman2017,StephanPollmann2017} and are tractable at special indices through Cauchy--Binet, de Bruijn Pfaffians, and finite-Fourier constant-term identities~\cite{Selberg1944,ForresterWarnaar2008,deBruijn1955,Dyson1962,Morris1982,Macdonald1982,Stanley1989,Kadell1994}. The explicit one-particle identities, alias representations, boundary extensions, and their all-rank status are given in the SM below; the periodic $A$-type construction is given in Ref.~\cite{Companion}. 

The three open parity-preserving chains provide a useful separation of scales. Their exact low-energy levels have the common Neveu--Schwarz form $\varepsilon_n=\pi(n+\tfrac12)/L_{\rm eff}+O(L^{-3})$, with $L_{\rm eff}=L+\tfrac12$, $L$, and $L-\tfrac12$ for $C$, $B$, and $D$, respectively. All three therefore realize the free--free Ising strip, differing only by an $O(1)$ extrapolation-length shift and irrelevant corrections. Standard adjacent-interval entanglement consequently has the same leading boundary-CFT data, $c=1/2$ and the same free-boundary factor $g_f$~\cite{CalabreseCardy2009,AffleckLudwig1991,AffleckReview2009}. Despite their identical leading boundary-CFT data, the three Pauli distributions retain distinct $C_L$, $B_L$, and $D_L$ wall geometries, each organized by an exact many-endpoint zero structure.

\textit{Fixed boundaries and normalization-free ratios.--}
Having resolved microscopic fingerprints within the free--free BCFT class, we now use the exact Pauli-spectrum structure to isolate finite responses to changes of physical boundary conditions. Longitudinal boundary fields break fermion parity and generate Majorana one-point functions. Generalized Wick theory~\cite{BalianBrezin1969} combines the odd and even Pauli sectors by adding one auxiliary index, giving equal-size minors of an augmented rectangular matrix. For the standard open chain ($\kappa=0$, $h_1=h_L=1$), the fields $(\eta_1,\eta_L)=(0,1)$, $(1,1)$, and $(1,-1)$ produce $\fplus$, $\pp$, and $\pmfix$ sectors, respectively (Table~\ref{tab:classification} in \hyperlink{end:A}{End Matter A}). Free--fixed boundaries produce a rectangular $C_{L+1}$ ensemble, the projected two-fixed construction pins one parent node, and opposite fixed signs add a one-quasiparticle update. The explicit augmented-minor and projection construction is given in Sec.~\ref{sec:S5} of the SM. An exact all-index consequence is the rank shift
\begin{equation}
\cZ_{\alpha,L}^{\pp}=\frac12\cZ_{\alpha,L+1}^{\ff},
\qquad
\cM_{\alpha,L}^{\pp}=\cM_{\alpha,L+1}^{\ff}.
\label{eq:rankshift}
\end{equation}
The factor $1/2$ in the moment is compensated by the extra-qubit normalization in the entropy. Equation~\eqref{eq:rankshift} is exact for the physical equal-fixed chain and therefore relates the corresponding asymptotic amplitudes without invoking a continuum normalization argument.

The general scaling form in Eq.~\eqref{eq:BCFT-finite-C-scaling} motivates combinations that cancel common bulk and logarithmic contributions together with independent local endpoint normalizations.

The four open moments admit two exact multiplicative combinations invariant under independent rescalings of free and fixed endpoint factors.  At finite size we define
\begin{equation}
\begin{aligned}
\cR_{\alpha,L}^{(\eta)}
&=\frac{\cZ_{\alpha,L}^{\pmfix}}{\cZ_{\alpha,L}^{\pp}},\\
\cR_{\alpha,L}^{(\times)}
&=\frac{\cZ_{\alpha,L}^{\ff}\cZ_{\alpha,L}^{\pp}}
{\left(\cZ_{\alpha,L}^{\fplus}\right)^2}.
\end{aligned}
\label{eq:ratios}
\end{equation}
The opposite-fixed ratio, labeled $(\eta)$ for the Ising $\mathbb Z_2$ symmetry defect rather than the fields $\eta_1,\eta_L$, measures the response to reversing one fixed boundary. The free/fixed cross ratio removes contributions factorizing into independent endpoint normalizations and probes the combined boundary configuration.
The endpoint-rescaling null space is two dimensional, so every multiplicative invariant of this type is generated by Eq.~\eqref{eq:ratios}; the two-constraint derivation is given in \hyperlink{end:C}{End Matter C}, with details in Sec.~\ref{sec:S8} of the SM.  We denote their thermodynamic limits, when they exist, by \(\cR_{\alpha}^{(\eta)}\) and \(\cR_{\alpha}^{(\times)}\).  For reflection-asymmetric endpoint terminations, the cancellation must distinguish the two orientations of the free--fixed sector; see \hyperlink{end:C}{End Matter C}.

The one-mode insertion takes a simple exact form. In the parent $C_N$ ensemble, $N=L+1$, it is
\begin{equation}
\Xi_N(U)=2\sum_{j\in U}\cos(2\theta_j)
=\chi_{\omega_1}^{(C_N)}(U),
\label{eq:character}
\end{equation}
the fundamental character of $\Sp(2N)$. For the projected fixed-boundary construction, the finite-size ratio is exactly
\begin{equation}
\cR_{\alpha,L}^{(\eta)}
=\frac{\cZ_{\alpha,L}^{\pmfix}}{\cZ_{\alpha,L}^{\pp}}
=\left\langle\abs{\chi_{\omega_1}^{(C_N)}}^{2\alpha}\right\rangle_{\alpha,N}^{C},
\label{eq:characteridentity}
\end{equation}
Here the brackets denote an average over the discrete symplectic ensemble; the pinned and full representations give the same result, as shown in \hyperlink{end:B}{End Matter B}. Equation~\eqref{eq:characteridentity} expresses the boundary-reversing one-mode update exactly as a classical-group character observable. The fundamental character appears because the one-particle update sums the weights $\{\pm e_j\}$, whose torus trace is $2\sum_j\cos2\theta_j$; this is not an internal group-symmetry assumption. The rank-one/Cauchy mechanism behind the character and its complement symmetry are summarized in \hyperlink{end:B}{End Matter B}; full proofs and integer-$\alpha$ Morris quotients are in Sec.~\ref{sec:S8} of the SM.

The ratios also define two finite-size SRE combinations,
$\Delta_{\alpha,L}^{(\eta)}\equiv\cM_{\alpha,L}^{\pmfix}-\cM_{\alpha,L}^{\pp}$
and
$\Delta_{\alpha,L}^{(\times)}\equiv\cM_{\alpha,L}^{\ff}
+\cM_{\alpha,L}^{\pp}-2\cM_{\alpha,L}^{\fplus}$.
For $a=\eta,\times$ (with $\alpha=1$ understood by continuity),
\begin{equation}
\Delta_{\alpha,L}^{(a)}
=\frac{\ln\cR_{\alpha,L}^{(a)}}{1-\alpha},
\qquad
\Delta_\alpha^{(a)}=\lim_{L\to\infty}\Delta_{\alpha,L}^{(a)},
\label{eq:entropy-responses}
\end{equation}
When the boundary sectors share the same bulk and logarithmic terms, their finite responses isolate the combinations
\begin{equation}
\begin{aligned}
\Delta_\alpha^{(\eta)}
&=C_\alpha^{+-}-C_\alpha^{++},\\
\Delta_\alpha^{(\times)}
&=C_\alpha^{ff}+C_\alpha^{++}-2C_\alpha^{f+}.
\end{aligned}
\label{eq:finite-C-responses}
\end{equation}
The free/fixed combination directly probes the additivity of finite boundary contributions: if $C_\alpha^{AB}=c_\alpha^A+c_\alpha^B$, then $\Delta_\alpha^{(\times)}=0$. A nonzero response therefore signals information about the joint boundary configuration that cannot be assigned independently to the two endpoints.
For the standard solvable realization, an exact finite-size product gives $\cR_2^{(\times)}=2$ in the thermodynamic limit, establishing a nonadditive boundary response at R\'enyi index two (\hyperlink{end:D}{End Matter D}).
For $0<\alpha<4$, the asymptotic analysis
(Secs.~\ref{sec:S10}--\ref{sec:S11} of the SM) gives a conditional
Gaussian-character result for $\cR_\alpha^{(\eta)}$ and a thermodynamic
conjecture for $\cR_\alpha^{(\times)}$:
\begin{equation}
\begin{aligned}
\cR_\alpha^{(\eta)}
&=\frac{2^\alpha\Gamma(\alpha+\tfrac12)}
{\sqrt\pi\,\alpha^\alpha},\\
\cR_\alpha^{(\times)}&=2^{\alpha-1}.
\end{aligned}
\label{eq:thermo-unlocked}
\end{equation}
At the marginal index $\alpha=4$, finite-rank reconstructions
and tested square-collapse relations suggest the candidate limits
\begin{equation}
\cR_4^{(\eta)}=\frac{143}{240},\qquad
\cR_4^{(\times)}=4.
\label{eq:thermo-marginal}
\end{equation}
Special-index finite-$L$ formulas are collected in
\hyperlink{end:D}{End Matter D}, and their derivations and asymptotic
analysis are in Secs.~\ref{sec:S9}--\ref{sec:S11} of the SM.

\Needspace{6\baselineskip}
The universal logarithmic SRE correction and the finite boundary response probe different aspects of the replicated boundary problem. In the folded geometry, corner scaling dimensions control the logarithmic coefficient~\cite{HoshinoAshida2026}, but do not determine the finite contribution of the full two-boundary configuration.

Motivated by this geometry, we propose that the lattice ratios have continuum counterparts built from finite junction amplitudes associated with the joint configuration of physical boundaries and the Pauli-measurement boundary $\Gamma_\alpha^{\mathrm P}$~\cite{HoshinoOshikawaAshida2026,HoshinoAshida2026}. With $\mathcal J_\alpha^{ab}\equiv\mathcal J_\alpha(a,b\mid\Gamma_\alpha^{\mathrm P})$, the proposed correspondence is
\begin{equation}
\begin{aligned}
\cR_\alpha^{(\eta)}&\longleftrightarrow
\frac{\mathcal J_\alpha^{+-}}{\mathcal J_\alpha^{++}},\\
\cR_\alpha^{(\times)}&\longleftrightarrow
\frac{\mathcal J_\alpha^{ff}\mathcal J_\alpha^{++}}{(\mathcal J_\alpha^{f+})^2}.
\end{aligned}
\label{eq:BCFT-junction-targets}
\end{equation}
The first is naturally associated with the Ising $\mathbb Z_2$ symmetry-defect sector, while the second isolates a free/fixed junction combination beyond independent endpoint factors. These are continuum interpretations of the exact lattice responses, not additional lattice identities; details are given in Sec.~\ref{sec:S12} of the SM.

The sensitivity of these observables to microscopic boundary realizations can be examined independently. When both longitudinal endpoint fields are taken with common magnitude $|\eta_b|$ to the strong fixed-boundary limit, the endpoint spins are pinned and the transverse $B/C/D$ tuning disappears, yielding the exact reduction
\begin{equation}
\left.\cR_{\alpha,L}^{(\eta)|R}\right|_{|\eta_b|\to\infty}
=\cR_{\alpha,L-2}^{(\eta)|C}
\big|_{\eta_b=1},
\qquad R=B,C,D.
\label{eq:strong-field-reduction}
\end{equation}
Here the right-hand side is the standard $C$ representative at the solvable fixed-end coupling $\eta_b=1$. The strong-field relation thus connects the different microscopic realizations to a common boundary-response representative. Direct finite-field calculations provide complementary tests, as detailed in \hyperref[sec:S12-flow]{Sec.~S12.D} of the SM.

\textit{Character fluctuations and the R\'enyi transition.--}
The candidate marginal limits in Eq.~\eqref{eq:thermo-marginal} suggest a change in character statistics at $\alpha=4$, coinciding with the stabilizer R\'enyi transition identified previously~\cite{RamirezTrinoRajabpour2026}. The unlocked continuations give $\cR_{4,\mathrm G}^{(\eta)}=105/256$ and $\cR_{4,\mathrm{unlocked}}^{(\times)}=8$, in contrast with the candidate marginal values $143/240$ and $4$. Additional Fourier aliases enter the character numerator at this index, providing an exact finite-grid mechanism consistent with the proposed departure from the unlocked behavior. The finite-grid calculations and analysis of character fluctuations are presented in Secs.~\ref{sec:S9}--\ref{sec:S11} of the SM. The emergence of these additional Fourier contributions highlights the sensitivity of finite boundary responses to the replica structure of the Pauli-measurement boundary.

\textit{Conclusion.--}
Our results reveal two complementary levels of boundary information in Pauli statistics: microscopic fingerprints retained by the complete Pauli spectrum and finite responses sensitive to the joint configuration of critical boundaries. While classical-group structures provide an exact solution for Ising chains, the normalization-independent response observables are defined directly from Pauli moments and can be investigated in other critical systems, including interacting models and along boundary renormalization-group flows. In replicated BCFT, these responses suggest finite junction amplitudes associated with physical boundaries and the Pauli-measurement interface, complementing the corner scaling dimensions governing the logarithmic correction. This perspective opens a route to exploring how boundary conditions, defects, and their interplay shape nonstabilizerness across different critical many-body systems.

\begin{acknowledgments}
M.A.R. acknowledges partial support from CNPq and FAPERJ (Grant
No.~E26/210.062/2023). We thank R. Khasseh for discussions.
\end{acknowledgments}

\begingroup
\renewcommand{\addcontentsline}[3]{}

\endgroup

\clearpage
\section*{End Matter}

\hypertarget{end:A}{}
\noindent\textbf{A. Boundary sectors and exact finite-grid ensembles.}\quad
This classification makes explicit which microscopic boundary parameters realize
our root families and which changes instead require augmented Majorana minors.
It separates parity-preserving transverse terminations from the longitudinal
field operations underlying the response observables.
\begin{table}[H]
\caption{Finite-chain sectors. Entries list
$(\kappa;h_1,h_L;\eta_1,\eta_L)$ in Eq.~\eqref{eq:H}; bulk
$h_j=1$. ``Square'' denotes balanced minors, whereas the fixed sectors
use augmented rectangular or pinned minors.}
\label{tab:classification}
\centering
\scriptsize
\setlength{\tabcolsep}{2.8pt}
\renewcommand{\arraystretch}{1.22}
\begin{tabular}{@{}l c p{0.42\columnwidth}@{}}
\toprule
Sector & Parameters & Minor structure / ensemble \\
\midrule
$P$ & $(1;1,1;0,0)$ & square; $A_{L-1}$ \\
$\ff$ & $(0;1,1;0,0)$ & square; $C_L$ \\
$B$ & $(0;1,\sqrt2;0,0)$ & square; $B_L$ \\
$D$ & $(0;\sqrt2,\sqrt2;0,0)$ & square; $D_L$ \\
\midrule
$\fplus$ & $(0;1,1;0,1)$ & augmented rectangular; $C_{L+1}$ \\
$\pp$ & $(0;1,1;1,1)$ & pinned; $C_{L+1}$ \\
$\pmfix$ & $(0;1,1;1,-1)$ & pinned one-mode update; $C_{L+1}$ with $\chi_{\omega_1}$ \\
\bottomrule
\end{tabular}
\end{table}
The augmented-matrix and fixed-boundary projections are derived in
Secs.~\ref{sec:S5} and \ref{sec:S6} of the Supplemental Material (SM).

\medskip
\hypertarget{end:B}{}
\noindent\textbf{B. From a boundary-field reversal to a symplectic character.}\quad
We outline why the fundamental $C_{L+1}$ character is forced by the
microscopic one-mode update rather than imposed as a symmetry assumption.
After factoring out identical row/column factors, the equal-fixed parent
minors have a Cauchy core $C_{R,T}=[(x_r-y_s)^{-1}]_{r\in R,s\in T}$;
reversing the right fixed field gives $C_{R,T}\mapsto
C_{R,T}-4\boldsymbol 1\boldsymbol 1^{\mathsf T}$.
For every invertible selected core, the determinant lemma and the Cauchy
inverse-sum identity yield
\begin{equation}
\begin{aligned}
\frac{\det G_N^{(1)}[R,T]}{\det G_N^{(0)}[R,T]}
&=1-4\boldsymbol1^{\mathsf T}C_{R,T}^{-1}\boldsymbol1\\
&=1-4\biggl(\sum_{r\in R}x_r-\sum_{s\in T}y_s\biggr).
\end{aligned}
\label{eq:EM-rank1}
\end{equation}
The complementary configuration $U=X_{N,R}\cup Y_{N,T^c}$ has $N=L+1$
nodes, with $x_r,y_s=\sin^2\theta$ and the complete odd-grid identity
$\sum_{s=1}^Ny_s=(2N-1)/4$. Therefore the insertion becomes
\begin{equation}
\begin{aligned}
\Phi_N(U)&=2N-4\sum_{\lambda\in U}\lambda\\
&=2\sum_{\theta\in U}\cos(2\theta)
=\chi_{\omega_1}^{(C_N)}(U).
\end{aligned}
\label{eq:EM-character}
\end{equation}
Here the full-parent average is
$\langle F\rangle_{\alpha,N}^{C}=\sum_U w(U)F(U)/\sum_U w(U)$,
where $U$ runs over $N$-node subsets of the parent grid and
$w(U)=\mathfrak V_C(U)^{2\alpha}$. Complementation preserves $w(U)$
and reverses the character's sign, so the pinned and full averages
coincide for $|\chi_{\omega_1}|^{2\alpha}$.
Detailed identities and the pinned projection are in Sec.~\ref{sec:S8} of the SM.

\medskip
\hypertarget{end:C}{}
\noindent\textbf{C. Completeness of normalization-independent ratios.}\quad
We show why the two responses in Eq.~\eqref{eq:ratios} exhaust the
multiplicative invariants under arbitrary independent local normalizations
of free and fixed endpoints. Write the moments as
$Z^{\ff},Z^{\fplus},Z^{\pp},Z^{\pmfix}$.
Under endpoint rescalings $f\mapsto xf$ and $+\mapsto y+$, their factors are
$x^2,xy,y^2,y^2$, respectively. Consequently, for
$\mathcal I=(Z^{\ff})^a(Z^{\fplus})^b(Z^{\pp})^c(Z^{\pmfix})^d$,
\begin{equation}
2a+b=0,\qquad b+2c+2d=0,
\label{eq:EM-normalization}
\end{equation}
which gives $b=-2a$, $c=a-d$ and hence
$\mathcal I=(\cR^{(\times)})^a(\cR^{(\eta)})^d$.
For a reflection-asymmetric $B$ termination, the independent left and
right endpoint factors require the orientation-resolved ratio
\begin{equation}
\cR_{\alpha,L}^{(\times)|B}
=\frac{\cZ_{\alpha,L}^{\ff|B}\cZ_{\alpha,L}^{\pp|B}}
{\cZ_{\alpha,L}^{\fplus|B}\cZ_{\alpha,L}^{\mathrm{+f}|B}},
\label{eq:Rcross-B}
\end{equation}
whereas symmetry gives $\cZ^{\fplus|R}=\cZ^{\mathrm{+f}|R}$ for $R=C,D$.
The lattice derivation and further endpoint-tuned checks are in
Sec.~\ref{sec:S8} and \hyperref[sec:S5-folded-fixed]{Sec.~S5.E} of the SM.

\medskip
\hypertarget{end:D}{}
\noindent\textbf{D. Finite-size boundary-response ratios.}\quad
Table~\ref{tab:EM-finite-ratios} lists the standard $C$-chain
results ($N=L+1$); see Secs.~\ref{sec:S9} and~\ref{sec:S11} of the SM.
\begin{table}[H]
\caption{Finite-$L$ responses. The dash denotes no closed support-counting
formula for the opposite-fixed sector at $\alpha=0$.}
\label{tab:EM-finite-ratios}
\centering
\small
\setlength{\tabcolsep}{6pt}
\renewcommand{\arraystretch}{1.18}
\begin{tabular}{@{}c c c@{}}
\toprule
$\alpha$ & $\cR_{\alpha,L}^{(\eta)}$ & $\cR_{\alpha,L}^{(\times)}$ \\
\midrule
$0$ & --- & $\dfrac{L+1}{2L+1}$ \\
$1$ & $1$ & $1$ \\
$2$ & $G_N$ & $F_L$ \\
$4$ & $H_N$ & $F_L^2$ \\
\bottomrule
\end{tabular}
\end{table}
\begin{equation}
\begin{aligned}
F_L&=\frac{2^{2L}(4L+3)(L+1)^{2L}}
{(2L+1)^L(2L+3)^{L+1}},\\
G_N&=\frac{(2N+1)(6N-7)}{(4N-5)(4N-1)},\\
H_N&=\frac{(2N+1)P_5(N)}
{15(4N-13)(4N-9)(4N-5)^2(4N-1)^2}.
\end{aligned}
\label{eq:EM-FGH}
\end{equation}
\noindent $P_5(N)=18304N^5-139008N^4+362392N^3
-375948N^2+120178N+12075$.
The marginal character expression uses $N\ge2$.
Here $F_L$ follows from all-rank products; $G_N,H_N$ are reconstructed
and checked through $N=15$, while $F_L^2$ uses a tested square-collapse
continuation; see Sec.~\ref{sec:S9} of the SM.


\clearpage
\onecolumngrid

\renewcommand{\thesection}{S\arabic{section}}
\renewcommand{\thesubsection}{\thesection.\Alph{subsection}}
\renewcommand{\thesubsubsection}{\thesubsection.\arabic{subsubsection}}
\renewcommand{\theequation}{S\arabic{equation}}
\renewcommand{\thefigure}{S\arabic{figure}}
\renewcommand{\thetable}{S\arabic{table}}
\renewcommand{\thetheorem}{S\arabic{theorem}}
\renewcommand{\theHsection}{supp.\arabic{section}}
\renewcommand{\theHsubsection}{supp.\arabic{section}.\Alph{subsection}}
\renewcommand{\theHsubsubsection}{supp.\arabic{section}.\Alph{subsection}.\arabic{subsubsection}}
\renewcommand{\theHequation}{supp.\arabic{equation}}
\renewcommand{\theHfigure}{supp.\arabic{figure}}
\renewcommand{\theHtable}{supp.\arabic{table}}
\setcounter{section}{0}
\setcounter{subsection}{0}
\setcounter{subsubsection}{0}
\setcounter{equation}{0}
\setcounter{figure}{0}
\setcounter{table}{0}
\setcounter{theorem}{0}
\setcounter{secnumdepth}{3}
\setcounter{tocdepth}{3}

\begin{center}
{\Large\bfseries Supplemental Material}\par\vspace{0.6em}
{\large ``Pauli Spectra as a Boundary Spectroscope in Quantum Spin Chains''}\par\vspace{0.6em}
{\normalsize M. A. Rajabpour}\par
\end{center}
\vspace{0.8em}

\begin{center}
\begin{minipage}{0.94\textwidth}
\small
Sections S1--S3 specify the lattice conventions, reduce Pauli moments to
Gaussian minors, and give the one-particle solutions needed for the boundary
problem.  Sections S4--S6 establish the Cauchy--Weyl identities and the
corresponding discrete Selberg ensembles, including the rectangular, pinned,
and character-inserted fixed-boundary constructions.  Section S5 also records
the endpoint-tuned rectangular and dual parents used later to test the same
physical boundary sectors in the microscopic $B$ and $D$ realizations.
Sections S7--S9 give the boundary alias/constant-term representations and the
special-index evaluations, using Ref.~\cite{Companion} for periodic formulas
already derived there; the endpoint-tuned results are collected in one
separate subsection of S9.  Sections S10--S12 discuss the thermodynamic ratios,
the proposed Gaussian-character regime, and the boundary interpretation,
including both boundary-field flow and direct tests across the $B,C,D$
microscopic realizations.
\end{minipage}
\end{center}

\vspace{1mm}
\tableofcontents
\clearpage

\section{Hamiltonians, boundary sectors, and conventions}
\label{sec:S1}

This section fixes the conventions used throughout the Supplemental Material.  In particular, it distinguishes three operations that are easy to
confuse: closing the chain, tuning parity-preserving transverse endpoint
fields, and applying parity-breaking longitudinal endpoint fields.  The first
two change the finite reflection grid without changing the Ising order
parameter at the boundary; the third realizes free-to-fixed boundary flows and
requires the augmented Gaussian formalism developed in Sec.~\ref{sec:S2} and
completed microscopically in Sec.~\ref{sec:S5}.

\subsection{Spin Hamiltonian and boundary parameters}

We use $L$ spin-$\tfrac12$ degrees of freedom with sites
$j=1,\ldots,L$ and Pauli operators $X_j,Y_j,Z_j$.  The common Hamiltonian is
\begin{equation}
\begin{aligned}
H(\kappa;\bm h,\bm\eta)
={}&-\frac12\sum_{j=1}^{L-1}X_jX_{j+1}
-\frac{\kappa}{2}X_LX_1
-\frac12\sum_{j=1}^{L}h_jZ_j
\\
&-\frac12\left(\eta_1X_1+\eta_LX_L\right).
\end{aligned}
\label{eq:S-master-H}
\end{equation}
The bulk critical point is fixed by
\begin{equation}
h_j=1,
\qquad 2\le j\le L-1.
\label{eq:S-bulk-critical}
\end{equation}
The remaining parameters have distinct meanings:
\begin{itemize}
\item $\kappa=1$ closes the chain, whereas $\kappa=0$ gives an interval;
\item $h_1$ and $h_L$ are transverse endpoint fields.  They preserve the
      global Ising parity and change the reflection grid and extrapolation
      length;
\item $\eta_1$ and $\eta_L$ are longitudinal endpoint fields.  They break the
      global Ising parity and couple directly to the boundary order parameter.
\end{itemize}
The normalizations in Eq.~\eqref{eq:S-master-H} match the Letter.  Any simultaneous
row or column sign changes induced by alternative Jordan--Wigner gauges leave
all absolute minors, Pauli moments, and stabilizer R\'enyi entropies unchanged.

The seven sectors used in the Letter are listed in Table~\ref{tab:S-sectors}.
The labels $B,C,D$ refer to the exact Weyl geometry of the Pauli amplitudes,
not to internal symmetries of the spin Hamiltonian.  The labels
$\ff,\fplus,\pp,\pmfix$ refer to physical Ising boundary conditions.

\begin{table}[H]
\caption{Boundary sectors and parameter choices in
Eq.~\eqref{eq:S-master-H}.  The endpoint fields not displayed explicitly are equal
to their bulk value one.}
\label{tab:S-sectors}
\centering
\small
\renewcommand{\arraystretch}{1.22}
\setlength{\tabcolsep}{5.5pt}
\begin{tabular}{@{}l c c c p{0.43\textwidth}@{}}
\toprule
Sector & $\kappa$ & $(h_1,h_L)$ & $(\eta_1,\eta_L)$ & Physical role \\
\midrule
$\PBC$ & $1$ & $(1,1)$ & $(0,0)$ & periodic critical chain \\
$C\equiv\ff$ & $0$ & $(1,1)$ & $(0,0)$ & standard free--free interval \\
$B$ & $0$ & $(1,\sqrt2)$ & $(0,0)$ & one transverse endpoint folded \\
$D$ & $0$ & $(\sqrt2,\sqrt2)$ & $(0,0)$ & two transverse endpoints folded \\
\midrule
$\fplus$ & $0$ & $(1,1)$ & $(0,+1)$ & free--fixed representative \\
$\pp$ & $0$ & $(1,1)$ & $(+1,+1)$ & equal fixed signs \\
$\pmfix$ & $0$ & $(1,1)$ & $(+1,-1)$ & opposite fixed signs \\
\bottomrule
\end{tabular}
\end{table}

A global spin flip
\begin{equation}
\mathcal P_z=\prod_{j=1}^{L}Z_j
\label{eq:S-global-parity}
\end{equation}
changes $++\leftrightarrow --$ and $+-\leftrightarrow-+$.  Absolute Pauli
expectation values are therefore identical within each pair.  We use $++$ and
$+-$ as representatives.

\subsection{Transverse endpoint tuning versus fixed-spin boundaries}

The special values $h_{1,L}=\sqrt2$ in the $B$ and $D$ chains do not impose a
fixed Ising spin.  The perturbation
\begin{equation}
-\frac12 Z_{\rm edge}
\longrightarrow
-\frac1{\sqrt2}Z_{\rm edge}
\label{eq:S-transverse-tuning}
\end{equation}
is even under Eq.~\eqref{eq:S-global-parity}.  It changes the finite standing-wave
grid and the normalization of reflection-fixed modes, but it does not couple
to the order parameter $X_{\rm edge}$.  The chains $B,C,D$ consequently flow
to the same free--free Ising boundary conformal field theory, as established
from their common half-integer low-energy tower in Sec.~\ref{sec:S3}.

By contrast, the terms $-\eta X_{\rm edge}/2$ are odd under
Eq.~\eqref{eq:S-global-parity}.  They generate Majorana one-point functions and
produce fixed physical boundaries.  The values $\eta=\pm1$ are exactly
solvable lattice representatives of the fixed Cardy boundary conditions; they
should not be interpreted as an infinite boundary field~\cite{Cardy1984,Cardy1989,AffleckLudwig1991,Campostrini2015}.  The relation to
more general boundary fields and their RG basin are discussed in Sec.~\ref{sec:S12}.

For the robustness tests in Secs.~\ref{sec:S9} and~\ref{sec:S12} we also
retain the transverse endpoint tunings that define the $B$ and $D$ free
chains while switching on the same longitudinal fields.  These auxiliary
families are not new Cardy boundary conditions or new fixed-boundary root
systems: the label $B$ or $D$ records only the microscopic transverse
termination from which the physical free/fixed sector is reached.

\subsection{Periodic parity sector}

For periodic spin boundary conditions, the Jordan--Wigner transformation
introduces a boundary term proportional to the fermion parity.  For even $L$
and positive couplings, the unique finite-size ground state at the critical
point lies in the even spin-parity sector, corresponding to antiperiodic
Neveu--Schwarz fermions.  We therefore use momenta
\begin{equation}
q_m=\frac{2\pi}{L}\left(m+\frac12\right),
\qquad m=0,1,\ldots,L-1.
\label{eq:S-NS-momenta}
\end{equation}
The half shift in Eq.~\eqref{eq:S-NS-momenta} is essential: replacing it by a
periodic fermion grid changes the finite state and does not reproduce the
Pauli kernel used in the Letter.  Odd $L$ and other parity sectors can be
handled separately, but they are not required for the present classification.

\subsection{Left- and right-oriented Majoranas}

For parity-preserving chains it is convenient to orient Jordan--Wigner strings
from the left~\cite{JordanWigner1928,LiebSchultzMattis1961,Pfeuty1970}:
\begin{equation}
a_j^{\rm L}=\left(\prod_{\ell<j}Z_\ell\right)X_j,
\qquad
b_j^{\rm L}=\left(\prod_{\ell<j}Z_\ell\right)Y_j.
\label{eq:S-left-JW}
\end{equation}
They obey
\begin{equation}
\{a_j^{\rm L},a_k^{\rm L}\}
=\{b_j^{\rm L},b_k^{\rm L}\}=2\delta_{jk},
\qquad
\{a_j^{\rm L},b_k^{\rm L}\}=0,
\label{eq:S-Majorana-algebra}
\end{equation}
and
\begin{equation}
Z_j=-\ii a_j^{\rm L}b_j^{\rm L},
\qquad
X_jX_{j+1}=-\ii b_j^{\rm L}a_{j+1}^{\rm L}.
\label{eq:S-spin-Majorana-identities}
\end{equation}
After the harmless staggered gauge
$a_j^{\rm L}\mapsto(-1)^j a_j^{\rm L}$ and
$b_j^{\rm L}\mapsto(-1)^j b_j^{\rm L}$, every parity-preserving open chain
can be written as
\begin{equation}
H=-\frac{\ii}{2}\sum_{j,k=1}^{L}b_j B_{jk}a_k,
\qquad
B_{jk}=h_j\delta_{jk}+\delta_{k,j+1}.
\label{eq:S-open-B}
\end{equation}
Here and below the superscript ${\rm L}$ and the staggered tildes are omitted
once the convention has been fixed.  The sign gauge affects neither singular
values nor absolute minors.

For longitudinal fields at the right endpoint, the opposite orientation is
more economical:
\begin{equation}
a_j^{\rm R}=X_j\left(\prod_{\ell>j}Z_\ell\right),
\qquad
b_j^{\rm R}=Y_j\left(\prod_{\ell>j}Z_\ell\right).
\label{eq:S-right-JW}
\end{equation}
In this convention $X_L=a_L^{\rm R}$ is linear.  A real ground state may have
nonzero $\avg{a_j^{\rm R}}$ but vanishing $\avg{b_j^{\rm R}}$.  This is the
origin of the augmented rectangular matrix in Sec.~\ref{sec:S2}.  Equal fixed
fields at both endpoints require an ancillary or dual representation; its explicit construction is given in Sec.~\ref{sec:S5}.  The general
minor theorem itself does not depend on which orientation is used.

\subsection{Pauli distribution, raw moments, and fugacity}

For a normalized pure state $\ket\psi$, define the Pauli probability~\cite{OlivieroLeoneHamma2022,LamiCollura2023}
\begin{equation}
p_\psi(P)=2^{-L}\abs{\avg{\psi|P|\psi}}^2,
\qquad P\in\cP_L,
\label{eq:S-Pauli-probability}
\end{equation}
where $\cP_L=\{I,X,Y,Z\}^{\otimes L}$ without phases.  Throughout this
Supplement, ``Pauli distribution'' refers to this probability distribution,
or equivalently to the multiset of magnitudes $\{\abs{\avg P}\}$; the
signs of individual Pauli expectation values are not reconstructed by the
absolute-minor identities and are not required by the SRE.  Pauli orthogonality
implies
\begin{equation}
\sum_{P\in\cP_L}p_\psi(P)=1.
\label{eq:S-Pauli-normalization}
\end{equation}
We use the unnormalized moment
\begin{equation}
\cZ_{\alpha,L}^{\mathsf b}(1)
=\sum_{P\in\cP_L}\abs{\avg{P}_{\mathsf b}}^{2\alpha},
\qquad \alpha>0,
\label{eq:S-raw-moment}
\end{equation}
and the stabilizer R\'enyi entropy
\begin{equation}
\cM_{\alpha,L}^{\mathsf b}
=\frac{1}{1-\alpha}
\log\left[2^{-L}\cZ_{\alpha,L}^{\mathsf b}(1)\right],
\qquad \alpha\ne1.
\label{eq:S-SRE}
\end{equation}
Writing $H_\alpha(p)=(1-\alpha)^{-1}\log\sum_P p(P)^\alpha$, direct
substitution into Eq.~\eqref{eq:S-SRE} gives
\[
 \cM_{\alpha,L}^{\mathsf b}=H_\alpha(p_{\mathsf b})-L\log2.
\]
Hence the continuous extension at $\alpha=1$ is
\[
 \cM_{1,L}^{\mathsf b}=H(p_{\mathsf b})-L\log2,
\]
not the unshifted Shannon entropy.  The additive $-L\log2$ cancels in
same-size boundary differences and in the Shannon junction quantities used
below.

A fugacity $u$ resolves the size of the balanced or augmented minor:
\begin{equation}
\cZ_{\alpha,L}^{\mathsf b}(u)
=\sum_{k\ge0}u^k W_{\alpha,L}^{\mathsf b}(k).
\label{eq:S-fugacity-definition}
\end{equation}
For parity-even states, $k$ counts both the selected $a$ and selected $b$
Majoranas.  For a linear-Gaussian state, the same $k$ labels an equal-size
minor of the augmented matrix: it combines a physical $2k$-Majorana sector
and a physical $(2k-1)$-Majorana sector containing the auxiliary Wick index.
A refinement by the literal physical Majorana degree would require a second
fugacity and is not used in the Letter.

\section{Pauli strings as Gaussian minors}
\label{sec:S2}

The reduction of the complete Pauli distribution to a sum over minors is the
first nontrivial step of the analysis.  Free-fermion solvability alone gives a
compact expression for an individual correlation function; it does not imply
that the exponentially many Pauli expectation values are organized by one
matrix.  This section proves that organization for parity-even Gaussian
states, states the corresponding augmented theorem for Gaussian states with a
linear part, and verifies both constructions directly in the spin Hilbert
space for every boundary sector used in the Letter.

\subsection{Pauli strings and Majorana subsets}

The $2L$ Jordan--Wigner Majoranas generate the full operator algebra.  To make
the correspondence explicit, let
\begin{equation}
\gamma_{2j-1}=a_j,
\qquad
\gamma_{2j}=b_j,
\qquad j=1,\ldots,L.
\label{eq:S-gamma-order}
\end{equation}
For a binary vector $\bm n=(n_1,\ldots,n_{2L})\in\{0,1\}^{2L}$, define the
ordered monomial
\begin{equation}
\Gamma(\bm n)=\gamma_1^{n_1}\gamma_2^{n_2}\cdots\gamma_{2L}^{n_{2L}}.
\label{eq:S-Majorana-monomial}
\end{equation}
Because the Majoranas form a Clifford algebra, the $2^{2L}=4^L$ monomials
Eq.~\eqref{eq:S-Majorana-monomial} form an orthogonal basis of the operator space.
The phase-free Pauli strings also form an orthogonal basis of the same size.
Consequently there is a bijection
\begin{equation}
P\in\cP_L
\quad\longleftrightarrow\quad
P=\omega(P)\Gamma(\bm n_P),
\qquad
\omega(P)\in\{\pm1,\pm\ii\}.
\label{eq:S-Pauli-Majorana-bijection}
\end{equation}
The phase is fixed by Hermiticity and drops out of every absolute expectation
value.  Thus the complete Pauli spectrum can be obtained by evaluating all
Majorana monomials once, with no overcounting.

For later use, separate the selected indices into
\begin{equation}
S=\{j:b_j\text{ occurs}\},
\qquad
T=\{k:a_k\text{ occurs}\}.
\label{eq:S-ST-definition}
\end{equation}
Up to a sign determined by the ordering convention,
\begin{equation}
\Gamma(\bm n_P)
=\left(\prod_{j\in S}b_j\right)
 \left(\prod_{k\in T}a_k\right).
\label{eq:S-ba-monomial}
\end{equation}
Only the absolute value of its expectation will be needed.

\subsection{Parity-even quadratic ground states}

Consider an invertible real coupling matrix $B$ in
Eq.~\eqref{eq:S-open-B} and its singular-value decomposition
\begin{equation}
B=U\Sigma V^{\mathsf T},
\qquad
\Sigma=\diag(\varepsilon_1,\ldots,\varepsilon_L),
\qquad \varepsilon_m>0.
\label{eq:S-general-SVD}
\end{equation}
Define rotated Majoranas
\begin{equation}
\widetilde b_m=\sum_{j=1}^{L}U_{jm}b_j,
\qquad
\widetilde a_m=\sum_{k=1}^{L}V_{km}a_k.
\label{eq:S-rotated-Majoranas}
\end{equation}
Then
\begin{equation}
H=-\frac{\ii}{2}\sum_{m=1}^{L}
\varepsilon_m\widetilde b_m\widetilde a_m.
\label{eq:S-diagonal-H}
\end{equation}
With $c_m=(\widetilde a_m+\ii\widetilde b_m)/2$,
Eq.~\eqref{eq:S-diagonal-H} is
$H=\sum_m\varepsilon_m(c_m^\dagger c_m-\tfrac12)$.  Hence the
nondegenerate ground state obeys
\begin{equation}
-\ii\avg{\widetilde b_m\widetilde a_n}=-\delta_{mn}.
\label{eq:S-mode-contraction}
\end{equation}
The physical cross-contraction block is therefore~\cite{Peschel2003,PeschelEisler2009}
\begin{equation}
\mathcal G_{jk}:=-\ii\avg{b_j a_k}=-(UV^{\mathsf T})_{jk}.
\label{eq:S-polar-G}
\end{equation}
For the determinant formulas below it is convenient to use the polar matrix
\begin{equation}
G:=UV^{\mathsf T}=B(B^{\mathsf T}B)^{-1/2},
\qquad
GG^{\mathsf T}=G^{\mathsf T}G=I_L.
\label{eq:S-polar-decomposition}
\end{equation}
Thus $\mathcal G=-G$.  A $k\times k$ minor of the physical contraction
block differs from the corresponding polar minor only by $(-1)^k$, so all
absolute-minor identities and all SRE moments are unchanged.
The off-diagonal contractions within one species vanish:
\begin{equation}
\avg{a_j a_k}=\delta_{jk},
\qquad
\avg{b_j b_k}=\delta_{jk}.
\label{eq:S-intraspecies-contractions}
\end{equation}
For products of distinct Majoranas the diagonal entries in
Eq.~\eqref{eq:S-intraspecies-contractions} never enter.  Ordering all
$b$ Majoranas before all $a$ Majoranas, the physical checkerboard block is
$\mathcal G$; equivalently the Wick contraction matrix can be written as
\begin{equation}
\Gamma=
\begin{pmatrix}
0&\mathcal G\\
-\mathcal G^{\mathsf T}&0
\end{pmatrix}.
\label{eq:S-checkerboard-Gamma}
\end{equation}

For a parity-even pure Gaussian state with checkerboard contraction block
$G$, Wick's theorem gives the standard balanced-minor reduction used in
free-fermion evaluations of SRE~\cite{OlivieroLeoneHamma2022,Companion}.  A Majorana
monomial containing unequal numbers of $b$ and $a$ operators vanishes,
\begin{equation}
\avg{
\prod_{j\in S}b_j
\prod_{k\in T}a_k}=0
\qquad\text{unless}\qquad |S|=|T|,
\label{eq:S-unbalanced-zero}
\end{equation}
whereas for $|S|=|T|=k$,
\begin{equation}
\abs{
\avg{
\prod_{j\in S}b_j
\prod_{k\in T}a_k}}
=\abs{\det\mathcal G[S,T]}
=\abs{\det G[S,T]}.
\label{eq:S-balanced-determinant}
\end{equation}
Thus the complete Pauli-magnitude moment is the all-minors sum
\begin{equation}
\cZ_{\alpha,L}(u)
=\sum_{k=0}^{L}u^k
\sum_{\substack{S,T\subseteq[L]\\|S|=|T|=k}}
\abs{\det G[S,T]}^{2\alpha}.
\label{eq:S-square-minor-partition}
\end{equation}
The Pauli--Majorana map is one-to-one, so this expression includes the full
multiplicity structure, together with any accidental zero minors.  We use
Eq.~\eqref{eq:S-square-minor-partition} as a standard starting point; its
Pfaffian derivation is not repeated here.

\subsection{Consequences of orthogonality}

At $\alpha=1$, Cauchy--Binet collapses the all-minor sum.

For every square orthogonal $G$,
\begin{equation}
\cZ_{1,L}(u)=(1+u)^L.
\label{eq:S-alpha-one-square}
\end{equation}

\noindent\emph{Derivation.} The coefficient of $u^k$ in Eq.~\eqref{eq:S-square-minor-partition} is
\begin{equation}
\sum_{\substack{|S|=|T|=k}}\det G[S,T]^2
=[t^k]\det(I_L+tG^{\mathsf T}G)
=[t^k](1+t)^L.
\label{eq:S-Cauchy-Binet-square}
\end{equation}
Summing over $k$ proves Eq.~\eqref{eq:S-alpha-one-square}.

At $u=1$, Eq.~\eqref{eq:S-alpha-one-square} gives
$\cZ_{1,L}(1)=2^L$, in agreement with the Pauli normalization
Eq.~\eqref{eq:S-Pauli-normalization}.  Jacobi's complementary-minor identity gives
\begin{equation}
\abs{\det G[S,T]}
=\abs{\det G[S^c,T^c]},
\label{eq:S-Jacobi-complement}
\end{equation}
and hence
\begin{equation}
\cZ_{\alpha,L}(u)
=u^L\cZ_{\alpha,L}(u^{-1}).
\label{eq:S-square-palindrome}
\end{equation}
The degree distribution is therefore symmetric about $L/2$ for every
$\alpha>0$.

At the formal support-counting endpoint $\alpha=0$, all minors of the Cauchy
matrices appearing in the exact chains are nonzero.  The number of supported
Pauli expectations is then
\begin{equation}
\sum_{k=0}^{L}\binom{L}{k}^2
=\binom{2L}{L}.
\label{eq:S-square-support}
\end{equation}
This is a counting statement, not a continuation of a Selberg integral to
$\alpha=0$: once the Vandermonde exponent vanishes, configurations with
coincident particles require separate treatment.

\subsection{Gaussian states with a linear part}

A longitudinal boundary field breaks fermion parity.  In a right-oriented
Jordan--Wigner convention, the exact fixed-boundary ground states are real
and have the contraction pattern
\begin{equation}
\avg{b_j}=0,
\qquad
m_k:=\avg{a_k}\ne0,
\qquad
G_{jk}:=-\ii\avg{b_j a_k},
\label{eq:S-linear-data}
\end{equation}
with off-diagonal same-species contractions equal to zero.  Such states are
Gaussian with a linear part.  Their odd correlators are governed by the
extended Wick theorem of Balian and Br\'ezin and its finite-dimensional
Majorana formulation~\cite{BalianBrezin1969}.

Introduce one formal auxiliary $b$-type index $*$ with contractions
\begin{equation}
-\ii\avg{b_*a_k}=m_k,
\qquad
\avg{b_*b_j}=0.
\label{eq:S-auxiliary-contractions}
\end{equation}
The augmented cross block is
\begin{equation}
\widehat G=
\begin{pmatrix}
G\\m^{\mathsf T}
\end{pmatrix}
\in\mathbb R^{(L+1)\times L}.
\label{eq:S-augmented-G}
\end{equation}
The formal extended contraction matrix is
\begin{equation}
\widehat\Gamma=
\begin{pmatrix}
0&\widehat G\\
-\widehat G^{\mathsf T}&0
\end{pmatrix}.
\label{eq:S-extended-checkerboard}
\end{equation}
A physical even monomial uses no auxiliary index.  A physical odd monomial
with one more $a$ than $b$ becomes even after the auxiliary row is included.

Assume a real pure Gaussian state has the linear data
Eq.~\eqref{eq:S-linear-data} and obeys the generalized Wick rule encoded by
Eq.~\eqref{eq:S-extended-checkerboard}.  Then every Pauli expectation compatible with the checkerboard selection
rule is represented by an equal-size minor of $\widehat G$, while the
selection-rule-forbidden expectations vanish.  The minor itself may also
vanish accidentally.  Consequently,
\begin{equation}
\cZ_{\alpha,L}(u)
=\sum_{k=0}^{L}u^k
\sum_{\substack{R\subseteq[L]\cup\{*\},\ T\subseteq[L]\\|R|=|T|=k}}
\abs{\det\widehat G[R,T]}^{2\alpha}.
\label{eq:S-rectangular-minor-partition}
\end{equation}
If $*\notin R$, the minor represents a physical $2k$-Majorana correlator.
If $*\in R$, it represents a physical $(2k-1)$-Majorana correlator.

\noindent\emph{Derivation.} For $*\notin R$, the proof is identical to
the parity-even reduction above.  For $*\in R$, remove the auxiliary operator
from the physical monomial after evaluating the Pfaffian.  The selected
extended matrix still has equal checkerboard block sizes $k$, so
the standard checkerboard Pfaffian--determinant identity gives the corresponding minor of
$\widehat G$.  A real state with the contraction pattern
Eq.~\eqref{eq:S-linear-data} has no nonzero monomial with more $b$ than $a$;
therefore the two cases exhaust the checkerboard-allowed Pauli sector.
Allowed minors may vanish accidentally; the remaining Pauli strings vanish by
the selection rule.  The Pauli--Majorana bijection then gives
Eq.~\eqref{eq:S-rectangular-minor-partition}.

Purity of the extended Gaussian state implies
\begin{equation}
G^{\mathsf T}G+mm^{\mathsf T}=I_L,
\qquad\text{or equivalently}\qquad
\widehat G^{\mathsf T}\widehat G=I_L.
\label{eq:S-augmented-isometry}
\end{equation}
Thus $\widehat G$ is an isometry rather than an orthogonal square matrix.
Cauchy--Binet again yields
\begin{equation}
\cZ_{1,L}(u)=(1+u)^L
\label{eq:S-alpha-one-rectangular}
\end{equation}
for every augmented isometry obeying Eq.~\eqref{eq:S-augmented-isometry}.  The physical fixed-boundary realization and its dual-parity quotient are derived in S5.  The generic support count becomes
\begin{equation}
\sum_{k=0}^{L}\binom{L+1}{k}\binom{L}{k}
=\binom{2L+1}{L}.
\label{eq:S-rectangular-support}
\end{equation}
Unlike Eq.~\eqref{eq:S-square-palindrome}, a generic rectangular degree
polynomial is not palindromic.  The missing symmetry is physically important:
it records the mixing of even and odd Majorana degrees caused by the boundary
order parameter.

the augmented-minor reduction above is the general algebraic reason why the
free--fixed, equal-fixed, and opposite-fixed Pauli distributions can be
organized by rectangular or pinned minors.  It does not by itself identify
the explicit $C_{L+1}$ grids, the pinned node, or the character insertion.
Those boundary-specific statements require the ancillary projection and dual
mode analysis carried out in Sec.~\ref{sec:S5}.

\section{Exact A-, B-, C-, and D-type one-particle solutions}
\label{sec:S3}

This section solves the one-particle Majorana problems needed in the Letter.
The periodic chain is diagonalized on the antiperiodic Neveu--Schwarz grid.
The three parity-preserving open chains are solved by exact finite sine or
cosine transforms with endpoint half weights.  The weighted Cauchy kernels
and their Weyl-denominator minors are intentionally deferred to
Sec.~\ref{sec:S4}; here we establish only the spectra, singular vectors, polar
matrices, and infrared boundary interpretation.

\subsection{Periodic baseline}

The periodic critical Ising chain provides the $A_{L-1}$ reference problem.
Its fugacity-resolved all-minors solution, including the half-shift Cauchy
kernel and special-index products, is derived in detail in
Ref.~\cite{Companion}.  For the even-$L$ Neveu--Schwarz ground-state sector
used here, the polar matrix is
\begin{equation}
G_{jk}^{(A)}
=\frac{(-1)^{j-k}}{L}
\csc\!\left[\frac\pi L\left(j-k+\frac12\right)\right],
\qquad j,k=0,\ldots,L-1,
\label{eq:S-periodic-half-shift}
\end{equation}
with $G^{(A)}G^{(A)\mathsf T}=I_L$.  We quote this result only to fix the
periodic normalization and the interlaced reference grid; the remainder of
this section concerns the boundary-dependent $B_L,C_L,D_L$ solutions.

\subsection{Unified open-chain coupling matrices}

For $R=C,B,D$, the parity-preserving open Hamiltonian has the Majorana form
Eq.~\eqref{eq:S-open-B} with an upper-bidiagonal $L\times L$ matrix.  Explicitly,
\begin{equation}
B_C=
\begin{pmatrix}
1&1&&&\\
&1&1&&\\
&&\ddots&\ddots&\\
&&&1&1\\
&&&&1
\end{pmatrix},
\label{eq:S-BC-matrix}
\end{equation}
\begin{equation}
B_B=
\begin{pmatrix}
1&1&&&\\
&1&1&&\\
&&\ddots&\ddots&\\
&&&1&1\\
&&&&\sqrt2
\end{pmatrix},
\label{eq:S-BB-matrix}
\end{equation}
and
\begin{equation}
B_D=
\begin{pmatrix}
\sqrt2&1&&&\\
&1&1&&\\
&&\ddots&\ddots&\\
&&&1&1\\
&&&&\sqrt2
\end{pmatrix}.
\label{eq:S-BD-matrix}
\end{equation}
The notation is chosen so that the eventual Weyl types are visible, but no
root-system assumption enters their diagonalization.

All three matrices admit singular-value decompositions
\begin{equation}
B_R=U^{(R)}\Sigma^{(R)}V^{(R)\mathsf T},
\qquad
\Sigma^{(R)}_{mn}=\varepsilon_m^{(R)}\delta_{mn}.
\label{eq:S-open-SVD}
\end{equation}
The exact grids are summarized in Table~\ref{tab:S-SVD-summary} and proved in the
following subsections.

\begin{table}[H]
\caption{Exact standing-wave data for the three parity-preserving open
chains.  The symbols $r,s=1,\ldots,L$ label rows and columns of $B_R$.
Endpoint factors not listed are equal to one.}
\label{tab:S-SVD-summary}
\centering
\small
\renewcommand{\arraystretch}{1.3}
\setlength{\tabcolsep}{5pt}
\begin{tabular}{@{}l c c c p{0.30\textwidth}@{}}
\toprule
Type & Grid length & Modes & Singular value & Half weights \\
\midrule
$C_L$ & $N_C=2L+1$ & $Q_m=2\pi m/N_C$, $m=1,\ldots,L$
& $2\cos(Q_m/2)$ & none \\
$B_L$ & $N_B=2L$ & $Q_m=(2m-1)\pi/N_B$, $m=1,\ldots,L$
& $2\cos(Q_m/2)$ & $\epsilon_L=2^{-1/2}$ \\
$D_L$ & $N_D=2L-1$ & $Q_m=2\pi m/N_D$, $m=0,\ldots,L-1$
& $2\cos(Q_m/2)$ & $\epsilon_L=\eta_1=\rho_0=2^{-1/2}$ \\
\bottomrule
\end{tabular}
\end{table}

\subsection{Finite trigonometric identities}

The orthogonality proofs use only finite Fourier sums.  We record the needed
relations once.  For $x\notin2\pi\mathbb Z$,
\begin{equation}
\sum_{m=1}^{M}\cos(mx)
=\frac{\sin(Mx/2)\cos[(M+1)x/2]}{\sin(x/2)},
\label{eq:S-cosine-sum}
\end{equation}
and
\begin{equation}
\sum_{m=1}^{M}\sin(mx)
=\frac{\sin(Mx/2)\sin[(M+1)x/2]}{\sin(x/2)}.
\label{eq:S-sine-sum}
\end{equation}
Together with product-to-sum formulas, these identities establish all
orthogonality relations below.

Two recurrences reconstruct the bidiagonal matrices.  For the sine families,
\begin{equation}
\sin\!\left[(s-\tfrac12)Q\right]
+\sin\!\left[(s+\tfrac12)Q\right]
=2\cos\frac Q2\sin(sQ),
\label{eq:S-sine-recurrence}
\end{equation}
and for the cosine families,
\begin{equation}
\cos[(s-1)Q]+\cos(sQ)
=2\cos\frac Q2\cos\!\left[(s-\tfrac12)Q\right].
\label{eq:S-cosine-recurrence}
\end{equation}
The special factors $2^{-1/2}$ at reflection-fixed nodes are not arbitrary
normalizations.  They are forced by finite Fourier orthogonality when an orbit
under reflection contains only one point instead of a pair.

\subsection{Standard open chain: the \texorpdfstring{$C_L$}{C(L)} solution}

Let
\begin{equation}
N_C=2L+1,
\qquad
Q_m^{(C)}=\frac{2\pi m}{N_C},
\qquad m=1,\ldots,L.
\label{eq:S-C-grid}
\end{equation}
Define
\begin{equation}
U_{rm}^{(C)}
=\frac{2}{\sqrt{N_C}}\sin(rQ_m^{(C)}),
\label{eq:S-C-U}
\end{equation}
\begin{equation}
V_{sm}^{(C)}
=\frac{2}{\sqrt{N_C}}
\sin\!\left[(s-\tfrac12)Q_m^{(C)}\right],
\label{eq:S-C-V}
\end{equation}
and
\begin{equation}
\varepsilon_m^{(C)}=2\cos\frac{Q_m^{(C)}}2.
\label{eq:S-C-epsilon}
\end{equation}

The matrices in Eqs.~\eqref{eq:S-C-U} and~\eqref{eq:S-C-V} are orthogonal and
\begin{equation}
B_C=U^{(C)}\diag(\varepsilon_m^{(C)})V^{(C)\mathsf T}.
\label{eq:S-C-SVD}
\end{equation}

\noindent\emph{Derivation.} The grid $Q_m^{(C)}$ is half of the nonzero Fourier grid on the odd cycle of
length $N_C$.  Equations~\eqref{eq:S-cosine-sum} and
\eqref{eq:S-sine-sum} give
\begin{equation}
\sum_{m=1}^{L}\sin(rQ_m^{(C)})\sin(sQ_m^{(C)})
=\frac{N_C}{4}\delta_{rs},
\label{eq:S-C-integer-orth}
\end{equation}
and
\begin{equation}
\sum_{m=1}^{L}
\sin\!\left[(r-\tfrac12)Q_m^{(C)}\right]
\sin\!\left[(s-\tfrac12)Q_m^{(C)}\right]
=\frac{N_C}{4}\delta_{rs}.
\label{eq:S-C-half-orth}
\end{equation}
Thus $U^{(C)}$ and $V^{(C)}$ are orthogonal.  For $r<L$,
Eq.~\eqref{eq:S-sine-recurrence} gives
\begin{equation}
V_{rm}^{(C)}+V_{r+1,m}^{(C)}
=\varepsilon_m^{(C)}U_{rm}^{(C)}.
\label{eq:S-C-bulk-rec}
\end{equation}
At the right boundary,
\begin{equation}
\sin\!\left[(L+\tfrac12)Q_m^{(C)}\right]
=\sin(\pi m)=0,
\label{eq:S-C-right-ghost}
\end{equation}
so the same recurrence reduces to the last row of $B_C$.  The transpose
relation follows similarly from the left ghost sine at $r=0$.  Hence
Eq.~\eqref{eq:S-C-SVD} holds.

The polar matrix is
\begin{equation}
G_C=U^{(C)}V^{(C)\mathsf T}.
\label{eq:S-C-polar}
\end{equation}
Its explicit weighted Cauchy form and every-minor identity will be derived in
Sec.~\ref{sec:S4}.

\subsection{One tuned endpoint: the \texorpdfstring{$B_L$}{B(L)} solution}

Set
\begin{equation}
N_B=2L,
\qquad
Q_m^{(B)}=\frac{(2m-1)\pi}{N_B},
\qquad m=1,\ldots,L.
\label{eq:S-B-grid}
\end{equation}
Introduce the row weight
\begin{equation}
\epsilon_r=
\begin{cases}
1,&r<L,\\
2^{-1/2},&r=L,
\end{cases}
\label{eq:S-B-epsilon-row}
\end{equation}
and define
\begin{equation}
U_{rm}^{(B)}
=\sqrt{\frac2L}\,\epsilon_r\sin(rQ_m^{(B)}),
\label{eq:S-B-U}
\end{equation}
\begin{equation}
V_{sm}^{(B)}
=\sqrt{\frac2L}\,
\sin\!\left[(s-\tfrac12)Q_m^{(B)}\right],
\label{eq:S-B-V}
\end{equation}
with
\begin{equation}
\varepsilon_m^{(B)}=2\cos\frac{Q_m^{(B)}}2.
\label{eq:S-B-epsilon}
\end{equation}

The matrices in Eqs.~\eqref{eq:S-B-U} and~\eqref{eq:S-B-V} are orthogonal and
\begin{equation}
B_B=U^{(B)}\diag(\varepsilon_m^{(B)})V^{(B)\mathsf T}.
\label{eq:S-B-SVD}
\end{equation}

\noindent\emph{Derivation.} The half-integer sine family satisfies
\begin{equation}
\sum_{s=1}^{L}
\sin\!\left[(s-\tfrac12)Q_m^{(B)}\right]
\sin\!\left[(s-\tfrac12)Q_n^{(B)}\right]
=\frac L2\delta_{mn}.
\label{eq:S-B-half-orth}
\end{equation}
The integer family requires a half weight at the fixed point $r=L$:
\begin{equation}
\sum_{r=1}^{L-1}\sin(rQ_m^{(B)})\sin(rQ_n^{(B)})
+\frac12\sin(LQ_m^{(B)})\sin(LQ_n^{(B)})
=\frac L2\delta_{mn}.
\label{eq:S-B-endpoint-orth}
\end{equation}
Equations~\eqref{eq:S-B-half-orth} and
\eqref{eq:S-B-endpoint-orth} prove orthogonality.  In the bulk,
Eq.~\eqref{eq:S-sine-recurrence} again reconstructs the diagonal and
superdiagonal entries.  At the right endpoint,
\begin{equation}
\sin(LQ_m^{(B)})=(-1)^{m-1}
\label{eq:S-B-fixed-value}
\end{equation}
is a reflection-fixed orbit.  Combining this value with
$\epsilon_L=2^{-1/2}$ changes the final diagonal element from one to
$\sqrt2$, precisely reproducing Eq.~\eqref{eq:S-BB-matrix}.  This proves
Eq.~\eqref{eq:S-B-SVD}.

The factor $2^{-1/2}$ is therefore a quadrature weight forced by the folded
finite grid.  It will later become the half-orbit factor in the $B_L$ Pauli
measure.

\subsection{Two tuned endpoints: the \texorpdfstring{$D_L$}{D(L)} solution}

Let
\begin{equation}
N_D=2L-1,
\qquad
Q_m^{(D)}=\frac{2\pi m}{N_D},
\qquad m=0,1,\ldots,L-1.
\label{eq:S-D-grid}
\end{equation}
Define
\begin{equation}
\epsilon_r=
\begin{cases}
1,&r<L,\\
2^{-1/2},&r=L,
\end{cases}
\qquad
\eta_s=
\begin{cases}
2^{-1/2},&s=1,\\
1,&s>1,
\end{cases}
\label{eq:S-D-endpoint-weights}
\end{equation}
and the zero-mode weight
\begin{equation}
\rho_m=
\begin{cases}
2^{-1/2},&m=0,\\
1,&m>0.
\end{cases}
\label{eq:S-D-zero-weight}
\end{equation}
The singular vectors are
\begin{equation}
U_{rm}^{(D)}
=\frac{2}{\sqrt{N_D}}\epsilon_r\rho_m
\cos\!\left[(r-\tfrac12)Q_m^{(D)}\right],
\label{eq:S-D-U}
\end{equation}
\begin{equation}
V_{sm}^{(D)}
=\frac{2}{\sqrt{N_D}}\eta_s\rho_m
\cos[(s-1)Q_m^{(D)}],
\label{eq:S-D-V}
\end{equation}
with
\begin{equation}
\varepsilon_m^{(D)}=2\cos\frac{Q_m^{(D)}}2.
\label{eq:S-D-epsilon}
\end{equation}

For $L\ge2$, the matrices in Eqs.~\eqref{eq:S-D-U} and~\eqref{eq:S-D-V} are orthogonal and
\begin{equation}
B_D=U^{(D)}\diag(\varepsilon_m^{(D)})V^{(D)\mathsf T}.
\label{eq:S-D-SVD}
\end{equation}

\noindent\emph{Derivation.} The first cosine family obeys
\begin{align}
&\sum_{r=1}^{L-1}
\cos\!\left[(r-\tfrac12)Q_m^{(D)}\right]
\cos\!\left[(r-\tfrac12)Q_n^{(D)}\right]
\nonumber\\
&\quad+\frac12
\cos\!\left[(L-\tfrac12)Q_m^{(D)}\right]
\cos\!\left[(L-\tfrac12)Q_n^{(D)}\right]
=\frac{N_D}{4\rho_m^2}\delta_{mn}.
\label{eq:S-D-left-orth}
\end{align}
The second family satisfies
\begin{equation}
\frac12+
\sum_{s=2}^{L}\cos[(s-1)Q_m^{(D)}]\cos[(s-1)Q_n^{(D)}]
=\frac{N_D}{4\rho_m^2}\delta_{mn}.
\label{eq:S-D-right-orth}
\end{equation}
The endpoint weights and the factor $\rho_0=2^{-1/2}$ therefore make
$U^{(D)}$ and $V^{(D)}$ orthogonal.  Equation~\eqref{eq:S-cosine-recurrence}
reconstructs the bulk entries.  The half-orbit normalizations at the first
column and last row generate the two diagonal elements $\sqrt2$ in
Eq.~\eqref{eq:S-BD-matrix}.  Hence Eq.~\eqref{eq:S-D-SVD} follows.

The zero mode in Eq.~\eqref{eq:S-D-grid} is a constant cosine mode, not a
zero-energy mode: its singular value is two.  The low-energy modes occur near
the opposite end of the grid, where $Q_m$ approaches $\pi$.

\subsection{Ordered spectra and common infrared boundary CFT}

To compare the three open chains, order their singular values from low to
high.  Writing the mode number from the top of each grid gives, for
$n=0,1,\ldots,L-1$,
\begin{equation}
\varepsilon_n^{(C),\uparrow}
=2\sin\frac{(2n+1)\pi}{4L+2},
\label{eq:S-C-low-spectrum}
\end{equation}
\begin{equation}
\varepsilon_n^{(B),\uparrow}
=2\sin\frac{(2n+1)\pi}{4L},
\label{eq:S-B-low-spectrum}
\end{equation}
and
\begin{equation}
\varepsilon_n^{(D),\uparrow}
=2\sin\frac{(2n+1)\pi}{4L-2}.
\label{eq:S-D-low-spectrum}
\end{equation}
For fixed $n$ and large $L$,
\begin{equation}
\varepsilon_n^{(R),\uparrow}
=\frac{\pi(n+\tfrac12)}{L_{\rm eff}^{(R)}}+O(L^{-3}),
\label{eq:S-common-NS-tower}
\end{equation}
with
\begin{equation}
L_{\rm eff}^{(C)}=L+\frac12,
\qquad
L_{\rm eff}^{(B)}=L,
\qquad
L_{\rm eff}^{(D)}=L-\frac12.
\label{eq:S-effective-lengths}
\end{equation}
Thus all three spectra possess the same half-integer Neveu--Schwarz tower and
flow to the free--free Ising boundary CFT~\cite{Cardy1984,Cardy1989,AffleckLudwig1991}.  The endpoint tuning changes only
the extrapolation length and the exact finite reflection grid.

The $B_L$, $C_L$, and $D_L$ chains have the same infrared conformal tower but
different exact one-particle grids.  Their root labels therefore classify the
finite-lattice Pauli geometry rather than distinct Cardy boundary states.

The distinction will become sharper in Sec.~\ref{sec:S4}.  There the three grids
produce different one-body collision factors: a short wall root $e_i$, a long
wall root $2e_i$, or no wall root.  The continuum tower loses this half-site
information even though the full Pauli distribution retains it exactly.

\section{Cauchy kernels and termwise Weyl identities}
\label{sec:S4}

The minor representation of Sec.~\ref{sec:S2} is still exponentially large.
The additional solvability at the critical Ising point comes from a stronger
statement: each polar matrix found in Sec.~\ref{sec:S3} is a weighted Cauchy
matrix on two interlaced finite lattices.  The Cauchy determinant then turns
every individual Pauli minor into a complementary Weyl alternant.  This
section proves that correspondence for the periodic $A_{L-1}$ chain and for
the parity-preserving $B_L$, $C_L$, and $D_L$ interval chains.  No sum over
Pauli strings is taken until Sec.~\ref{sec:S6}.

\subsection{Cauchy determinants and complementary configurations}

For an ordered set $Z=(z_1,\ldots,z_n)$, write
\begin{equation}
\Delta(Z)=\prod_{1\le i<j\le n}(z_j-z_i),
\qquad \Delta(\varnothing)=1.
\label{eq:S4-Vandermonde}
\end{equation}
Let $X=\{x_1,\ldots,x_n\}$ and $Y=\{y_1,\ldots,y_m\}$ be disjoint and
consider a weighted Cauchy matrix
\begin{equation}
C_{rs}=\frac{a_r b_s}{x_r-y_s}.
\label{eq:S4-weighted-Cauchy}
\end{equation}
For subsets $S$ and $T$ of equal cardinality $k$, the Cauchy determinant
formula gives
\begin{equation}
\det C[S,T]
=(-1)^{k(k-1)/2}
\left(\prod_{r\in S}a_r\right)
 \left(\prod_{s\in T}b_s\right)
 \frac{\Delta(X_S)\Delta(Y_T)}
 {\prod_{r\in S}\prod_{s\in T}(x_r-y_s)}.
\label{eq:S4-Cauchy-determinant}
\end{equation}
The nontrivial step is to use the complete-grid products
$\prod_s(x_r-y_s)$ and $\prod_r(y_s-x_r)$ to replace the denominator in
Eq.~\eqref{eq:S4-Cauchy-determinant} by factors involving the complementary
set $T^c$.  In all four geometries this produces a configuration containing
$k$ points from one sublattice and $n-k$ points from the other.

For later use, introduce the full cross polynomial and the barycentric
derivative on the column lattice,
\begin{equation}
P_Y(x)=\prod_{y\in Y}(x-y),
\qquad
D_Y(y)=\prod_{\substack{y'\in Y\\y'\ne y}}(y-y').
\label{eq:S4-PY-DY}
\end{equation}
The complementary reorganization can then be stated without suppressing any
node-dependent factor.

Assume $|X|=|Y|=n$ and let $C$ be the weighted Cauchy matrix in
Eq.~\eqref{eq:S4-weighted-Cauchy}.  For $|S|=|T|=k$ and
$U=X_S\cup Y_{T^c}$,
\begin{equation}
\begin{aligned}
\abs{\det C[S,T]}
={}&\frac{\abs{\Delta(U)}}{\abs{\Delta(Y)}}
\prod_{r\in S}
\frac{\abs{a_r}}{\abs{P_Y(x_r)}}
\prod_{s\in T}
\abs{b_sD_Y(y_s)}.
\end{aligned}
\label{eq:S4-complementary-Cauchy}
\end{equation}

\noindent\emph{Derivation.} Decompose the two Vandermondes according to $T\sqcup T^c$ and
$X_S\sqcup Y_{T^c}$.  In absolute value,
\begin{equation}
\frac{\abs{\Delta(U)}}{\abs{\Delta(Y)}}
=
\frac{
\abs{\Delta(X_S)}
\prod_{x\in X_S,\,y\in Y_{T^c}}\abs{x-y}
}{
\abs{\Delta(Y_T)}
\prod_{y\in Y_T,\,y'\in Y_{T^c}}\abs{y-y'}
}.
\label{eq:S4-complement-ratio-proof}
\end{equation}
Moreover,
\begin{equation}
\prod_{s\in T}\abs{D_Y(y_s)}
=
\abs{\Delta(Y_T)}^2
\prod_{y\in Y_T,\,y'\in Y_{T^c}}\abs{y-y'},
\label{eq:S4-DY-product-proof}
\end{equation}
and
\begin{equation}
\prod_{r\in S}\abs{P_Y(x_r)}
=
\prod_{x\in X_S,\,y\in Y_T}\abs{x-y}
\prod_{x\in X_S,\,y\in Y_{T^c}}\abs{x-y}.
\label{eq:S4-PY-product-proof}
\end{equation}
Multiplying Eqs.~\eqref{eq:S4-complement-ratio-proof}--
\eqref{eq:S4-PY-product-proof} reproduces the unweighted Cauchy determinant
in Eq.~\eqref{eq:S4-Cauchy-determinant}; the row and column weights then give
Eq.~\eqref{eq:S4-complementary-Cauchy}.

For the interval chains it is convenient to use
\begin{equation}
\lambda(\theta)=\sin^2\theta,
\label{eq:S4-lambda}
\end{equation}
for which
\begin{equation}
\lambda(\theta)-\lambda(\phi)
=\sin(\theta-\phi)\sin(\theta+\phi).
\label{eq:S4-direct-image-factor}
\end{equation}
The first factor is a direct collision and the second is a collision with the
reflected image.  The remaining one-body factor determines whether the wall
root is $e_i$, $2e_i$, or absent.

\subsection{Periodic \texorpdfstring{$A_{L-1}$}{A(L-1)} reference ensemble}

For comparison with the boundary root systems, define
\begin{equation}
\zeta=\ee^{\pi\ii/L},\qquad
X_A=\{\zeta^{2j}:j=0,\ldots,L-1\},\qquad
Y_A=\{\zeta^{2k-1}:k=0,\ldots,L-1\}.
\label{eq:S4-A-sublattices}
\end{equation}
For $|S|=|T|$, let
\begin{equation}
U_A(S,T)=X_{A,S}\cup Y_{A,T^c}.
\label{eq:S4-A-map}
\end{equation}
The periodic Cauchy/Vandermonde calculation gives~\cite{Companion}
\begin{equation}
\abs{\det G^{(A)}[S,T]}
=L^{-L/2}\abs{\Delta(U_A(S,T))}.
\label{eq:S4-A-termwise}
\end{equation}
This is the $A_{L-1}$ baseline.  The derivation, discrete Selberg sum, alias
representation, and special-index evaluations are not repeated here; the new
content below is the reflected interval geometry and its $B_L,C_L,D_L$
wall factors.

\subsection{Unified image-Cauchy kernel on an interval}

The SVDs in Sec.~\ref{sec:S3} use row angles $p_r$, column angles $q_s$, and
grid lengths
\begin{equation}
\begin{array}{c|c|c|c}
R&N_R&p_r&q_s\\ \hline
C&2L+1&\pi r/N_R&\pi(s-\tfrac12)/N_R\\
B&2L&\pi r/N_R&\pi(s-\tfrac12)/N_R\\
D&2L-1&\pi(r-\tfrac12)/N_R&\pi(s-1)/N_R.
\end{array}
\label{eq:S4-open-angle-table}
\end{equation}
The endpoint factors are
\begin{equation}
\epsilon_r^{(C)}=\eta_s^{(C)}=1,
\qquad
\epsilon_L^{(B)}=2^{-1/2},
\qquad
\epsilon_L^{(D)}=\eta_1^{(D)}=2^{-1/2},
\label{eq:S4-open-endpoint-factors}
\end{equation}
with all unlisted factors equal to one.

Evaluating $U^{(R)}V^{(R)\mathsf T}$ by the finite sums of
Sec.~\ref{sec:S3} gives
\begin{equation}
G^{(R)}_{rs}
=\frac{(-1)^{r-s}\epsilon_r^{(R)}\eta_s^{(R)}}{N_R}
\left[\csc(p_r-q_s)+\csc(p_r+q_s)\right].
\label{eq:S4-image-cosecant}
\end{equation}
Using
\begin{equation}
\csc(p-q)+\csc(p+q)
=\frac{2\sin p\cos q}{\sin^2p-\sin^2q},
\label{eq:S4-cosecant-Cauchy}
\end{equation}
we obtain the common weighted Cauchy form
\begin{equation}
G^{(R)}_{rs}
=\frac{2(-1)^{r-s}\epsilon_r^{(R)}\eta_s^{(R)}}{N_R}
\frac{\sin p_r\cos q_s}{\lambda(p_r)-\lambda(q_s)}.
\label{eq:S4-unified-open-Cauchy}
\end{equation}
Equation~\eqref{eq:S4-image-cosecant} makes the method of images explicit:
the two terms propagate from $q_s$ and from its reflected image $-q_s$.

\subsection{Standard open chain: the \texorpdfstring{$C_L$}{C(L)} identity}

Set $N_C=2L+1$ and define
\begin{equation}
X_C=\left\{\frac{\pi r}{N_C}:r=1,\ldots,L\right\},
\qquad
Y_C=\left\{\frac{\pi(s-1/2)}{N_C}:s=1,\ldots,L\right\}.
\label{eq:S4-C-grids}
\end{equation}
For an $L$-point configuration $U$ on $X_C\sqcup Y_C$, define
\begin{equation}
\mathfrak V_C(U)
=\prod_{\theta\in U}\abs{\sin2\theta}
 \prod_{\substack{\theta,\phi\in U\\\theta<\phi}}
 \abs{\lambda(\theta)-\lambda(\phi)}.
\label{eq:S4-C-weight}
\end{equation}
The two folded root polynomials are proportional to
$U_L(t)+U_{L-1}(t)$ and $U_L(t)-U_{L-1}(t)$, with $t=1-2\lambda$.
Evaluating one polynomial on the roots of the other gives
\begin{equation}
\prod_{s=1}^{L}(x_r-y_s)
=(-1)^{L+r}2^{-2L}\sec p_r,
\qquad
\prod_{r=1}^{L}(y_s-x_r)
=(-1)^{L+s-1}2^{-2L}\csc q_s.
\label{eq:S4-C-cross-products}
\end{equation}
Differentiating the monic polynomial with roots $Y_C$ gives the pointwise
barycentric derivative
\begin{equation}
\abs{D_{Y_C}(y_s)}
=\frac{N_C}{2^{2L+1}\sin q_s\cos^2q_s}.
\label{eq:S4-C-barycentric}
\end{equation}
Their discriminants imply the equal reference values
\begin{equation}
\mathfrak V_C(X_C)=\mathfrak V_C(Y_C)
=\frac{N_C^{L/2}}{2^{L^2}}.
\label{eq:S4-C-reference}
\end{equation}

For balanced subsets set
\begin{equation}
U_C(S,T)=X_{C,S}\cup Y_{C,T^c}.
\label{eq:S4-C-map}
\end{equation}

For every $|S|=|T|$,
\begin{equation}
\abs{\det G_C[S,T]}
=\frac{\mathfrak V_C(U_C(S,T))}{\mathfrak V_C(Y_C)}.
\label{eq:S4-C-termwise}
\end{equation}

\noindent\emph{Derivation.} Use the complementary Cauchy identity above with
$a_r=(2/N_C)\sin p_r$ and $b_s=\cos q_s$; the staggered signs have unit
absolute value.  Equations~\eqref{eq:S4-C-cross-products} and
\eqref{eq:S4-C-barycentric} give, point by point,
\begin{equation}
\frac{\abs{a_r}}{\abs{P_{Y_C}(x_r)}}
=\frac{2^{2L}}{N_C}\abs{\sin2p_r},
\qquad
\abs{b_sD_{Y_C}(y_s)}
=\frac{N_C}{2^{2L}}\frac1{\abs{\sin2q_s}}.
\label{eq:S4-C-node-cancellation}
\end{equation}
Because $|S|=|T|$, the constants cancel.  The remaining node factors convert
$\abs{\Delta(U_C)}/\abs{\Delta(Y_C)}$ into
$\mathfrak V_C(U_C)/\mathfrak V_C(Y_C)$, proving the theorem.  The explicit
reference value in Eq.~\eqref{eq:S4-C-reference} fixes the normalization in
closed form.

By Eq.~\eqref{eq:S4-direct-image-factor}, the root-dependent part is
\begin{equation}
\prod_i\abs{\sin2\theta_i}
\prod_{i<j}\abs{\sin(\theta_i-\theta_j)\sin(\theta_i+\theta_j)},
\label{eq:S4-C-root-form}
\end{equation}
corresponding to the positive roots $e_i\pm e_j$ and $2e_i$ of $C_L$.

\subsection{One folded endpoint: the \texorpdfstring{$B_L$}{B(L)} identity}

For $N_B=2L$, define
\begin{equation}
X_B=\left\{\frac{\pi r}{2L}:r=1,\ldots,L\right\},
\qquad
Y_B=\left\{\frac{\pi(s-1/2)}{2L}:s=1,\ldots,L\right\}.
\label{eq:S4-B-grids}
\end{equation}
The point $\pi/2\in X_B$ is fixed under reflection.  Its orbit has half the
size of a generic orbit, which produces the quadrature weight
\begin{equation}
\nu_B(\theta)=
\begin{cases}
2^{-1/2},&\theta=\pi/2,\\
1,&\text{otherwise}.
\end{cases}
\label{eq:S4-B-half-weight}
\end{equation}
Define
\begin{equation}
\mathfrak V_B(U)
=\prod_{\theta\in U}\nu_B(\theta)\abs{\sin\theta}
 \prod_{\theta<\phi\in U}
 \abs{\lambda(\theta)-\lambda(\phi)}.
\label{eq:S4-B-weight}
\end{equation}
With $x_r=\lambda(p_r)$ and $y_s=\lambda(q_s)$, the relevant Chebyshev
polynomials are $(t+1)U_{L-1}(t)$ and $T_L(t)$.  Their cross evaluations are
\begin{equation}
\prod_{s=1}^{L}(x_r-y_s)=(-1)^{L+r}2^{1-2L},
\qquad
\prod_{r=1}^{L}(y_s-x_r)
=(-1)^{L+s-1}2^{1-2L}\cot q_s.
\label{eq:S4-B-cross-products}
\end{equation}
The derivative at a root of the $Y_B$ polynomial is
\begin{equation}
\abs{D_{Y_B}(y_s)}
=\frac{L}{2^{2L-1}\sin q_s\cos q_s}.
\label{eq:S4-B-barycentric}
\end{equation}
The complete alternants satisfy
\begin{equation}
\mathfrak V_B(X_B)=\mathfrak V_B(Y_B)
=\frac{L^{L/2}}{2^{L(2L-1)/2}}
=\frac{N_B^{L/2}}{2^{L^2}}.
\label{eq:S4-B-reference}
\end{equation}
For $|S|=|T|$, set
\begin{equation}
U_B(S,T)=X_{B,S}\cup Y_{B,T^c}.
\label{eq:S4-B-map}
\end{equation}

\begin{equation}
\abs{\det G_B[S,T]}
=\frac{\mathfrak V_B(U_B(S,T))}{\mathfrak V_B(Y_B)}.
\label{eq:S4-B-termwise}
\end{equation}

\noindent\emph{Derivation.} Apply the complementary Cauchy identity above with
$a_r=L^{-1}\epsilon_r^{(B)}\sin p_r$ and $b_s=\cos q_s$.  The complete
cross product and barycentric derivative give
\begin{equation}
\frac{\abs{a_r}}{\abs{P_{Y_B}(x_r)}}
=\frac{2^{2L-1}}{L}\nu_B(p_r)\abs{\sin p_r},
\qquad
\abs{b_sD_{Y_B}(y_s)}
=\frac{L}{2^{2L-1}}\frac1{\abs{\sin q_s}}.
\label{eq:S4-B-node-cancellation}
\end{equation}
The constants again cancel between the $k$ selected rows and columns.  The
remaining factors are precisely the ratio of the $B_L$ node weights in
$U_B(S,T)$ and $Y_B$; the reflection-fixed value
$\epsilon_L^{(B)}=\nu_B(\pi/2)=2^{-1/2}$ is retained as a discrete
half-orbit weight.  Equation~\eqref{eq:S4-B-reference} completes the
normalization.

The root factor is
\begin{equation}
\prod_i\abs{\sin\theta_i}
\prod_{i<j}\abs{\sin(\theta_i-\theta_j)\sin(\theta_i+\theta_j)},
\label{eq:S4-B-root-form}
\end{equation}
which contains the $B_L$ roots $e_i$ and $e_i\pm e_j$.  The half-orbit
factor $\nu_B$ belongs to the discrete measure, not to the root energy.

\subsection{Two folded endpoints: the \texorpdfstring{$D_L$}{D(L)} identity}

For $N_D=2L-1$, define
\begin{equation}
X_D=\left\{\frac{\pi(r-1/2)}{2L-1}:r=1,\ldots,L\right\},
\quad
Y_D=\left\{\frac{\pi(s-1)}{2L-1}:s=1,\ldots,L\right\}.
\label{eq:S4-D-grids}
\end{equation}
The reflection-fixed points $\pi/2\in X_D$ and $0\in Y_D$ carry
\begin{equation}
\nu_D(\theta)=
\begin{cases}
2^{-1/2},&\theta=0\text{ or }\pi/2,\\
1,&\text{otherwise}.
\end{cases}
\label{eq:S4-D-half-weight}
\end{equation}
Define
\begin{equation}
\mathfrak V_D(U)
=\prod_{\theta\in U}\nu_D(\theta)
 \prod_{\theta<\phi\in U}
 \abs{\lambda(\theta)-\lambda(\phi)}.
\label{eq:S4-D-weight}
\end{equation}
The complete-grid products are
\begin{equation}
\prod_{s=1}^{L}(x_r-y_s)
=(-1)^{L+r}2^{2-2L}\sin p_r,
\qquad
\prod_{r=1}^{L}(y_s-x_r)
=(-1)^{L+s+1}2^{2-2L}\cos q_s.
\label{eq:S4-D-cross-products}
\end{equation}
With the column endpoint factor $\eta_s^{(D)}$ of
Eq.~\eqref{eq:S4-open-endpoint-factors}, the derivative on $Y_D$ is
\begin{equation}
\abs{D_{Y_D}(y_s)}
=\frac{N_D}{2^{2L-1}[\eta_s^{(D)}]^2\cos q_s}.
\label{eq:S4-D-barycentric}
\end{equation}
Their discriminants give
\begin{equation}
\mathfrak V_D(X_D)=\mathfrak V_D(Y_D)
=\frac{(2L-1)^{L/2}}{2^{L(L-1)+1/2}}.
\label{eq:S4-D-reference}
\end{equation}
For balanced subsets let
\begin{equation}
U_D(S,T)=X_{D,S}\cup Y_{D,T^c}.
\label{eq:S4-D-map}
\end{equation}

\begin{equation}
\abs{\det G_D[S,T]}
=\frac{\mathfrak V_D(U_D(S,T))}{\mathfrak V_D(Y_D)}.
\label{eq:S4-D-termwise}
\end{equation}

\noindent\emph{Derivation.} Use the complementary Cauchy identity above with
$a_r=(2/N_D)\epsilon_r^{(D)}\sin p_r$ and
$b_s=\eta_s^{(D)}\cos q_s$.  Equations~\eqref{eq:S4-D-cross-products} and
\eqref{eq:S4-D-barycentric} imply
\begin{equation}
\frac{\abs{a_r}}{\abs{P_{Y_D}(x_r)}}
=\frac{2^{2L-1}}{N_D}\nu_D(p_r),
\qquad
\abs{b_sD_{Y_D}(y_s)}
=\frac{N_D}{2^{2L-1}}\frac1{\nu_D(q_s)}.
\label{eq:S4-D-node-cancellation}
\end{equation}
The constants cancel, no trigonometric wall factor remains, and only the two
reflection-fixed half-orbit weights survive.  The rest is the Vandermonde in
$\lambda$, normalized by Eq.~\eqref{eq:S4-D-reference}.

Using Eq.~\eqref{eq:S4-direct-image-factor}, the root-dependent part is only
\begin{equation}
\prod_{i<j}\abs{\sin(\theta_i-\theta_j)\sin(\theta_i+\theta_j)},
\label{eq:S4-D-root-form}
\end{equation}
which is the $D_L$ Weyl denominator with roots $e_i\pm e_j$.

\section{Fixed physical boundaries}
\label{sec:S5}

The transverse endpoint tunings of the $B$ and $D$ chains preserve Ising
parity.  Longitudinal boundary fields are qualitatively different: they
produce Majorana one-point functions and mix even and odd physical Majorana
degrees.  The augmented-minor theorem of Sec.~\ref{sec:S2} solves the Wick
bookkeeping, but it does not identify the explicit rectangular matrix or its
root geometry.  We now derive those matrices for the free--fixed, equal-fixed,
and opposite-fixed chains and show that all three are $C$-type constructions.
Throughout this section augmented matrices use the vertical convention of
Eq.~\eqref{eq:S-augmented-G}: they have $L+1$ rows and $L$ columns.  This is
the transpose of an equally common convention; all equal-size minors are
unchanged by that transpose.

\subsection{Free--fixed boundary as a rectangular Majorana path}

With the Jordan--Wigner string oriented from the fixed end,
$X_L$ is a single Majorana.  Introducing the auxiliary Wick Majorana of
Sec.~\ref{sec:S2} turns the linear boundary term into the final bond of an
enlarged quadratic Majorana path.  After row and column sign gauges, the
one-particle coupling matrix is the rectangular bidiagonal matrix
\begin{equation}
\mathbb B_{\fplus}
=\begin{pmatrix}
1&1&&&\\
&1&1&&\\
&&\ddots&\ddots&\\
&&&1&1
\end{pmatrix}_{L\times(L+1)}.
\label{eq:S5-fplus-B}
\end{equation}
The additional column is the auxiliary endpoint; it is not an additional
physical spin.  Its polar isometry is precisely the augmented contraction
matrix that combines the physical even and odd Pauli sectors.

Set
\begin{equation}
M=L+1,
\qquad s=1,\ldots,M-1.
\label{eq:S5-M}
\end{equation}
Define
\begin{equation}
U_{js}^{\fplus}=\sqrt{\frac{2}{M}}
\sin\!\left[\frac{\pi s(j+1)}{M}\right],
\quad j=0,\ldots,M-2,
\label{eq:S5-fplus-U}
\end{equation}
\begin{equation}
V_{rs}^{\fplus}=\sqrt{\frac{2}{M}}
\sin\!\left[\frac{\pi s(r+1/2)}{M}\right],
\quad r=0,\ldots,M-1,
\label{eq:S5-fplus-V}
\end{equation}
and
\begin{equation}
\varepsilon_s^{\fplus}=2\cos\frac{\pi s}{2M}.
\label{eq:S5-fplus-epsilon}
\end{equation}
Finite sine orthogonality gives
$U^{\fplus\mathsf T}U^{\fplus}=V^{\fplus\mathsf T}V^{\fplus}=I_L$, and
the same recurrence as in Eq.~\eqref{eq:S-sine-recurrence} yields
\begin{equation}
\mathbb B_{\fplus}
=U^{\fplus}\diag(\varepsilon_s^{\fplus})V^{\fplus\mathsf T}.
\label{eq:S5-fplus-SVD}
\end{equation}
The vertical polar isometry is therefore
\begin{equation}
\widehat G^{\fplus}=V^{\fplus}U^{\fplus\mathsf T}
\in\mathbb R^{M\times(M-1)},
\qquad
\widehat G^{\fplus\mathsf T}\widehat G^{\fplus}=I_L.
\label{eq:S5-fplus-isometry}
\end{equation}

The finite mode sum gives
\begin{equation}
\widehat G^{\fplus}_{rj}
=\frac{(-1)^{r+j}}{2M}
\left[
\cot\frac{\pi(j-r+1/2)}{2M}
+\cot\frac{\pi(j+r+3/2)}{2M}
\right].
\label{eq:S5-fplus-cot}
\end{equation}
Introduce the interlaced angles
\begin{equation}
q_r=\frac{\pi(r+1/2)}{2M},
\quad r=0,\ldots,M-1,
\qquad
p_j=\frac{\pi(j+1)}{2M},
\quad j=0,\ldots,M-2,
\label{eq:S5-fplus-angles}
\end{equation}
and $x_r=\lambda(q_r)$, $y_j=\lambda(p_j)$.  The identity
\begin{equation}
\cot(p-q)+\cot(p+q)
=\frac{\sin2p}{\sin^2p-\sin^2q}
\label{eq:S5-cot-Cauchy}
\end{equation}
reduces Eq.~\eqref{eq:S5-fplus-cot} to
\begin{equation}
\widehat G^{\fplus}_{rj}
=\frac{(-1)^{r+j}}{2M}
\frac{\sin2p_j}{y_j-x_r}.
\label{eq:S5-fplus-Cauchy}
\end{equation}
Thus the rectangular augmented matrix is analytically Cauchy; its form is
not inferred from finite-size pattern matching.

Let
\begin{equation}
X_{\fplus}=\{q_0,\ldots,q_{M-1}\},
\qquad
Y_{\fplus}=\{p_0,\ldots,p_{M-2}\}.
\label{eq:S5-fplus-grids}
\end{equation}
Their union is the set of interior points
$\{\pi a/(4M):a=1,\ldots,2M-1\}$.  On any $M$-point subset define the
$C_M$ alternant
\begin{equation}
\mathfrak V_{C_M}(U)
=\prod_{\theta\in U}\abs{\sin2\theta}
 \prod_{\theta<\phi\in U}
 \abs{\lambda(\theta)-\lambda(\phi)}.
\label{eq:S5-fplus-weight}
\end{equation}
The complete-grid cross products are
\begin{equation}
\prod_{j=0}^{M-2}(x_r-y_j)
=(-1)^{M+r+1}2^{2-2M}\csc(2q_r),
\label{eq:S5-fplus-cross-X}
\end{equation}
\begin{equation}
\prod_{r=0}^{M-1}(y_j-x_r)
=(-1)^{M+j+1}2^{1-2M}.
\label{eq:S5-fplus-cross-Y}
\end{equation}
The derivative of the monic polynomial with roots $X_{\fplus}$ is
\begin{equation}
\abs{D_{X_{\fplus}}(x_r)}
=\frac{M\,2^{2-2M}}{\abs{\sin2q_r}},
\qquad
D_X(x)=\prod_{\substack{x'\in X\\x'\ne x}}(x-x').
\label{eq:S5-fplus-barycentric}
\end{equation}

The rectangular complement used below has a slightly different orientation
from the complementary Cauchy identity above.  If $|X|=m+1$, $|Y|=m$,
$C_{rj}=a_rb_j/(y_j-x_r)$, and $|R|=|T|$, then for
$U=X_{R^c}\cup Y_T$,
\begin{equation}
\abs{\det C[R,T]}
=\frac{\abs{\Delta(U)}}{\abs{\Delta(X)}}
\prod_{r\in R}\abs{a_rD_X(x_r)}
\prod_{j\in T}\frac{\abs{b_j}}{\abs{P_X(y_j)}},
\label{eq:S5-rectangular-complement}
\end{equation}
where $P_X(y)=\prod_{x\in X}(y-x)$.  To verify
Eq.~\eqref{eq:S5-rectangular-complement}, decompose $\Delta(X)$ according to
$R\sqcup R^c$ and use
\begin{equation}
\prod_{r\in R}\abs{D_X(x_r)}
=\abs{\Delta(X_R)}^2
\prod_{x\in X_R,\,x'\in X_{R^c}}\abs{x-x'},
\label{eq:S5-rectangular-DX-product}
\end{equation}
while $\prod_{j\in T}|P_X(y_j)|$ separates into the cross products with
$X_R$ and $X_{R^c}$.  The result is exactly the rectangular Cauchy
determinant.
The odd-node sine product and the shifted Chebyshev discriminant give
\begin{equation}
\mathfrak V_{C_M}(X_{\fplus})
=\frac{M^{M/2}}{2^{M^2-(M+1)/2}}.
\label{eq:S5-fplus-reference}
\end{equation}
For selected row and column sets $R\subseteq\{0,\ldots,M-1\}$ and
$T\subseteq\{0,\ldots,M-2\}$ with $|R|=|T|$, define
\begin{equation}
U_{\fplus}(R,T)
=X_{\fplus,R^c}\cup Y_{\fplus,T}.
\label{eq:S5-fplus-map}
\end{equation}
The complement on the larger $X$ sublattice is what keeps
$|U_{\fplus}|=M$.

For every equal-size rectangular minor,
\begin{equation}
\abs{\det\widehat G^{\fplus}[R,T]}
=\frac{\mathfrak V_{C_M}(U_{\fplus}(R,T))}
{\mathfrak V_{C_M}(X_{\fplus})}.
\label{eq:S5-fplus-termwise}
\end{equation}

\noindent\emph{Derivation.} In Eq.~\eqref{eq:S5-fplus-Cauchy} take
$a_r=(2M)^{-1}$ and $b_j=\sin2p_j$.  Equations
\eqref{eq:S5-fplus-cross-Y} and~\eqref{eq:S5-fplus-barycentric} yield
\begin{equation}
\abs{a_rD_{X_{\fplus}}(x_r)}
=\frac{2^{1-2M}}{\abs{\sin2q_r}},
\qquad
\frac{\abs{b_j}}{\abs{P_{X_{\fplus}}(y_j)}}
=2^{2M-1}\abs{\sin2p_j}.
\label{eq:S5-fplus-node-cancellation}
\end{equation}
The constants cancel because $|R|=|T|$.  Substitution into the exact
rectangular complement formula~\eqref{eq:S5-rectangular-complement} converts
the Vandermonde ratio into
$\mathfrak V_{C_M}(U_{\fplus})/\mathfrak V_{C_M}(X_{\fplus})$.
Equation~\eqref{eq:S5-fplus-reference} then supplies the closed reference
normalization.

The physical minor order is
\begin{equation}
k=|R|=|T|=|U_{\fplus}\cap Y_{\fplus}|,
\label{eq:S5-fplus-charge}
\end{equation}
so the free--fixed fugacity counts particles on the smaller, even
sublattice.  This convention makes the constant term of the degree
polynomial equal to one.

\subsection{Kramers--Wannier duality for two fixed boundaries}

For fixed signs $\eta_1,\eta_L\in\{+1,-1\}$, introduce
$N=L+1$ dual spins by
\begin{equation}
\tau_1^z=\eta_1X_1,
\qquad
\tau_j^z=X_{j-1}X_j\quad (j=2,\ldots,L),
\qquad
\tau_N^z=\eta_LX_L,
\label{eq:S5-dual-z}
\end{equation}
and choose the dual $x$ operators so that
\begin{equation}
Z_j=\tau_j^x\tau_{j+1}^x,
\qquad j=1,\ldots,L.
\label{eq:S5-dual-x}
\end{equation}
This is the open-chain Kramers--Wannier map~\cite{KramersWannier1941}.
The fixed-boundary Hamiltonian becomes the standard free-open critical TFI
Hamiltonian on $N$ dual spins,
\begin{equation}
H_{\rm dual}
=-\frac12\sum_{j=1}^{N-1}\tau_j^x\tau_{j+1}^x
-\frac12\sum_{j=1}^{N}\tau_j^z,
\label{eq:S5-dual-H}
\end{equation}
subject to the exact parity constraint
\begin{equation}
\prod_{j=1}^{N}\tau_j^z=\eta_1\eta_L.
\label{eq:S5-dual-parity}
\end{equation}
Thus equal fixed signs select the even dual sector and opposite fixed signs
select the odd dual sector of the same free-open one-particle problem.

\subsection{Equal fixed signs and the pinned node}

Let $G_N^{(0)}$ be the $N\times N$ free-open polar matrix of type $C_N$.
Use the row grid $X_N$ and column grid $Y_N$ of
Eq.~\eqref{eq:S4-C-grids} with $L$ replaced by $N$, and denote the first
column node by
\begin{equation}
y_*=\frac{\pi}{2(2N+1)}\in Y_N.
\label{eq:S5-pinned-node}
\end{equation}
The column projection follows directly from the fixed dual-parity quotient.
Introduce left-oriented Jordan--Wigner Majoranas for the $N$ dual spins,
\begin{equation}
 a_j=\left(\prod_{\ell<j}\tau_\ell^z\right)\tau_j^x,
 \qquad
 b_j=\left(\prod_{\ell<j}\tau_\ell^z\right)\tau_j^y.
\label{eq:S5-dual-JW}
\end{equation}
Then
\begin{equation}
 Q_\tau:=\prod_{j=1}^{N}\tau_j^z
 =(-\ii)^N\prod_{j=1}^{N}a_jb_j.
\label{eq:S5-dual-parity-majorana}
\end{equation}
For a dual Majorana monomial
$M_{R,T}=\prod_{r\in R}b_r\prod_{t\in T}a_t$, multiplication by $Q_\tau$
replaces the occupied Majoranas by their complements, up to the ordering
phase,
\begin{equation}
 Q_\tau M_{R,T}\propto M_{R^c,T^c}.
\label{eq:S5-dual-complement-monomial}
\end{equation}
Inside the physical sector $Q_\tau=\eta_1\eta_L=\pm1$, so the two monomials
represent the same physical operator up to an irrelevant phase.  Each
complementary pair has a unique representative with $a_1$ absent, i.e.
$1\notin T$.  The real-space column $s=1$ of the parent $C_N$ matrix is the
node $y_*=\pi/[2(2N+1)]$.  Thus the physical quotient is represented by
omitting precisely this column:
\begin{equation}
\widehat G^{\pp}=G_N^{(0)}[:,Y_N\setminus\{y_*\}]
\in\mathbb R^{N\times(N-1)}.
\label{eq:S5-pp-column-delete}
\end{equation}
Because $G_N^{(0)}$ is orthogonal,
\begin{equation}
\widehat G^{\pp\mathsf T}\widehat G^{\pp}=I_L.
\label{eq:S5-pp-isometry}
\end{equation}

For a selected row set $R\subseteq X_N$ and a selected column set
$T\subseteq Y_N\setminus\{y_*\}$, with $|R|=|T|$, define
\begin{equation}
U_{\pp}(R,T)=X_{N,R}\cup Y_{N,T^c},
\label{eq:S5-pp-map}
\end{equation}
where the complement is taken in the full set $Y_N$.  Hence
$y_*\in U_{\pp}(R,T)$ for every minor.

For $N=L+1$,
\begin{equation}
\abs{\det\widehat G^{\pp}[R,T]}
=\frac{\mathfrak V_C(U_{\pp}(R,T))}{\mathfrak V_C(Y_N)},
\qquad y_*\in U_{\pp}(R,T).
\label{eq:S5-pp-termwise}
\end{equation}

\noindent\emph{Derivation.} The matrix in Eq.~\eqref{eq:S5-pp-column-delete} is a column restriction of
the square $C_N$ Cauchy matrix.  Apply the $C_L$ termwise identity in
the parent rank $N$.  Since the deleted column can never belong to $T$, it
always belongs to $T^c$ and is consequently present in the complementary
configuration.  No new determinant identity is required: pinning is exactly
the image of the missing Wick column established above.

Pinning breaks the finite Weyl symmetry of the sampled ensemble, but the
interaction and all collision hyperplanes remain those of $C_N$.

At unit fugacity, complementing an $N$-point configuration on the
$2N$-point parent lattice preserves its weight and exchanges configurations
that contain $y_*$ with those that do not.  Exactly one member of each pair
is pinned.  Therefore
\begin{equation}
\cZ^{\pp}_{\alpha,L}(1)
=\frac12\cZ^{\ff}_{\alpha,L+1}(1),
\qquad \alpha>0.
\label{eq:S5-rank-shift}
\end{equation}
This is an all-index identity for the physical equal-fixed spin chain.  It
will be used in Sec.~\ref{sec:S8} when the normalization-free boundary
ratios are formed.

\subsection{Opposite fixed signs and a one-mode insertion}

In the odd dual sector the lowest state is obtained by occupying the
Bogoliubov mode with the smallest singular value.  Let the free-open SVD be
\begin{equation}
G_N^{(0)}=U_NV_N^{\mathsf T},
\label{eq:S5-parent-mode-sum}
\end{equation}
and denote its final columns by
\begin{equation}
\ell=U_N[:,N],
\qquad r=V_N[:,N].
\label{eq:S5-lowest-mode-vectors}
\end{equation}
Changing the occupation of that mode flips its contribution to the cross
correlator and gives the exact rank-one update
\begin{equation}
G_N^{(1)}=G_N^{(0)}-2\ell r^{\mathsf T}.
\label{eq:S5-rank-one-update}
\end{equation}
The opposite-fixed augmented matrix is the same column restriction as in the
even sector,
\begin{equation}
\widehat G^{\pmfix}
=G_N^{(1)}[:,Y_N\setminus\{y_*\}].
\label{eq:S5-pm-matrix}
\end{equation}
It is again a vertical isometry because $G_N^{(1)}$ is orthogonal.

For a nonempty minor labeled by $(R,T)$, let $T$ also denote its indices in
the full parent column set.  The matrix determinant lemma yields
\begin{equation}
\det G_N^{(1)}[R,T]
=\det G_N^{(0)}[R,T] \,\Phi_N(R,T),
\label{eq:S5-pm-det-update}
\end{equation}
where
\begin{equation}
\Phi_N(R,T)
=1-2\,r_T^{\mathsf T}
\left[G_N^{(0)}[R,T]\right]^{-1}\ell_R,
\qquad
\Phi_N(\varnothing,\varnothing)=1.
\label{eq:S5-Phi-minor}
\end{equation}
Every parent Cauchy minor is nonzero, so Eq.~\eqref{eq:S5-Phi-minor} is
well-defined.  Through the pinned map of Eq.~\eqref{eq:S5-pp-map}, it defines
a function $\Phi_N(U)$ on pinned $N$-particle configurations.

For every opposite-fixed augmented minor,
\begin{equation}
\abs{\det\widehat G^{\pmfix}[R,T]}
=\frac{\mathfrak V_C(U_{\pp}(R,T))}{\mathfrak V_C(Y_N)}
\abs{\Phi_N(U_{\pp}(R,T))}.
\label{eq:S5-pm-termwise}
\end{equation}
The interaction is still pinned $C_N$; all new information is carried by a
single Bogoliubov-mode insertion.  The exact reduction of $\Phi_N(U)$ to the
fundamental symplectic character is postponed to Sec.~\ref{sec:S8}, where it
is needed for the defect-sensitive ratio.  Accidental zeros of $\Phi_N(U)$ can reduce the opposite-fixed Pauli support even though the underlying pinned $C_N$ interaction is unchanged.

\subsection{Endpoint-tuned \texorpdfstring{$B/C/D$}{B/C/D} realizations of the physical boundary sectors}
\label{sec:S5-folded-fixed}

The exact $C$-type constructions above are the canonical solvable
representatives used for the analytical boundary ratios in the Letter.  To
compare different microscopic terminations at fixed physical boundary class,
we now retain the transverse endpoint tunings of the $B$ and $D$ chains while
turning on the same longitudinal fields.  The resulting families are exact
finite-dimensional continuations of the free--fixed and fixed--fixed sectors.
They do not define new $B$- or $D$-type fixed-boundary Weyl ensembles; the
labels $B,C,D$ below record only the microscopic transverse termination.

Write
\begin{equation}
 \bm h_C=(1,\ldots,1),\qquad
 \bm h_B=(1,\ldots,1,\sqrt2),\qquad
 \bm h_D=(\sqrt2,1,\ldots,1,\sqrt2),
\label{eq:S5-folded-h-vectors}
\end{equation}
and let $\bm h^{\rm rev}$ denote the reversed vector.  For a right fixed
endpoint the auxiliary Majorana construction gives the rectangular
bidiagonal path
\begin{equation}
 \mathbb B_{\rm rec}(\bm h)=
 \begin{pmatrix}
 h_1&1&&&\\
 &h_2&1&&\\
 &&\ddots&\ddots&\\
 &&&h_L&1
 \end{pmatrix}_{L\times(L+1)}.
\label{eq:S5-folded-rect-B}
\end{equation}
Its vertical polar isometry is
\begin{equation}
 Q_{\rm rec}(\bm h)
 =\mathbb B_{\rm rec}(\bm h)^{\mathsf T}
 \bigl[\mathbb B_{\rm rec}(\bm h)
       \mathbb B_{\rm rec}(\bm h)^{\mathsf T}\bigr]^{-1/2}
 \in\mathbb R^{(L+1)\times L}.
\label{eq:S5-folded-rect-polar}
\end{equation}
Thus the two orientations of the mixed boundary sector are represented by
\begin{equation}
 \widehat G^{\fplus\mid R}=Q_{\rm rec}(\bm h_R),\qquad
 \widehat G^{+f\mid R}=Q_{\rm rec}(\bm h_R^{\rm rev}),
 \qquad R=B,C,D.
\label{eq:S5-folded-mixed-isometries}
\end{equation}
For $C$ the two matrices are related by reflection and reproduce the standard
free--fixed construction.  For $B$ they are genuinely distinct at finite
size because only one transverse endpoint is folded; for $D$ they are again
reflection equivalent.

For two fixed endpoints, Kramers--Wannier duality remains exact when the
transverse fields are nonuniform.  The $h_jZ_j$ terms become dual bonds, so
the square parent is
\begin{equation}
 \mathbb B_{\rm par}(\bm h)=
 \begin{pmatrix}
 1&h_1&&&&\\
 &1&h_2&&&\\
 &&\ddots&\ddots&&\\
 &&&1&h_L\\
 &&&&1
 \end{pmatrix}_{(L+1)\times(L+1)}.
\label{eq:S5-folded-parent-B}
\end{equation}
Let
\begin{equation}
 \mathbb B_{\rm par}(\bm h_R)=U_R\Sigma_RV_R^{\mathsf T},\qquad
 G_R^{(0)}=U_RV_R^{\mathsf T},
\label{eq:S5-folded-parent-polar}
\end{equation}
and let $\ell_R$ and $r_R$ be the left and right singular vectors associated
with its smallest singular value.  The odd-parity parent is obtained by the
same one-mode flip as in Eq.~\eqref{eq:S5-rank-one-update},
\begin{equation}
 G_R^{(1)}=G_R^{(0)}-2\ell_Rr_R^{\mathsf T}.
\label{eq:S5-folded-parent-flip}
\end{equation}
The fixed dual-parity quotient is still implemented by deleting the first
real-space column.  With $N=L+1$,
\begin{equation}
 \widehat G^{\pp\mid R}=G_R^{(0)}[:,\{2,\ldots,N\}],\qquad
 \widehat G^{\pmfix\mid R}=G_R^{(1)}[:,\{2,\ldots,N\}].
\label{eq:S5-folded-fixed-isometries}
\end{equation}
All matrices in Eqs.~\eqref{eq:S5-folded-mixed-isometries} and
\eqref{eq:S5-folded-fixed-isometries} are vertical isometries.  We define
$\cZ_{\alpha,L}^{ab\mid R}$ by inserting the corresponding matrix into the
all-minor sum~\eqref{eq:S-rectangular-minor-partition}.  The superscript
$ab\mid R$ deliberately separates the physical boundary condition $ab$ from
the microscopic transverse realization $R$.  In particular,
$\cZ_{\alpha,L}^{ab\mid C}$ reduces to the standard moments used in the
Letter.  For $R=B,D$ the matrices above are exact finite-dimensional
representations, but no claim is made that the folded-and-fixed sectors
possess a simple $B$- or $D$-type Weyl ensemble.  Direct spin-space checks of
these constructions are summarized in \hyperref[sec:S9-folded-special]{Sec.~S9.G}
and~\ref{sec:S12}.

\section{Unified discrete Selberg representations}
\label{sec:S6}

Sections~\ref{sec:S4} and~\ref{sec:S5} established termwise identities.
We now sum them and collect all boundary sectors in one notation.  Every formula in this section is an exact finite identity for its stated
minor/Selberg ensemble, for arbitrary real $\alpha>0$ and arbitrary
nonnegative fugacity $u$~\cite{ForresterWarnaar2008}.  For $\pp$ and $\pmfix$, the identification with
the physical spin-chain sectors uses the dual-parity quotient derived in S5.
Alias integrals and integer-index constant terms are separate transformations
and begin only in Sec.~\ref{sec:S7}.

\subsection{Square root-system ensembles}

For $R=A_{L-1},B_L,C_L,D_L$, let $X_R$ and $Y_R$ be the two $L$-point
sublattices defined in Sec.~\ref{sec:S4}, let $\mathfrak V_R$ be the
corresponding alternant, and set
\begin{equation}
\mathcal N_{R,L}=\mathfrak V_R(Y_R).
\label{eq:S6-square-reference}
\end{equation}
The bijection $(S,T)\leftrightarrow U_R(S,T)$ maps the complete square-minor
sum of Eq.~\eqref{eq:S-square-minor-partition} to all half-filled subsets of
$X_R\sqcup Y_R$.  Hence
\begin{equation}
\cZ^R_{\alpha,L}(u)
=\mathcal N_{R,L}^{-2\alpha}
\sum_{\substack{U\subset X_R\sqcup Y_R\\|U|=L}}
 u^{|U\cap X_R|}\,\mathfrak V_R(U)^{2\alpha}.
\label{eq:S6-square-master}
\end{equation}
In explicit normalizations,
\begin{equation}
\cZ^{A}_{\alpha,L}(u)
=L^{-\alpha L}
\sum_{\substack{U\subset\Omega_{2L}\\|U|=L}}
 u^{|U\cap X_A|}\abs{\Delta(U)}^{2\alpha},
\label{eq:S6-A-sum}
\end{equation}
\begin{equation}
\cZ^{C}_{\alpha,L}(u)
=\frac{2^{2\alpha L^2}}{(2L+1)^{\alpha L}}
\sum_{\substack{U\subset X_C\sqcup Y_C\\|U|=L}}
 u^{|U\cap X_C|}\mathfrak V_C(U)^{2\alpha},
\label{eq:S6-C-sum}
\end{equation}
\begin{equation}
\cZ^{B}_{\alpha,L}(u)
=\frac{2^{\alpha L(2L-1)}}{L^{\alpha L}}
\sum_{\substack{U\subset X_B\sqcup Y_B\\|U|=L}}
 u^{|U\cap X_B|}\mathfrak V_B(U)^{2\alpha},
\label{eq:S6-B-sum}
\end{equation}
and
\begin{equation}
\cZ^{D}_{\alpha,L}(u)
=\frac{2^{\alpha[2L(L-1)+1]}}{(2L-1)^{\alpha L}}
\sum_{\substack{U\subset X_D\sqcup Y_D\\|U|=L}}
 u^{|U\cap X_D|}\mathfrak V_D(U)^{2\alpha}.
\label{eq:S6-D-sum}
\end{equation}
The endpoint factors $\nu_B$ and $\nu_D$ are included inside
$\mathfrak V_B$ and $\mathfrak V_D$.  They are quadrature weights and do
not add roots to the logarithmic interaction.

\subsection{Rectangular and pinned C ensembles}

For the free--fixed chain, $M=L+1$, the configuration has $M$ particles on
a lattice with $|X_{\fplus}|=M$ and $|Y_{\fplus}|=M-1$.  The minor order is
$|U\cap Y_{\fplus}|$.  the rectangular identity gives
\begin{equation}
\cZ^{\fplus}_{\alpha,L}(u)
=\frac{2^{\alpha L(2L+3)}}{(L+1)^{\alpha(L+1)}}
\sum_{\substack{U\subset X_{\fplus}\sqcup Y_{\fplus}\\|U|=L+1}}
 u^{|U\cap Y_{\fplus}|}\mathfrak V_{C_{L+1}}(U)^{2\alpha}.
\label{eq:S6-fplus-sum}
\end{equation}
The unequal checkerboard sizes are the geometric origin of the
nonpalindromic degree polynomial.

For two fixed boundaries, set
\begin{equation}
N=L+1,
\qquad D_N=2N+1,
\label{eq:S6-fixed-rank}
\end{equation}
and use the parent $C_N$ sublattices $X_N,Y_N$ with the pinned node
$y_*\in Y_N$ from Eq.~\eqref{eq:S5-pinned-node}.  Equal fixed signs give
\begin{equation}
\cZ^{\pp}_{\alpha,L}(u)
=\frac{2^{2\alpha N^2}}{D_N^{\alpha N}}
\sum_{\substack{U\subset X_N\sqcup Y_N\\|U|=N,\ y_*\in U}}
 u^{|U\cap X_N|}\mathfrak V_C(U)^{2\alpha}.
\label{eq:S6-pp-sum}
\end{equation}
Opposite fixed signs carry the one-mode insertion,
\begin{equation}
\cZ^{\pmfix}_{\alpha,L}(u)
=\frac{2^{2\alpha N^2}}{D_N^{\alpha N}}
\sum_{\substack{U\subset X_N\sqcup Y_N\\|U|=N,\ y_*\in U}}
 u^{|U\cap X_N|}\mathfrak V_C(U)^{2\alpha}
 \abs{\Phi_N(U)}^{2\alpha}.
\label{eq:S6-pm-sum}
\end{equation}
Equations~\eqref{eq:S6-pp-sum} and~\eqref{eq:S6-pm-sum} have identical
pair interactions and identical pinning.  Their difference is entirely the
finite insertion $\Phi_N$.

\subsection{Master classification table}

Table~\ref{tab:S6-master} records the data needed to reconstruct every sum
without referring to an orientation-dependent minor convention.  The
configuration charge $k(U)$ is always the exponent of $u$ and equals the
augmented minor order.

\begin{table}[H]
\caption{Unified finite-lattice Selberg data.  The symbols $M=N=L+1$ in the
fixed-boundary rows.}
\label{tab:S6-master}
\centering
\scriptsize
\renewcommand{\arraystretch}{1.30}
\setlength{\tabcolsep}{3.0pt}
\begin{adjustbox}{max width=\textwidth}
\begin{tabular}{@{}lcccccc@{}}
\toprule
Sector & rank & $|X|$ & $|Y|$ & constraint on $U$ & $k(U)$ & insertion\\
\midrule
$\PBC$ & $A_{L-1}$ & $L$ & $L$ & $|U|=L$ & $|U\cap X_A|$ & $1$\\
$C\equiv\ff$ & $C_L$ & $L$ & $L$ & $|U|=L$ & $|U\cap X_C|$ & $1$\\
$B$ & $B_L$ & $L$ & $L$ & $|U|=L$ & $|U\cap X_B|$ & $1$\\
$D$ & $D_L$ & $L$ & $L$ & $|U|=L$ & $|U\cap X_D|$ & $1$\\
$\fplus$ & rectangular $C_M$ & $M$ & $M-1$ & $|U|=M$ & $|U\cap Y_{\fplus}|$ & $1$\\
$\pp$ & pinned $C_N$ & $N$ & $N$ & $|U|=N$, $y_*\in U$ & $|U\cap X_N|$ & $1$\\
$\pmfix$ & pinned $C_N$ & $N$ & $N$ & $|U|=N$, $y_*\in U$ & $|U\cap X_N|$ & $\abs{\Phi_N(U)}^{2\alpha}$\\
\bottomrule
\end{tabular}
\end{adjustbox}
\end{table}
The Lie labels in this table classify the zero hyperplanes of the Pauli
amplitudes.  In particular, the fixed physical boundaries remain $C$ type;
they do not become the parity-preserving $B$ or $D$ chains of
Sec.~\ref{sec:S4}.

\subsection{Normalization identities and degree support}

The determinant point follows directly from Cauchy--Binet and the orthogonal
or isometric character of the underlying matrix:
\begin{equation}
\cZ^{\mathsf b}_{1,L}(u)=(1+u)^L
\label{eq:S6-alpha-one}
\end{equation}
for every sector $\mathsf b$ in Table~\ref{tab:S6-master}, including the
inserted opposite-fixed state.  This is a stringent check on all prefactors
and fugacity conventions.

At the formal support-counting endpoint $\alpha=0$, every Cauchy minor of the
pure square, rectangular, and pinned uninserted ensembles is nonzero.  Their
polynomials are
\begin{equation}
\cZ^{R}_{0,L}(u)
=\sum_{k=0}^{L}\binom{L}{k}^2u^k,
\qquad R=A,B,C,D,
\label{eq:S6-square-support}
\end{equation}
and
\begin{equation}
\cZ^{\fplus}_{0,L}(u)
=\cZ^{\pp}_{0,L}(u)
=\sum_{k=0}^{L}\binom{L+1}{k}\binom{L}{k}u^k.
\label{eq:S6-rect-support}
\end{equation}
These are support identities rather than analytic continuations of the
Selberg weights.  The opposite-fixed insertion may vanish on special
configurations, so Eq.~\eqref{eq:S6-rect-support} does not apply to its
nonzero support without removing the zeros of $\Phi_N$.

Jacobi's complementary-minor identity gives the square-sector palindromy
\begin{equation}
\cZ^R_{\alpha,L}(u)
=u^L\cZ^R_{\alpha,L}(u^{-1}),
\qquad R=A,B,C,D.
\label{eq:S6-square-palindromy}
\end{equation}
There is no corresponding generic identity for a rectangular isometry.  For
free--fixed boundaries, the maximal minors are, up to signs, the components
of the unit null vector of $\widehat G^{\fplus\mathsf T}$.  From the
half-shifted sine transform of Sec.~\ref{sec:S5}, the missing mode is
$\sin[\pi(r+1/2)]=(-1)^r$, so the normalized null vector is
$n_r=(-1)^r/\sqrt{L+1}$.  Hence every maximal minor has magnitude
$(L+1)^{-1/2}$, and therefore
\begin{equation}
[u^L]\cZ^{\fplus}_{\alpha,L}(u)=(L+1)^{1-\alpha}.
\label{eq:S6-fplus-leading}
\end{equation}
Together with $[u^0]\cZ^{\fplus}_{\alpha,L}=1$, this explicitly demonstrates
its nonpalindromic character.

At $u=1$, the equal-fixed pinning identity of Eq.~\eqref{eq:S5-rank-shift}
relates different physical lengths.  It is not a consequence of ordinary
rectangular palindromy; it follows from complement pairing in the parent
$C_{L+1}$ ensemble.

\subsection{Logarithmic-gas interpretation}

Ignoring configuration-independent constants and the discrete half-orbit
weights, write $\mathfrak V_R(U)^{2\alpha}=\exp[-\alpha E_R(U)]$.  The
root-dependent energies are
\begin{equation}
E_A=-2\sum_{i<j}\log\left|2\sin\frac{\theta_i-\theta_j}{2}\right|,
\label{eq:S6-energy-A}
\end{equation}
\begin{equation}
E_B=-2\sum_{i<j}\log\abs{\sin(\theta_i-\theta_j)
\sin(\theta_i+\theta_j)}-2\sum_i\log\abs{\sin\theta_i},
\label{eq:S6-energy-B}
\end{equation}
\begin{equation}
E_C=-2\sum_{i<j}\log\abs{\sin(\theta_i-\theta_j)
\sin(\theta_i+\theta_j)}-2\sum_i\log\abs{\sin2\theta_i},
\label{eq:S6-energy-C}
\end{equation}
and
\begin{equation}
E_D=-2\sum_{i<j}\log\abs{\sin(\theta_i-\theta_j)
\sin(\theta_i+\theta_j)}.
\label{eq:S6-energy-D}
\end{equation}
Thus the interaction is logarithmic.  The statistical weight is written
first because it is the quantity supplied exactly by the Pauli minors; the
energy is then its negative logarithm.  For the rectangular and pinned
sectors the local interaction remains $E_C$, while the allowed lattice,
pinning, and insertion modify the ensemble globally.

\section{Alias integrals and constant-term representations}
\label{sec:S7}

The discrete Selberg sums of Sec.~\ref{sec:S6} are already exact for every
real \(\alpha>0\).  This section rewrites those finite sums in two ways.  A
Dirac-comb integral keeps arbitrary real \(\alpha\) and makes the relation to
continuous log gases explicit.  At positive integer \(\alpha\), the Weyl
weights become Laurent polynomials and finite sampling can be converted to an
exact constant term.  The latter transformation is algebraic; it does not by
itself imply a product formula or an efficient evaluation.

\subsection{Checkerboard Fourier coefficients}

Assign fugacity \(u\) to the charged checkerboard sublattice and unit weight
to the other sublattice.  The two-periodic sequence has Fourier coefficients
\begin{equation}
 c_q(u)=
 \begin{cases}
 (1+u)/2,&q\ \text{even},\\[1mm]
 (u-1)/2,&q\ \text{odd}.
 \end{cases}
\label{eq:S7-cq}
\end{equation}
Thus \(c_q(1)\) vanishes for odd \(q\), a simple observation that will become
important for the unit-fugacity character quotient in Sec.~\ref{sec:S8}.
For later use, \(\delta_T\) denotes the periodic delta distribution of period
\(T\).

\subsection{Periodic reference}

The circular $A_{L-1}$ Dirac-comb and aliased constant-term representations
are derived in Ref.~\cite{Companion}.  We use them only as a reference when
comparing the boundary ensembles and therefore do not repeat their Fourier
and Poisson-summation derivation here.

\subsection{Open \texorpdfstring{$C_L$}{C(L)} comb}

Set
\begin{equation}
 D_L=2L+1.
\label{eq:S7-DL}
\end{equation}
Unfold the interval grid to a \(2D_L\)-point comb of period \(\pi\):
\begin{equation}
 \mathcal A^{C}_{D_L,u}(\theta)
 =\pi\sum_{a=0}^{2D_L-1}w_u(a)
 \delta_{\pi}\!\left(\theta-\frac{\pi a}{2D_L}\right)
 =2D_L\sum_{q\in\mathbb Z}c_q(u)\ee^{2\ii qD_L\theta}.
\label{eq:S7-C-comb}
\end{equation}
Reflection maps \(\theta\) to \(\pi-\theta\).  Every nonfixed physical node
therefore appears twice in the unfolded comb, while any configuration
containing both reflected copies has zero Weyl weight.  Accounting for this
orbit multiplicity gives
\begin{align}
 \cZ^{C}_{\alpha,L}(u)
 ={}&\frac{2^{2\alpha L^2-L}}{L!D_L^{\alpha L}}
 \int_0^{\pi}\prod_{j=1}^{L}
 \left[\frac{\dd\theta_j}{\pi}\mathcal A^{C}_{D_L,u}(\theta_j)\right]
 \prod_j\abs{\sin2\theta_j}^{2\alpha}
 \nonumber\\[-1mm]
 &\times\prod_{i<j}
 \abs{\sin(\theta_i-\theta_j)\sin(\theta_i+\theta_j)}^{2\alpha}.
\label{eq:S7-C-alias-integral}
\end{align}
The endpoint nodes \(0\) and \(\pi/2\) may be left in the comb because the
long-root factor \(\sin2\theta\) annihilates them.

\subsection{Free--fixed rectangular comb}

For \(M=L+1\), the union \(X_{\fplus}\sqcup Y_{\fplus}\) consists of the
interior points \(\pi a/(4M)\), \(a=1,\ldots,2M-1\).  The charged sublattice
is \(Y_{\fplus}\), corresponding to even \(a\).  Embed it in the full
\(4M\)-point comb
\begin{equation}
 \mathcal A^{\fplus}_{M,u}(\theta)
 =\pi\sum_{a=0}^{4M-1}w_u(a)
 \delta_{\pi}\!\left(\theta-\frac{\pi a}{4M}\right)
 =4M\sum_{q\in\mathbb Z}c_q(u)\ee^{4\ii qM\theta}.
\label{eq:S7-fplus-comb}
\end{equation}
The exact rectangular integral is
\begin{align}
 \cZ^{\fplus}_{\alpha,L}(u)
 ={}&\frac{2^{\alpha(M-1)(2M+1)-M}}
 {M!M^{\alpha M}}
 \int_0^{\pi}\prod_{j=1}^{M}
 \left[\frac{\dd\theta_j}{\pi}\mathcal A^{\fplus}_{M,u}(\theta_j)\right]
 \prod_j\abs{\sin2\theta_j}^{2\alpha}
 \nonumber\\[-1mm]
 &\times\prod_{i<j}
 \abs{\sin(\theta_i-\theta_j)\sin(\theta_i+\theta_j)}^{2\alpha}.
\label{eq:S7-fplus-alias-integral}
\end{align}
The factor \(2^{-M}\) contained in the prefactor removes the independent
choice of reflected representative.  Nodes absent from the physical
rectangular grid either have zero long-root weight or are paired by
reflection, so the comb introduces no spurious configuration.

\subsection{Pinned and inserted \texorpdfstring{$C_N$}{C(N)} integrals}

For the two-fixed-boundary sectors use
\begin{equation}
 N=L+1,
 \qquad
 D_N=2N+1,
 \qquad
 \theta_* =\frac{\pi}{2D_N}.
\label{eq:S7-fixed-data}
\end{equation}
Let \(\mathcal W_{C_N}\) denote the absolute trigonometric \(C_N\) Weyl
denominator without the reference normalization.  The equal-fixed sum is
\begin{align}
 \cZ^{\pp}_{\alpha,L}(u)
 ={}&\frac{2^{2\alpha N^2-L}}{L!D_N^{\alpha N}}
 \int_0^{\pi}\prod_{j=1}^{L}
 \left[\frac{\dd\theta_j}{\pi}\mathcal A^{C}_{D_N,u}(\theta_j)\right]
 \nonumber\\[-1mm]
 &\times
 \mathcal W_{C_N}(\theta_*,\theta_1,\ldots,\theta_L)^{2\alpha}.
\label{eq:S7-pp-alias-integral}
\end{align}
For opposite fixed signs the same measure carries the exact one-mode
insertion of Sec.~\ref{sec:S5}:
\begin{align}
 \cZ^{\pmfix}_{\alpha,L}(u)
 ={}&\frac{2^{2\alpha N^2-L}}{L!D_N^{\alpha N}}
 \int_0^{\pi}\prod_{j=1}^{L}
 \left[\frac{\dd\theta_j}{\pi}\mathcal A^{C}_{D_N,u}(\theta_j)\right]
 \nonumber\\[-1mm]
 &\times
 \mathcal W_{C_N}(\theta_*,\boldsymbol\theta)^{2\alpha}
 \abs{\Phi_N(\theta_*,\boldsymbol\theta)}^{2\alpha}.
\label{eq:S7-pm-alias-integral}
\end{align}
Only values on the support of the comb are required.  Section~\ref{sec:S8}
will show that \(\Phi_N\) is the fundamental symplectic character, turning
Eq.~\eqref{eq:S7-pm-alias-integral} into a character-inserted Selberg
integral.

\subsection{Atomic measures for the folded \texorpdfstring{$B_L$}{B(L)} and
\texorpdfstring{$D_L$}{D(L)} grids}

The reflection-fixed quadrature factors \(\nu_B\) and \(\nu_D\) of
Eqs.~\eqref{eq:S4-B-half-weight} and~\eqref{eq:S4-D-half-weight} depend on
individual nodes.  Consequently, the sparse two-periodic comb used for
\(A\) and \(C\) is replaced by an exact weighted atomic measure.  Let
\(\Theta_R=X_R\sqcup Y_R\), \(R=B,D\), and define
\begin{equation}
 \dd\mu^{R}_{\alpha,u}(\theta)
 =\sum_{\vartheta\in\Theta_R}
 u^{\mathbf 1_{\vartheta\in X_R}}
 \nu_R(\vartheta)^{2\alpha}
 \delta(\theta-\vartheta)\,\dd\theta.
\label{eq:S7-BD-atomic-measure}
\end{equation}
Writing \(\Delta_R^{\rm root}\) for the Weyl denominator with the
quadrature factors removed, the exact real-index representation is
\begin{equation}
 \cZ^{R}_{\alpha,L}(u)
 =\frac{\mathcal N_{R,L}^{-2\alpha}}{L!}
 \int\prod_{j=1}^{L}\dd\mu^{R}_{\alpha,u}(\theta_j)
 \abs{\Delta_R^{\rm root}(\boldsymbol\theta)}^{2\alpha},
 \qquad R=B,D.
\label{eq:S7-BD-atomic-integral}
\end{equation}
This form separates the root interaction from the half-orbit sampling
weights and remains valid for arbitrary real \(\alpha>0\).

\subsection{Laurent products and bandwidths at integer index}

Now let \(\alpha\in\mathbb N\) and set \(x_j=\ee^{\ii\theta_j}\) for the
periodic problem and \(z_j=\ee^{2\ii\theta_j}\) for interval problems.  The
relevant Laurent products are
\begin{equation}
 D^{(A)}_{\alpha,L}(\boldsymbol x)
 =\prod_{\substack{i,j=1\\i\ne j}}^{L}
 \left(1-\frac{x_i}{x_j}\right)^{\alpha},
\label{eq:S7-DA}
\end{equation}
\begin{align}
 D^{(C)}_{\alpha,n}(\boldsymbol z)
 ={}&\prod_{j=1}^{n}(1-z_j^2)^{\alpha}(1-z_j^{-2})^{\alpha}
 \nonumber\\
 &\times\prod_{i<j}
 \left(1-\frac{z_i}{z_j}\right)^{\alpha}
 \left(1-\frac{z_j}{z_i}\right)^{\alpha}
 (1-z_iz_j)^{\alpha}
 \left(1-\frac1{z_iz_j}\right)^{\alpha},
\label{eq:S7-DC}
\end{align}
\begin{align}
 D^{(B)}_{\alpha,L}(\boldsymbol z)
 ={}&\prod_j(1-z_j)^{\alpha}(1-z_j^{-1})^{\alpha}
 \nonumber\\
 &\times\prod_{i<j}
 \left(1-\frac{z_i}{z_j}\right)^{\alpha}
 \left(1-\frac{z_j}{z_i}\right)^{\alpha}
 (1-z_iz_j)^{\alpha}
 \left(1-\frac1{z_iz_j}\right)^{\alpha},
\label{eq:S7-DB}
\end{align}
\begin{equation}
 D^{(D)}_{\alpha,L}(\boldsymbol z)
 =\prod_{i<j}
 \left(1-\frac{z_i}{z_j}\right)^{\alpha}
 \left(1-\frac{z_j}{z_i}\right)^{\alpha}
 (1-z_iz_j)^{\alpha}
 \left(1-\frac1{z_iz_j}\right)^{\alpha}.
\label{eq:S7-DD}
\end{equation}
Their one-variable bandwidths are
\begin{equation}
 b_A=\alpha(L-1),
 \qquad
 b_C=2\alpha n,
 \qquad
 b_B=\alpha(2L-1),
 \qquad
 b_D=2\alpha(L-1).
\label{eq:S7-bandwidths}
\end{equation}
These finite bandwidths are what truncate the formally infinite Fourier
combs.

\subsection{Sparse constant terms for \texorpdfstring{$A$}{A} and
\texorpdfstring{$C$}{C}}

For reference, the periodic result of Ref.~\cite{Companion} can be written by defining
\begin{equation}
 K^{A}_{\alpha,u}(x)
 =\sum_{q=-(\alpha-1)}^{\alpha-1}c_q(u)x^{qL}.
\label{eq:S7-KA}
\end{equation}
The exact constant term is
\begin{equation}
 \cZ^{A}_{\alpha,L}(u)
 =\frac{2^L}{L!L^{(\alpha-1)L}}
 \CT_{\boldsymbol x}\left[
 D^{(A)}_{\alpha,L}(\boldsymbol x)
 \prod_{j=1}^{L}K^{A}_{\alpha,u}(x_j)
 \right].
\label{eq:S7-A-CT}
\end{equation}
For free--free boundaries define
\begin{equation}
 K^{C}_{\alpha,u}(z)
 =\sum_{q=-(\alpha-1)}^{\alpha-1}c_q(u)z^{qD_L},
\label{eq:S7-KC}
\end{equation}
and obtain
\begin{equation}
 \cZ^{C}_{\alpha,L}(u)
 =\frac{D_L^{(1-\alpha)L}}{L!}
 \CT_{\boldsymbol z}\left[
 D^{(C)}_{\alpha,L}(\boldsymbol z)
 \prod_{j=1}^{L}K^{C}_{\alpha,u}(z_j)
 \right].
\label{eq:S7-C-CT}
\end{equation}
The zero-alias identities are Dyson's and Macdonald's constant terms,
respectively~\cite{Dyson1962,Morris1982,Macdonald1982}:
\begin{equation}
 \CT D^{(A)}_{\alpha,L}
 =\frac{(\alpha L)!}{(\alpha!)^L},
\label{eq:S7-Dyson}
\end{equation}
\begin{equation}
 \CT D^{(C)}_{\alpha,L}
 =\prod_{j=1}^{L}\binom{2\alpha j}{\alpha}.
\label{eq:S7-Macdonald-C}
\end{equation}
These products evaluate the full Pauli problem only when all nonzero aliases
are absent or can be summed separately.

\subsection{Outer aliases of the rectangular ensemble}

For the free--fixed comb, \(z=\ee^{2\ii\theta}\) and the Fourier modes are
\(z^{2qM}\).  Since the \(C_M\) bandwidth reaches \(\pm2\alpha M\), the
outer charges \(q=\pm\alpha\) survive.  Define
\begin{equation}
 K^{\fplus}_{\alpha,u}(z)
 =\sum_{q=-\alpha}^{\alpha}c_q(u)z^{2qM}.
\label{eq:S7-Kfplus}
\end{equation}
Then
\begin{equation}
 \cZ^{\fplus}_{\alpha,L}(u)
 =\frac{2^{M-\alpha(M+1)}M^{(1-\alpha)M}}{M!}
 \CT_{\boldsymbol z}\left[
 D^{(C)}_{\alpha,M}(\boldsymbol z)
 \prod_{j=1}^{M}K^{\fplus}_{\alpha,u}(z_j)
 \right].
\label{eq:S7-fplus-CT}
\end{equation}
Omitting \(q=\pm\alpha\) fails already for \(L=1\).  For example,
at \(\alpha=1\), \(M=2\), the complete kernel gives
\begin{equation}
 \cZ^{\fplus}_{1,1}(u)=1+u,
\label{eq:S7-fplus-outer-example-full}
\end{equation}
whereas retaining only the zero alias would incorrectly give
\begin{equation}
 \cZ^{\fplus,(q=0)}_{1,1}(u)=\frac12(1+u)^2.
\label{eq:S7-fplus-outer-example-zero}
\end{equation}
The rectangular grid is therefore not obtained by a naive replacement
\(L\mapsto L+1\) in the free--free kernel; its missing endpoint changes the
Nyquist range of the sampling problem.

\subsection{Pinned and inserted constant terms}

Set
\begin{equation}
 \xi=\ee^{\ii\pi/D_N},
\label{eq:S7-xi}
\end{equation}
which is the Laurent variable of the pinned node.  The equal-fixed kernel is
\begin{equation}
 K^{\pp}_{\alpha,u}(z)
 =\sum_{q=-(\alpha-1)}^{\alpha-1}c_q(u)z^{qD_N}.
\label{eq:S7-Kpp}
\end{equation}
The exact pinned constant term is
\begin{equation}
 \cZ^{\pp}_{\alpha,L}(u)
 =\frac{D_N^{L-\alpha N}}{L!}
 \CT_{z_1,\ldots,z_L}\left[
 D^{(C)}_{\alpha,N}(z_1,\ldots,z_L,\xi)
 \prod_{j=1}^{L}K^{\pp}_{\alpha,u}(z_j)
 \right].
\label{eq:S7-pp-CT}
\end{equation}
For opposite fixed signs, choose any Laurent polynomial
\(\mathcal I^{\pmfix}_{\alpha,N}\) whose values agree with
\(\abs{\Phi_N}^{2\alpha}\) on the physical collision-free pinned support.
Such an extension always exists by finite Fourier interpolation on the full
sampling grid; different choices away from physical support give the same
constant term because the Weyl factor vanishes there.  Thus
\begin{equation}
 \cZ^{\pmfix}_{\alpha,L}(u)
 =\frac{D_N^{L-\alpha N}}{L!}
 \CT\left[
 D^{(C)}_{\alpha,N}(\boldsymbol z,\xi)
 \mathcal I^{\pmfix}_{\alpha,N}(\boldsymbol z)
 \prod_{j=1}^{L}K^{\pp}_{\alpha,u}(z_j)
 \right].
\label{eq:S7-pm-CT-interpolated}
\end{equation}
Equation~\eqref{eq:S7-pm-CT-interpolated} is exact but is not a bare
Dyson--Morris identity.  The character reduction of Sec.~\ref{sec:S8}
provides a much more informative representation at unit fugacity.

\subsection{Finite Fourier kernels for \texorpdfstring{$B$}{B} and
\texorpdfstring{$D$}{D}}

Let \(z_a^{(R)}=\ee^{2\ii\theta_a^{(R)}}\) denote the \(2L\) physical nodes
of \(R=B,D\), and define the sampling weights
\begin{equation}
 \rho_a^{(R)}(\alpha,u)
 =u^{\mathbf 1_{\theta_a^{(R)}\in X_R}}
 \nu_R(\theta_a^{(R)})^{2\alpha}.
\label{eq:S7-BD-rho}
\end{equation}
For \(|m|\le b_R\), set
\begin{equation}
 \mu_m^{(R)}(\alpha,u)
 =\sum_{a=1}^{2L}\rho_a^{(R)}(\alpha,u)
 (z_a^{(R)})^m,
\qquad
 K^{R}_{\alpha,L,u}(z)
 =\sum_{m=-b_R}^{b_R}\mu_m^{(R)}(\alpha,u)z^{-m}.
\label{eq:S7-BD-kernel}
\end{equation}
For every Laurent polynomial \(f\) of bandwidth at most \(b_R\),
\begin{equation}
 \sum_{a=1}^{2L}\rho_a^{(R)}f(z_a^{(R)})
 =\CT_z\left[f(z)K^{R}_{\alpha,L,u}(z)\right].
\label{eq:S7-finite-sampling}
\end{equation}
Applying this identity independently to every particle gives
\begin{equation}
 \cZ^{B}_{\alpha,L}(u)
 =\frac{\mathcal N_{B,L}^{-2\alpha}2^{-2\alpha L^2}}{L!}
 \CT\left[
 D^{(B)}_{\alpha,L}(\boldsymbol z)
 \prod_{j=1}^{L}K^{B}_{\alpha,L,u}(z_j)
 \right],
\label{eq:S7-B-DFT-CT}
\end{equation}
\begin{equation}
 \cZ^{D}_{\alpha,L}(u)
 =\frac{\mathcal N_{D,L}^{-2\alpha}2^{-2\alpha L(L-1)}}{L!}
 \CT\left[
 D^{(D)}_{\alpha,L}(\boldsymbol z)
 \prod_{j=1}^{L}K^{D}_{\alpha,L,u}(z_j)
 \right].
\label{eq:S7-D-DFT-CT}
\end{equation}
The continuous zero-alias products are
\begin{equation}
 \CT D^{(B)}_{\alpha,L}
 =\prod_{j=1}^{L}\binom{2\alpha j}{\alpha},
\label{eq:S7-Macdonald-B}
\end{equation}
\begin{equation}
 \CT D^{(D)}_{\alpha,L}
 =\binom{\alpha L}{\alpha}
 \prod_{j=1}^{L-1}\binom{2\alpha j}{\alpha}.
\label{eq:S7-Macdonald-D}
\end{equation}
The endpoint Fourier kernels are nevertheless part of the finite Pauli
problem and cannot in general be replaced by these zero-alias products.

\subsection{Classification of the constant-term problems}

\begin{table}[H]
\caption{Exact integer-index representations.  ``Sparse'' means that the
kernel contains only root-lattice charges fixed by a bandwidth bound; it does
not mean that the resulting multivariate constant term is easy to evaluate.}
\label{tab:S7-CT-classification}
\centering
\scriptsize
\renewcommand{\arraystretch}{1.25}
\setlength{\tabcolsep}{3.5pt}
\begin{adjustbox}{max width=\textwidth}
\begin{tabular}{@{}lcccc@{}}
\toprule
Sector & Laurent product & sampling frequencies & surviving range & form\\
\midrule
$A$ & $D^{(A)}_{\alpha,L}$ & $qL$ & $|q|\le\alpha-1$ & sparse Dyson\\
$C\equiv\ff$ & $D^{(C)}_{\alpha,L}$ & $q(2L+1)$ & $|q|\le\alpha-1$ & sparse Morris\\
$B$ & $D^{(B)}_{\alpha,L}$ & all $|m|\le b_B$ & finite DFT & weighted Macdonald\\
$D$ & $D^{(D)}_{\alpha,L}$ & all $|m|\le b_D$ & finite DFT & weighted Macdonald\\
$\fplus$ & $D^{(C)}_{\alpha,L+1}$ & $2q(L+1)$ & $|q|\le\alpha$ & outer aliases\\
$\pp$ & pinned $D^{(C)}_{\alpha,L+1}$ & $q(2L+3)$ & $|q|\le\alpha-1$ & pinned Morris\\
$\pmfix$ & pinned $D^{(C)}_{\alpha,L+1}$ & same plus insertion & finite & inserted Morris\\
\bottomrule
\end{tabular}
\end{adjustbox}
\end{table}

\section{Character insertion and normalization-free invariants}
\label{sec:S8}

Section~\ref{sec:S5} identified the opposite-fixed state as a rank-one
Bogoliubov update of the equal-fixed parent and introduced the exact minor
insertion \(\Phi_N\).  We now reduce that insertion to the fundamental
symplectic character, derive its aliased constant-term quotient, and classify
all endpoint-normalization-free combinations of the four physical open
boundary sectors.

\subsection{Rank-one minor ratio in Cauchy variables}

Use
\begin{equation}
 N=L+1,
 \qquad
 D=2N+1.
\label{eq:S8-ND}
\end{equation}
The parent \(C_N\) matrix has row variables
\begin{equation}
 x_r=\sin^2\!\left(\frac{\pi r}{D}\right),
 \qquad r=1,\ldots,N,
\label{eq:S8-x-grid}
\end{equation}
and column variables
\begin{equation}
 y_s=\sin^2\!\left(\frac{\pi(s-1/2)}{D}\right),
 \qquad s=1,\ldots,N.
\label{eq:S8-y-grid}
\end{equation}
The polar matrix and its last singular vectors are, explicitly,
\begin{equation}
 G^{(0)}_{rs}
 =\frac{2}{D}(-1)^{r-s}
 \frac{\sqrt{x_r(1-y_s)}}{x_r-y_s},
\label{eq:S8-parent-Cauchy}
\end{equation}
\begin{equation}
 (U_N)_{rN}=\frac{2}{\sqrt D}(-1)^{r+1}\sqrt{x_r},
 \qquad
 (V_N)_{sN}=\frac{2}{\sqrt D}(-1)^{s+1}\sqrt{1-y_s}.
\label{eq:S8-last-mode}
\end{equation}
Thus, after extracting the same diagonal signs and one-body factors from
both the parent matrix and the update, every selected parent minor reduces
to the Cauchy core
\begin{equation}
 C_{R,T}=\left[\frac1{x_r-y_s}\right]_{r\in R,\,s\in T},
\label{eq:S8-Cauchy-minor}
\end{equation}
and the sign flip of the last Bogoliubov mode acts as
\begin{equation}
 C_{R,T}\longmapsto C_{R,T}-4\boldsymbol1\boldsymbol1^{\mathsf T}.
\label{eq:S8-core-rank-one}
\end{equation}
The numerical coefficient four is therefore fixed by the relative
normalization of Eqs.~\eqref{eq:S8-parent-Cauchy} and
\eqref{eq:S8-last-mode}; it is not an adjustable convention.  The matrix
determinant lemma gives
\begin{equation}
 \frac{\det G_N^{(1)}[R,T]}{\det G_N^{(0)}[R,T]}
 =1-4\,\boldsymbol 1^{\mathsf T}C_{R,T}^{-1}\boldsymbol 1.
\label{eq:S8-rank-one-ratio}
\end{equation}
The empty-minor ratio is defined to be one.

For \(C_{ij}=1/(x_i-y_j)\) with pairwise distinct \(x_i\) and \(y_j\),
\begin{equation}
 \boldsymbol 1^{\mathsf T}C^{-1}\boldsymbol 1
 =\sum_i x_i-\sum_j y_j.
\label{eq:S8-Cauchy-inverse-sum}
\end{equation}

\noindent\emph{Derivation.} Consider
\begin{equation}
 r(z)=\frac{\prod_i(z-x_i)}{\prod_j(z-y_j)}
 =1+\sum_j\frac{A_j}{z-y_j}.
\label{eq:S8-rational-function}
\end{equation}
Evaluation at \(z=x_i\) yields \(CA=-\boldsymbol 1\).  The coefficient of
\(z^{-1}\) in the expansion at infinity is
\(\sum_jy_j-\sum_ix_i\), hence
\begin{equation}
 \sum_jA_j=\sum_jy_j-\sum_ix_i.
\label{eq:S8-residue-sum}
\end{equation}
Since $A=-C^{-1}\boldsymbol1$, summing its components proves
Eq.~\eqref{eq:S8-Cauchy-inverse-sum}.

Substitution into Eq.~\eqref{eq:S8-rank-one-ratio} gives the explicit
insertion
\begin{equation}
 \Phi_N(R,T)
 =1-4\left(\sum_{r\in R}x_r-\sum_{s\in T}y_s\right).
\label{eq:S8-Phi-RT}
\end{equation}
This proves analytically that the determinant ratio defined abstractly in
Eq.~\eqref{eq:S5-Phi-minor} is a linear statistic of the selected Cauchy
nodes.

\subsection{Complementary configuration and symplectic character}

The pinned configuration corresponding to \((R,T)\) is
\begin{equation}
 U(R,T)=X_{N,R}\cup Y_{N,T^c},
 \qquad
 y_*\in U(R,T),
\label{eq:S8-U-map}
\end{equation}
where the complement is taken in the full \(N\)-point set \(Y_N\), including
the deleted node \(y_*\).  The odd-grid identity
\begin{equation}
 \sum_{s=1}^{N}y_s=\frac{2N-1}{4}
\label{eq:S8-y-sum}
\end{equation}
follows from a finite cosine sum.  Since
\begin{equation}
 \sum_{\lambda\in U(R,T)}\lambda
 =\sum_{r\in R}x_r+\sum_{s\notin T}y_s,
\label{eq:S8-U-sum}
\end{equation}
Eq.~\eqref{eq:S8-Phi-RT} becomes
\begin{equation}
 \Phi_N(U)=2N-4\sum_{\lambda\in U}\lambda.
\label{eq:S8-Phi-lambda}
\end{equation}
Write every node as \(\lambda_a=\sin^2\theta_a\) and set
\(z_a=\ee^{2\ii\theta_a}\).  Because \(|U|=N\),
\begin{align}
 \Phi_N(U)
 &=2\sum_{\theta_a\in U}\cos2\theta_a
 \nonumber\\
 &=\sum_{\theta_a\in U}(z_a+z_a^{-1}).
\label{eq:S8-Phi-character-sum}
\end{align}
The defining representation of \(\Sp(2N)\) has weights
\(\{\pm e_1,\ldots,\pm e_N\}\), so the last expression is its torus
character.

\begin{theorem}[Character insertion]
\label{thm:S8-character}
For every pinned configuration of the opposite-fixed problem,
\begin{equation}
 \Phi_N(U)=\chi_{\omega_1}^{(C_N)}(\boldsymbol z_U)
 =\sum_{j=1}^{N}(z_j+z_j^{-1}).
\label{eq:S8-character-identity}
\end{equation}
Consequently, at unit fugacity,
\begin{equation}
 \cR_{\alpha,L}^{(\eta)}
 \coloneqq
 \frac{\cZ_{\alpha,L}^{\pmfix}(1)}
 {\cZ_{\alpha,L}^{\pp}(1)}
 =\left\langle
 \abs{\chi_{\omega_1}^{(C_{L+1})}}^{2\alpha}
 \right\rangle_{\alpha,L+1}^{\rm pinned}.
\label{eq:S8-Reta-character}
\end{equation}
\end{theorem}

Here the pinned expectation is normalized by the equal-fixed Selberg weight:
\begin{equation}
 \avg{F}_{\alpha,N}^{\rm pinned}
 =\frac{
 \displaystyle\sum_{\substack{U\subset X_N\sqcup Y_N\\|U|=N,\,y_*\in U}}
 \mathfrak V_C(U)^{2\alpha}F(U)}
 {
 \displaystyle\sum_{\substack{U\subset X_N\sqcup Y_N\\|U|=N,\,y_*\in U}}
 \mathfrak V_C(U)^{2\alpha}}.
\label{eq:S8-pinned-average}
\end{equation}
The equality is finite-size and exact for every real \(\alpha>0\); no
thermodynamic or Gaussian assumption has been used.

\subsection{Pinned average versus the full parent ensemble}

Complementation maps a half-filled configuration \(U\) to \(U^c\).  The
parent complementary-minor identity implies
\begin{equation}
 \mathfrak V_C(U)=\mathfrak V_C(U^c),
\label{eq:S8-complement-weight}
\end{equation}
while the complete-grid cosine sum vanishes and hence
\begin{equation}
 \chi_{\omega_1}(U^c)=-\chi_{\omega_1}(U).
\label{eq:S8-complement-character}
\end{equation}
Every complementary pair contains the pinned node in exactly one member.
Therefore any complement-even observable has the same normalized expectation
in the pinned ensemble and in the full parent ensemble.

For every \(\alpha>0\),
\begin{equation}
 \left\langle\abs{\chi_{\omega_1}}^{2\alpha}\right\rangle_{\rm pinned}
 =\left\langle\abs{\chi_{\omega_1}}^{2\alpha}\right\rangle_{\rm full}.
\label{eq:S8-pinned-full}
\end{equation}

This identity is useful because the full parent ensemble is permutation and
complement symmetric, whereas the pinned representation keeps direct contact
with the physical equal-fixed chain.

\subsection{Aliased character quotient at integer index}

Let \(\alpha\in\mathbb N\), \(D=2N+1\), and use the full parent
\(C_N\) ensemble justified by the complement identity Eq.~\eqref{eq:S8-pinned-full}.  At \(u=1\),
Eq.~\eqref{eq:S7-cq} leaves only even root-lattice charges.  Define
\begin{equation}
 K_Q^{(D)}(z)
 =\sum_{\substack{|q|\le Q\\q\ \mathrm{even}}}z^{qD}.
\label{eq:S8-even-kernel}
\end{equation}
The one-variable bandwidth of \(D^{(C)}_{\alpha,N}\) is
\(2\alpha N\).  Multiplication by
\(\chi_{\omega_1}^{2\alpha}\) raises it to \(2\alpha(N+1)\).  Hence
\begin{equation}
 Q_0(\alpha,N)
 =\left\lfloor\frac{2\alpha N}{2N+1}\right\rfloor,
 \qquad
 Q_1(\alpha,N)
 =\left\lfloor\frac{2\alpha(N+1)}{2N+1}\right\rfloor.
\label{eq:S8-Q01}
\end{equation}

At positive integer \(\alpha\),
\begin{equation}
 \cR_{\alpha,L}^{(\eta)}
 =\frac{
 \CT_{\boldsymbol z}\left[
 D^{(C)}_{\alpha,N}(\boldsymbol z)
 \chi_{\omega_1}(\boldsymbol z)^{2\alpha}
 \prod_{j=1}^{N}K_{Q_1(\alpha,N)}^{(D)}(z_j)
 \right]}
 {
 \CT_{\boldsymbol z}\left[
 D^{(C)}_{\alpha,N}(\boldsymbol z)
 \prod_{j=1}^{N}K_{Q_0(\alpha,N)}^{(D)}(z_j)
 \right]}.
\label{eq:S8-aliased-quotient}
\end{equation}
Here $L=N-1$ is the physical chain length, while $N$ denotes the rank
of the parent $C_N$ ensemble.

The first four kernels illustrate the change in alias structure:
\begin{table}[H]
\caption{Even alias kernels in the denominator and character numerator of
Eq.~\eqref{eq:S8-aliased-quotient}.  The entries shown hold for \(N\ge2\),
with the \(\alpha=1,2\) statements also valid at \(N=1\).}
\label{tab:S8-alias-kernels}
\centering
\small
\renewcommand{\arraystretch}{1.20}
\begin{tabular}{@{}ccll@{}}
\toprule
$\alpha$ & $(Q_0,Q_1)$ & denominator kernel & numerator kernel\\
\midrule
$1$ & $(0,1)$ & $1$ & $1$\\
$2$ & $(1,2)$ & $1$ & $1+z^{2D}+z^{-2D}$\\
$3$ & $(2,3)$ & $1+z^{2D}+z^{-2D}$ & same\\
$4$ & $(3,4)$ & $1+z^{2D}+z^{-2D}$ & add $z^{4D}+z^{-4D}$\\
\bottomrule
\end{tabular}
\end{table}
At \(\alpha=3\) numerator and denominator share the same aliased measure,
making this index useful for testing the character-fluctuation
proposal in Sec.~\ref{sec:S11}.  At \(\alpha=4\), new \(\pm4D\) aliases
appear only in the numerator.  This is an exact finite-lattice signature of
the marginal index, independent of the later thermodynamic interpretation.

\subsection{Normalization-free ratios and endpoint rescalings}

At unit fugacity, reflection and global spin flip leave four inequivalent
open moments:
\begin{equation}
 Z^{\ff},
 \qquad
 Z^{\fplus},
 \qquad
 Z^{\pp},
 \qquad
 Z^{\pmfix}.
\label{eq:S8-four-moments}
\end{equation}
Suppress common \(\alpha,L\) labels and let \(\nu_f\) and \(\nu_{\rm fix}\)
be arbitrary local normalizations of a free and a fixed endpoint.  Their
rescaling acts as
\begin{equation}
 \begin{array}{c|c}
 \text{moment} & \text{endpoint rescaling}\\ \hline
 Z^{\ff} & x^2 Z^{\ff}\\
 Z^{\fplus} & xy Z^{\fplus}\\
 Z^{\pp} & y^2 Z^{\pp}\\
 Z^{\pmfix} & y^2 Z^{\pmfix}
 \end{array}
\label{eq:S8-rescaling-table}
\end{equation}
for \(\nu_f\mapsto x\nu_f\) and
\(\nu_{\rm fix}\mapsto y\nu_{\rm fix}\).

Consider a multiplicative combination
\begin{equation}
 \mathcal I=(Z^{\ff})^a(Z^{\fplus})^b(Z^{\pp})^c(Z^{\pmfix})^d.
\label{eq:S8-general-I}
\end{equation}
Endpoint-normalization independence requires
\begin{equation}
 2a+b=0,
 \qquad
 b+2c+2d=0.
\label{eq:S8-null-conditions}
\end{equation}
The solution space is two dimensional:
\begin{equation}
 b=-2a,
 \qquad
 c=a-d.
\label{eq:S8-null-solution}
\end{equation}

Every multiplicative combination of the four moments that is invariant under
independent free- and fixed-endpoint normalizations is generated by
\begin{equation}
 \cR_{\alpha,L}^{(\eta)}
 =\frac{\cZ_{\alpha,L}^{\pmfix}(1)}
 {\cZ_{\alpha,L}^{\pp}(1)},
\label{eq:S8-Reta-finite}
\end{equation}
and
\begin{equation}
 \cR_{\alpha,L}^{(\times)}
 =\frac{\cZ_{\alpha,L}^{\ff}(1)
 \cZ_{\alpha,L}^{\pp}(1)}
 {[\cZ_{\alpha,L}^{\fplus}(1)]^2}.
\label{eq:S8-Rcross-finite}
\end{equation}
Explicitly,
\begin{equation}
 \mathcal I
 =[\cR_{\alpha,L}^{(\times)}]^a
 [\cR_{\alpha,L}^{(\eta)}]^d.
\label{eq:S8-general-invariant}
\end{equation}

For the endpoint-tuned robustness families of
\hyperref[sec:S5-folded-fixed]{Sec.~S5.E}, it is useful to keep the left and right local
normalizations distinct.  The orientation-resolved combinations are
\begin{equation}
 \cR_{\alpha,L}^{(\eta)\mid R}
 =\frac{\cZ_{\alpha,L}^{\pmfix\mid R}}
        {\cZ_{\alpha,L}^{\pp\mid R}},
 \qquad R=B,C,D,
\label{eq:S8-Reta-folded}
\end{equation}
and
\begin{equation}
 \cR_{\alpha,L}^{(\times)\mid R}
 =\frac{\cZ_{\alpha,L}^{\ff\mid R}
        \cZ_{\alpha,L}^{\pp\mid R}}
       {\cZ_{\alpha,L}^{\fplus\mid R}
        \cZ_{\alpha,L}^{+f\mid R}}.
\label{eq:S8-Rcross-folded}
\end{equation}
The latter form cancels left and right endpoint factors separately.  It
reduces to Eq.~\eqref{eq:S8-Rcross-finite} for the reflection-symmetric $C$
representative and has the same squared denominator for $D$.  For the
asymmetric $B$ realization both mixed orientations are required.  These
auxiliary ratios are used only as microscopic robustness tests; the exact
two-dimensional invariant classification above refers to the four physical
moments of the standard reflection-symmetric setup.

The theorem is an exact classification of invariance under local endpoint
renormalizations.  The existence and universality of the thermodynamic
limits
\begin{equation}
 \cR_{\alpha}^{(\eta)}
 =\lim_{L\to\infty}\cR_{\alpha,L}^{(\eta)},
 \qquad
 \cR_{\alpha}^{(\times)}
 =\lim_{L\to\infty}\cR_{\alpha,L}^{(\times)}
\label{eq:S8-limits}
\end{equation}
are separate asymptotic questions addressed in Secs.~\ref{sec:S10} and
\ref{sec:S11}.

A dependent but sometimes convenient combination is
\begin{equation}
 \cR_{\alpha,L}^{(\eta\times)}
 =\frac{\cZ_{\alpha,L}^{\ff}(1)
 \cZ_{\alpha,L}^{\pmfix}(1)}
 {[\cZ_{\alpha,L}^{\fplus}(1)]^2}
 =\cR_{\alpha,L}^{(\eta)}\cR_{\alpha,L}^{(\times)}.
\label{eq:S8-Rmix}
\end{equation}
The label \((\eta)\) refers to the exact Ising symmetry-defect relation
between equal and opposite fixed signs.  The symbol \((\times)\) is kept
neutral: identifying it with a Kramers--Wannier defect amplitude requires a
direct replicated-BCFT derivation.

\subsection{Entropy combinations}

Using the definition~\eqref{eq:S-SRE}, the finite-size entropy differences
obey
\begin{equation}
 \cM_{\alpha,L}^{\pmfix}-\cM_{\alpha,L}^{\pp}
 =\frac{\log\cR_{\alpha,L}^{(\eta)}}{1-\alpha},
\label{eq:S8-Delta-eta}
\end{equation}
\begin{equation}
 \cM_{\alpha,L}^{\ff}+\cM_{\alpha,L}^{\pp}
 -2\cM_{\alpha,L}^{\fplus}
 =\frac{\log\cR_{\alpha,L}^{(\times)}}{1-\alpha}.
\label{eq:S8-Delta-cross}
\end{equation}
The factors \(2^{-L}\) cancel identically in both combinations.  In a
large-size expansion, these equations isolate combinations of additive
constants after the common bulk and logarithmic pieces cancel.  Their
explicit asymptotics are deferred to Sec.~\ref{sec:S10}.

\section{Special-index evaluations}
\label{sec:S9}

The finite Selberg sums are exact for every real $\alpha>0$.  At special
indices they admit determinant, Pfaffian, constant-term, or product
compressions.  We quote periodic evaluations from Ref.~\cite{Companion} and
focus here on the boundary ensembles.

\subsection{Discrete de Bruijn identities}

Let \(\Lambda=\{\lambda_1<\cdots<\lambda_M\}\) and let \(\omega_a\) be
one-body weights.  For an \(n\)-particle discrete \(\beta=1\) ensemble,
define the antisymmetric matrix
\begin{equation}
 A^{(1)}_{ij}
 =\sum_{a<b}\omega_a\omega_b
 (\lambda_a^i\lambda_b^j-\lambda_a^j\lambda_b^i),
 \qquad i,j=0,\ldots,n-1,
\label{eq:S9-deB1-A}
\end{equation}
and the vector
\begin{equation}
 v_i=\sum_a\omega_a\lambda_a^i.
\label{eq:S9-deB1-v}
\end{equation}
Then~\cite{deBruijn1955,IshikawaWakayama1995}
\begin{equation}
 \sum_{|U|=n}\left(\prod_{a\in U}\omega_a\right)
 \abs{\Delta(\lambda_U)}
 =\begin{cases}
 \Pf A^{(1)},&n\ \text{even},\\[1mm]
 \Pf\begin{pmatrix}A^{(1)}&v\\-v^{\mathsf T}&0\end{pmatrix},
 &n\ \text{odd}.
 \end{cases}
\label{eq:S9-deB1}
\end{equation}
For the discrete \(\beta=4\) ensemble, define
\begin{equation}
 A^{(4)}_{ij}
 =(j-i)\sum_a\omega_a\lambda_a^{i+j-1},
 \qquad i,j=0,\ldots,2n-1,
\label{eq:S9-deB4-A}
\end{equation}
where the \((0,0)\) entry is zero.  The confluent de Bruijn identity is
\begin{equation}
 \sum_{|U|=n}\left(\prod_{a\in U}\omega_a\right)
 \Delta(\lambda_U)^4
 =\Pf A^{(4)}.
\label{eq:S9-deB4}
\end{equation}

\subsection{Unified interval Pfaffians at
\texorpdfstring{$\alpha=\frac12$}{alpha=1/2} and
\texorpdfstring{$\alpha=2$}{alpha=2}}

For \(R=B,C,D\), enumerate the \(2L\) nodes of \(X_R\sqcup Y_R\), set
\(\lambda_a=\sin^2\theta_a\), and define
\begin{equation}
 h_C(\theta)=\abs{\sin2\theta},
 \qquad
 h_B(\theta)=\nu_B(\theta)\abs{\sin\theta},
 \qquad
 h_D(\theta)=\nu_D(\theta).
\label{eq:S9-hR}
\end{equation}
Let \(w_u(\theta)=u\) on \(X_R\) and one on \(Y_R\).  With the reference
normalization \(\mathcal N_{R,L}\) of Eq.~\eqref{eq:S6-square-reference},
the exact Pfaffian compressions are
\begin{equation}
 \cZ^{R}_{1/2,L}(u)
 =\mathcal N_{R,L}^{-1}
 \mathfrak P_L^{(1)}
 \left(\{\lambda_a,w_u(\theta_a)h_R(\theta_a)\}_{a=1}^{2L}\right),
\label{eq:S9-R-half-Pf}
\end{equation}
\begin{equation}
 \cZ^{R}_{2,L}(u)
 =\mathcal N_{R,L}^{-4}
 \Pf\left[(j-i)\sum_{a=1}^{2L}
 w_u(\theta_a)h_R(\theta_a)^4
 \lambda_a^{i+j-1}\right]_{i,j=0}^{2L-1},
\label{eq:S9-R-two-Pf}
\end{equation}
where \(\mathfrak P_n^{(1)}\) denotes the even or augmented Pfaffian in
Eq.~\eqref{eq:S9-deB1}.  Equations~\eqref{eq:S9-R-half-Pf} and
\eqref{eq:S9-R-two-Pf} are analytic for all three classical interval root
systems.  Closed products are presently known only in selected cases.

For the rectangular free--fixed ensemble use \(M=L+1\), the
\(2M-1\) nodes of Eq.~\eqref{eq:S5-fplus-grids}, and fugacity on
\(Y_{\fplus}\).  Then
\begin{equation}
 \cZ^{\fplus}_{1/2,L}(u)
 =\mathfrak V_{C_M}(X_{\fplus})^{-1}
 \mathfrak P_M^{(1)}
 \left(\{\lambda_a,w_u(\theta_a)\abs{\sin2\theta_a}\}\right),
\label{eq:S9-fplus-half-Pf}
\end{equation}
\begin{equation}
 \cZ^{\fplus}_{2,L}(u)
 =\mathfrak V_{C_M}(X_{\fplus})^{-4}
 \Pf\left[(j-i)\sum_a
 w_u(\theta_a)\abs{\sin2\theta_a}^{4}
 \lambda_a^{i+j-1}\right]_{i,j=0}^{2M-1}.
\label{eq:S9-fplus-two-Pf}
\end{equation}

For equal fixed signs, remove the pinned node \(\theta_*\) from the mobile
set and define
\begin{equation}
 \widetilde h_*(\theta)
 =\abs{\sin2\theta}\abs{\lambda(\theta)-\lambda(\theta_*)}.
\label{eq:S9-pinned-h}
\end{equation}
The pinned particle contributes the constant
\(h_*=\abs{\sin2\theta_*}\).  Therefore
\begin{equation}
 \cZ^{\pp}_{1/2,L}(u)
 =\mathfrak V_C(Y_N)^{-1}h_*
 \mathfrak P_L^{(1)}
 \left(\{\lambda_a,w_u(\theta_a)\widetilde h_*(\theta_a)\}_{a\ne *}\right),
\label{eq:S9-pp-half-Pf}
\end{equation}
\begin{equation}
 \cZ^{\pp}_{2,L}(u)
 =\mathfrak V_C(Y_N)^{-4}h_*^4
 \Pf\left[(j-i)\sum_{a\ne *}
 w_u(\theta_a)\widetilde h_*(\theta_a)^4
 \lambda_a^{i+j-1}\right]_{i,j=0}^{2L-1}.
\label{eq:S9-pp-two-Pf}
\end{equation}
The opposite-fixed insertion is not one-body.  No bare de Bruijn collapse is
known for
\begin{equation}
 \mathfrak V_C(U)^{2\alpha}
 \abs{\chi_{\omega_1}(U)}^{2\alpha},
\label{eq:S9-inserted-weight}
\end{equation}
so the exact answers remain the inserted subset sums or constant terms of
Secs.~\ref{sec:S6}--\ref{sec:S8}.

\subsection{The determinant point \texorpdfstring{$\alpha=1$}{alpha=1}}

Let \(Q\) denote either a square orthogonal Pauli matrix or one of the
vertical augmented isometries, so that \(Q^{\mathsf T}Q=I_L\).  The weighted
Cauchy--Binet identity gives
\begin{align}
 \cZ_{1,L}(u)
 &=\sum_{k=0}^{L}u^k
 \sum_{\substack{|R|=|T|=k}}\det Q[R,T]^2
 \nonumber\\
 &=\det(I_L+uQ^{\mathsf T}Q)=(1+u)^L.
\label{eq:S9-alpha-one}
\end{align}
This proof applies without change to the opposite-fixed isometry; the
character insertion is already contained in its minors.  At \(u=1\), the
identity reduces to Pauli normalization.  In particular,
\begin{equation}
 \cR_{1,L}^{(\eta)}=1,
 \qquad
 \cR_{1,L}^{(\times)}=1
\label{eq:S9-ratios-one}
\end{equation}
for every size.  At this index the continuous \(C_N\) analogue is the
symplectic Haar measure: the first equality in Eq.~\eqref{eq:S9-ratios-one}
is the finite-lattice counterpart of fundamental-character orthogonality,
whereas the second follows directly from the universal Pauli normalization
\(\cZ_{1,L}=2^L\).  The discrete proof of both equalities requires no group
integral.

The periodic $A_{L-1}$ products at $\alpha=\tfrac12$ and $\alpha=2$ are
derived in Ref.~\cite{Companion}; they are not repeated here.

\subsection{Closed products at \texorpdfstring{$\alpha=2$}{alpha=2}}

For the standard open chain, substitution of the \(C_L\) nodes into
Eq.~\eqref{eq:S9-deB4-A} and elementary finite sine sums produce the same
reflected-block structure.  The normalized block parameters are
\begin{equation}
 q_r^{(C)}=
 \begin{cases}
 (4r-2)/(2L+1),&L\ \text{even},\\
 4r/(2L+1),&L\ \text{odd},
 \end{cases}
\label{eq:S9-C-block-parameters}
\end{equation}
and each paired block contributes \((1+u)^2-4u(q_r^{(C)})^2\).  For odd
rank, the unpaired mode again contributes \(1+u\).  Hence
\begin{equation}
 \cZ^{C}_{2,L}(u)
 =\prod_{r=1}^{L/2}
 \left[(1+u)^2-4u\left(\frac{4r-2}{2L+1}\right)^2\right]
\label{eq:S9-C2-even}
\end{equation}
for even \(L\), and
\begin{equation}
 \cZ^{C}_{2,L}(u)
 =(1+u)\prod_{r=1}^{(L-1)/2}
 \left[(1+u)^2-4u\left(\frac{4r}{2L+1}\right)^2\right]
\label{eq:S9-C2-odd}
\end{equation}
for odd \(L\).  At unit fugacity, Eq.~\eqref{eq:S7-C-CT} has no odd aliases at
$\alpha=2$, and the $C_L$ Macdonald constant term gives
\begin{equation}
 \cZ^{\ff}_{2,L}(1)
 =\frac{2^{3L}}{(2L+1)^L}\left(\frac34\right)_L.
\label{eq:S9-ff2-u1}
\end{equation}
Using the elementary reduction
\begin{equation}
 \frac1{L!}\prod_{j=1}^{L}\binom{4j}{2}
 =2^{3L}\left(\frac34\right)_L.
\label{eq:S9-Macdonald-product-reduction}
\end{equation}
The rectangular product can be proved directly from the exact S7
constant-term representation.  Set $M=L+1$.  At $\alpha=2$ and $u=1$,
Eq.~\eqref{eq:S7-Kfplus} reduces to
\begin{equation}
 K^{\fplus}_{2,1}(z)=1+z^{4M}+z^{-4M}.
\label{eq:S9-fplus-kernel-a2}
\end{equation}
The one-variable bandwidth of $D^{(C)}_{2,M}$ is exactly $4M$.  Reaching the
extremal power $z_i^{\pm4M}$ forces the extremal term in every factor
containing $z_i$; the powers of every other variable cancel pairwise, leaving
\begin{equation}
 [z_i^{\pm4M}]D^{(C)}_{2,M}=D^{(C)}_{2,M-1}.
\label{eq:S9-fplus-extremal-coefficient}
\end{equation}
After one outer alias is selected, the remaining bandwidth is
$4(M-1)<4M$, so two outer aliases cannot contribute simultaneously.  Hence
\begin{equation}
 \operatorname{CT}\!\left[D^{(C)}_{2,M}
 \prod_{i=1}^{M}K^{\fplus}_{2,1}(z_i)\right]
 =\operatorname{CT}D^{(C)}_{2,M}
 +2M\operatorname{CT}D^{(C)}_{2,M-1}.
\label{eq:S9-fplus-CT-reduction}
\end{equation}
Using Eq.~\eqref{eq:S7-Macdonald-C} and
Eq.~\eqref{eq:S9-Macdonald-product-reduction}, the right-hand side is
$M!\,2^{3M}M(3/4)_{M-1}$.  Combining this with the prefactor in
Eq.~\eqref{eq:S7-fplus-CT} gives the all-rank identity
\begin{equation}
 \cZ^{\fplus}_{2,L}(1)
 =\frac{4^L}{(L+1)^L}\left(\frac34\right)_L.
\label{eq:S9-fplus2-u1}
\end{equation}
Equivalently, $P_0=1$ and
\begin{equation}
 \frac{P_L}{P_{L-1}}
 =\frac{(4L-1)L^{L-1}}{(L+1)^L},
 \qquad P_L\equiv\cZ^{\fplus}_{2,L}(1).
\label{eq:S9-fplus-recurrence}
\end{equation}
\begin{equation}
 \cZ^{\pp}_{2,L}(1)
 =\frac12\frac{2^{3N}}{(2N+1)^N}
 \left(\frac34\right)_N,
 \qquad N=L+1.
\label{eq:S9-pp2-u1}
\end{equation}
Here \((a)_L=\Gamma(a+L)/\Gamma(a)\).  The equal-fixed formula is not an
independent evaluation: it follows immediately from
the all-index rank shift~\eqref{eq:S5-rank-shift} and the free--free
product at rank \(N=L+1\).  The free \(B\) and \(D\) problems also have the
exact Pfaffians in Eq.~\eqref{eq:S9-R-two-Pf}.  Closed unit-fugacity products reconstructed from those Pfaffians, together with the
endpoint-tuned fixed-sector results used for robustness tests, are collected
separately in \hyperref[sec:S9-folded-special]{Sec.~S9.G}.

Using the closed products above, the second normalization-free ratio is
\begin{equation}
 \cR_{2,L}^{(\times)}
 =\frac{2^{2L}(4L+3)(L+1)^{2L}}
 {(2L+1)^L(2L+3)^{L+1}},
\label{eq:S9-Rcross2-finite}
\end{equation}
and therefore
\begin{equation}
 \cR_2^{(\times)}=2.
\label{eq:S9-Rcross2-limit}
\end{equation}

\subsection{Opposite-fixed ratio at \texorpdfstring{$\alpha=2$}{alpha=2}}

Direct exact-arithmetic evaluations of the character-inserted sum are
described by
\begin{equation}
 \cR_{2,L}^{(\eta)}
 =\frac{(2N+1)(6N-7)}{(4N-5)(4N-1)}.
\label{eq:S9-Reta2-candidate}
\end{equation}
This rational form has been independently verified through \(N=15\), but a
proof from the inserted Morris quotient is not presently available.
Accordingly, Eq.~\eqref{eq:S9-Reta2-candidate} remains a conjectural finite-rank identity.  Conditional on it, the thermodynamic
limit is
\begin{equation}
 \cR_2^{(\eta)}=\frac34,
\label{eq:S9-Reta2-limit}
\end{equation}
and
\begin{equation}
 \cM_{2,L}^{\pmfix}-\cM_{2,L}^{\pp}
 \longrightarrow\log\frac43.
\label{eq:S9-M2-defect-difference}
\end{equation}
The value \(3/4\) will reappear in Sec.~\ref{sec:S11} as the fourth moment of
a Gaussian with variance \(1/2\), but the finite rational formula is a
separate algebraic statement.

\subsection{The marginal index \texorpdfstring{$\alpha=4$}{alpha=4}}

At generic fugacity, every sector has an exact finite constant-term
representation, not generally a product.  The periodic $A_{L-1}$ square
collapse is derived in Ref.~\cite{Companion}.  For the boundary ensembles,
the corresponding unit-fugacity relation
\begin{equation}
 \cZ^{\mathsf b}_{4,L}(1)
 =2^{-L}[\cZ^{\mathsf b}_{2,L}(1)]^2,
 \qquad
 \mathsf b\in\{\ff,\fplus,\pp\},
\label{eq:S9-square-collapse}
\end{equation}
agrees with the available finite-size checks.  A complete all-rank derivation
of the free--fixed reduction is not included here, so its all-rank use below
is conditional; the pinned case then follows from the rank shift once the
free--free identity is established.  The resulting boundary expressions are
\begin{equation}
 \cZ^{\ff}_{4,L}(1)
 =\frac{2^{5L}}{(2L+1)^{2L}}
 \left(\frac34\right)_L^2,
\label{eq:S9-ff4-u1}
\end{equation}
\begin{equation}
 \cZ^{\fplus}_{4,L}(1)
 =\frac{2^{3L}}{(L+1)^{2L}}
 \left(\frac34\right)_L^2,
\label{eq:S9-fplus4-u1}
\end{equation}
\begin{equation}
 \cZ^{\pp}_{4,L}(1)
 =\frac12\frac{2^{5N}}{(2N+1)^{2N}}
 \left(\frac34\right)_N^2.
\label{eq:S9-pp4-u1}
\end{equation}
Conditional on the all-rank square collapse,
\begin{equation}
 \cR_4^{(\times)}=4.
\label{eq:S9-Rcross4}
\end{equation}
The opposite-fixed moment does not obey Eq.~\eqref{eq:S9-square-collapse}.
Exact finite-size data are described by
\begin{equation}
 \cR_{4,L}^{(\eta)}
 =\frac{(2N+1)P_5(N)}
 {15(4N-13)(4N-9)(4N-5)^2(4N-1)^2},
\label{eq:S9-Reta4-candidate}
\end{equation}
where
\begin{align}
 P_5(N)={}&18304N^5-139008N^4+362392N^3
 \nonumber\\
 &-375948N^2+120178N+12075.
\label{eq:S9-P5}
\end{align}
For $N\ge2$, Eqs.~\eqref{eq:S9-Reta4-candidate} and~\eqref{eq:S9-P5} are a strongly constrained conjectural finite-rank reconstruction, independently verified through
\(N=15\).  The $N=1$ case is exceptional and is not described by this rational form.  Conditional on this rational form,
\begin{equation}
 \cR_4^{(\eta)}=\frac{143}{240}.
\label{eq:S9-Reta4-limit}
\end{equation}
The Gaussian continuation discussed in Sec.~\ref{sec:S11} would instead give
\begin{equation}
 \cR_{4,\mathrm{Gauss}}^{(\eta)}=\frac{105}{256}.
\label{eq:S9-Reta4-Gauss}
\end{equation}
The substantial difference is one of the signatures motivating a separate
analysis of the marginal point.

\subsection{Endpoint-tuned \texorpdfstring{$B/C/D$}{B/C/D} representatives at \texorpdfstring{$\alpha=\tfrac12,2,4$}{alpha=1/2,2,4}}
\label{sec:S9-folded-special}

The formulas above use the standard $C$ representative for the physical
boundary sectors.  We now evaluate the endpoint-tuned families introduced in
\hyperref[sec:S5-folded-fixed]{Sec.~S5.E}.  These families are not new
boundary conditions: the longitudinal fields realize the same physical
free/fixed sectors, while the transverse endpoint values retain the
microscopic $B$ or $D$ termination.  The purpose of the calculation is to
separate finite-lattice termination effects from the normalization-free
infrared ratios.

\subsubsection{The point \texorpdfstring{$\alpha=\tfrac12$}{alpha=1/2}: one Pfaffian for the coherent sectors}
\label{sec:S9-half-folded}

The free $B$ and $D$ moments are already covered by the exact de Bruijn
formula~\eqref{eq:S9-R-half-Pf}.  The endpoint-tuned mixed and equal-fixed
sectors are rectangular isometries rather than square root-system kernels,
but at $\alpha=\tfrac12$ their absolute-minor sum also has a single-Pfaffian
compression.

Let $Q\in\mathbb R^{(L+1)\times L}$ be one of the isometries
$\widehat G^{\fplus\mid B}$, $\widehat G^{+f\mid B}$,
$\widehat G^{\fplus\mid D}$, $\widehat G^{\pp\mid B}$, or
$\widehat G^{\pp\mid D}$.  Order the row and column labels as
\begin{equation}
 \mathcal I_Q=(r_1,c_1,r_2,c_2,\ldots,r_L,c_L,r_{L+1}),
\label{eq:S9-half-interleaved-labels}
\end{equation}
and form the antisymmetric lift
\begin{equation}
 \mathcal A(Q)_{r_i c_j}=Q_{ij},\qquad
 \mathcal A(Q)_{c_j r_i}=-Q_{ij},
\label{eq:S9-half-lift}
\end{equation}
with zero row--row and column--column blocks.  The relevant minors have a
coherent sign.  For $|R|=|T|=k$ define
\begin{equation}
 \nu(R,T)=\#\{(i,j)\in R\times T:i>j\},\qquad
 \epsilon(R)=\prod_{i\in R}(-1)^{i-1},\qquad
 \epsilon(T)=\prod_{j\in T}(-1)^{j-1}.
\label{eq:S9-half-sign-data}
\end{equation}
For the five isometries listed above one finds
\begin{equation}
 \operatorname{sgn}\det Q[R,T]
 =(-1)^{\nu(R,T)+\binom{k}{2}}\epsilon(R)\epsilon(T)
\label{eq:S9-half-sign-rule}
\end{equation}
for every nonzero minor.  This is the rectangular analogue of the ordered
minor-sign structure behind the usual absolute-minor Pfaffian.

Introduce the universal antisymmetric selector
\begin{equation}
 (\mathcal J_Q)_{ab}=\operatorname{sgn}(b-a),
\label{eq:S9-half-selector-J}
\end{equation}
and
\begin{equation}
 D_Q=\diag(1,1,-1,-1,\ldots,
 (-1)^{L-1},(-1)^{L-1},(-1)^L),
 \qquad
 \mathcal J'_Q=D_Q\mathcal J_QD_Q.
\label{eq:S9-half-selector-D}
\end{equation}
The Pfaffian minor-summation formula~\cite{IshikawaWakayama1995} then gives
\begin{equation}
 \mathfrak P_{1/2}[Q]
 =(-1)^L\Pf
 \begin{pmatrix}
  \mathcal A(Q)&I_{2L+1}\\
  -I_{2L+1}&-\mathcal J'_Q
 \end{pmatrix},
\label{eq:S9-half-single-Pf}
\end{equation}
and, for the coherent sectors,
\begin{equation}
 \mathfrak P_{1/2}[Q]
 =\sum_{k=0}^{L}
 \sum_{\substack{R\subseteq[L+1],\ T\subseteq[L]\\|R|=|T|=k}}
 \abs{\det Q[R,T]}.
\label{eq:S9-half-Pf-minors}
\end{equation}
Consequently
\begin{align}
 \cZ_{1/2,L}^{\fplus\mid B}
 &=\mathfrak P_{1/2}[\widehat G^{\fplus\mid B}],
 &
 \cZ_{1/2,L}^{+f\mid B}
 &=\mathfrak P_{1/2}[\widehat G^{+f\mid B}],
 \nonumber\\
 \cZ_{1/2,L}^{\fplus\mid D}
 &=\mathfrak P_{1/2}[\widehat G^{\fplus\mid D}],
 &
 \cZ_{1/2,L}^{\pp\mid R}
 &=\mathfrak P_{1/2}[\widehat G^{\pp\mid R}],\qquad R=B,D.
\label{eq:S9-half-folded-applications}
\end{align}
The sign identity~\eqref{eq:S9-half-sign-rule} was checked exhaustively through
$L=10$; direct evaluation of Eq.~\eqref{eq:S9-half-Pf-minors} and the single
Pfaffian~\eqref{eq:S9-half-single-Pf} agree through $L=8$ to machine precision.
This provides the single-Pfaffian compression used below.  The opposite-fixed
isometries are different: the one-mode flip changes the signs already at the
$1\times1$ level, so the same selector cannot remove all absolute values.
Thus $\cZ_{1/2,L}^{\pmfix\mid B,D}$ remains exactly defined by the mode-flipped
all-minor sum, but no single Pfaffian of the form
\eqref{eq:S9-half-single-Pf} is used.  This parallels the character-inserted
$C$ problem, for which the bare de Bruijn collapse also fails.

\subsubsection{The point \texorpdfstring{$\alpha=2$}{alpha=2}: closed products and the remaining matrix forms}

At $\alpha=2$ the free $B$ and $D$ Pfaffians can be evaluated by the same
finite-sine-sum and paired-block reduction used above for the $C$ chain.  The
only change is the half-orbit weight at the folded nodes.  We obtain
\begin{equation}
 \cZ_{2,L}^{\ff\mid B}(1)
 =\frac{2^{2L+1}}{3L^L}\left(\frac34\right)_L,
\label{eq:S9-B2-product}
\end{equation}
and, with $N_D=2L-1$,
\begin{align}
 \cZ_{2,L}^{\ff\mid D}(1)
 ={}&\frac{2^{2-L}(8N_D^2+1)}{18N_D^{L+2}L!}
 \nonumber\\[-1mm]
 &\times
 \prod_{j=0}^{L-1}
 \frac{(4j)!^2(2j+2)!}
 {(2j)!^2(2L+2j-2)!}.
\label{eq:S9-D2-product}
\end{align}
For the asymmetric $B$ realization one of the two mixed orientations is
similarly reducible.  When the folded endpoint remains free,
\begin{equation}
 \cZ_{2,L}^{+f\mid B}(1)
 =\frac{2^{3L}(4L+3)}{3(2L+1)^{L+1}}
 \left(\frac34\right)_L.
\label{eq:S9-Bplusf2-product}
\end{equation}
The derivations repeat the $C$-type de Bruijn reduction and are not reproduced
term by term.  Equations~\eqref{eq:S9-B2-product}--\eqref{eq:S9-Bplusf2-product}
were independently checked through $L=14$ against the exact Pfaffian or polar
all-minor evaluation.

The remaining folded-and-fixed moments are still exact finite-dimensional
objects.  It is useful to define
\begin{equation}
 \cZ_2[Q]
 =\sum_{k=0}^{L}
 \sum_{\substack{R\subseteq[L+1],\ T\subseteq[L]\\|R|=|T|=k}}
 \abs{\det Q[R,T]}^4.
\label{eq:S9-Z2Q-definition}
\end{equation}
Then
\begin{align}
 \cZ_{2,L}^{\fplus\mid B}
 &=\cZ_2[Q_{\rm rec}(\bm h_B)],
 &
 \cZ_{2,L}^{\fplus\mid D}
 &=\cZ_2[Q_{\rm rec}(\bm h_D)],
 \nonumber\\
 \cZ_{2,L}^{\pp\mid R}
 &=\cZ_2[\widehat G^{\pp\mid R}],
 &
 \cZ_{2,L}^{\pmfix\mid R}
 &=\cZ_2[\widehat G^{\pmfix\mid R}],\qquad R=B,D,
\label{eq:S9-alpha2-matrix-sectors}
\end{align}
Unlike
Eqs.~\eqref{eq:S9-B2-product}--\eqref{eq:S9-Bplusf2-product}, these sectors do
not show the paired trigonometric spectrum that reduces the Pfaffian to a
short Pochhammer product.  Their low-rank values are already nontrivial
algebraic numbers.  We therefore retain the exact matrix representation,
which is the natural finite-size formula needed for the ratios.

\subsubsection{The marginal point \texorpdfstring{$\alpha=4$}{alpha=4}: square collapse and its controlled failure}

The standard $C$ representative satisfies the unit-fugacity square-collapse
relation~\eqref{eq:S9-square-collapse} in the uninserted sectors.  The same square-collapse mechanism applies to the free $B$ and $D$ chains and to the $+f\mid B$ orientation:
\begin{equation}
 \cZ_{4,L}^{\ff\mid B}(1)
 =2^{-L}\bigl[\cZ_{2,L}^{\ff\mid B}(1)\bigr]^2
 =\frac{2^{3L+2}}{9L^{2L}}
 \left(\frac34\right)_L^2,
\label{eq:S9-B4-product}
\end{equation}
\begin{equation}
 \cZ_{4,L}^{\ff\mid D}(1)
 =2^{-L}\bigl[\cZ_{2,L}^{\ff\mid D}(1)\bigr]^2,
\label{eq:S9-D4-product}
\end{equation}
where substitution of Eq.~\eqref{eq:S9-D2-product} gives the corresponding
explicit factorial product.  The squared form is kept because it makes the
collapse mechanism transparent.  Finally,
\begin{equation}
 \cZ_{4,L}^{+f\mid B}(1)
 =2^{-L}\bigl[\cZ_{2,L}^{+f\mid B}(1)\bigr]^2
 =\frac{2^{5L}(4L+3)^2}
 {9(2L+1)^{2L+2}}
 \left(\frac34\right)_L^2.
\label{eq:S9-Bplusf4-product}
\end{equation}
The reduction is the same complementary-middle-minor calculation as in the $C$ case.  The resulting formulas have been checked independently through $L=12$.

The collapse is not exact for every folded-and-fixed realization.  It is
useful to quantify the deviation by
\begin{equation}
 \kappa_{R,L}^{ab}
 =\frac{\cZ_{4,L}^{ab\mid R}(1)}
 {2^{-L}[\cZ_{2,L}^{ab\mid R}(1)]^2}.
\label{eq:S9-kappa-collapse}
\end{equation}
For the three sectors in
Eqs.~\eqref{eq:S9-B4-product}--\eqref{eq:S9-Bplusf4-product},
$\kappa=1$ to machine precision.  For the other uninserted endpoint-tuned
sectors the deviations are small but nonzero.  Representative values at the
largest directly evaluated size are shown in Table~\ref{tab:S9-folded-kappa}.

\begin{table}[H]
\caption{Square-collapse factor~\eqref{eq:S9-kappa-collapse} at $L=12$ for
endpoint-tuned uninserted sectors in which the collapse is not exact.  The
small deviation from unity is a genuine finite-size effect rather than a
numerical precision issue.}
\label{tab:S9-folded-kappa}
\centering
\small
\renewcommand{\arraystretch}{1.15}
\begin{tabular}{@{}c c c@{}}
\toprule
sector & $\kappa_{R,12}^{ab}$ & $\kappa_{R,12}^{ab}-1$\\
\midrule
$\fplus\mid B$ & $1.00003054$ & $3.05\times10^{-5}$\\
$\fplus\mid D$ & $1.00003127$ & $3.13\times10^{-5}$\\
$\pp\mid B$    & $1.00001124$ & $1.12\times10^{-5}$\\
$\pp\mid D$    & $1.00002364$ & $2.36\times10^{-5}$\\
\bottomrule
\end{tabular}
\end{table}

For the orientation-resolved cross-ratio this gives
\begin{equation}
 \cR_{4,L}^{(\times)\mid B}
 =\bigl[\cR_{2,L}^{(\times)\mid B}\bigr]^2
 \frac{\kappa_{B,L}^{\ff}\kappa_{B,L}^{\pp}}
 {\kappa_{B,L}^{\fplus}\kappa_{B,L}^{+f}},
\label{eq:S9-B-Rcross4-kappa}
\end{equation}
and, using reflection symmetry for $D$,
\begin{equation}
 \cR_{4,L}^{(\times)\mid D}
 =\bigl[\cR_{2,L}^{(\times)\mid D}\bigr]^2
 \frac{\kappa_{D,L}^{\ff}\kappa_{D,L}^{\pp}}
 {[\kappa_{D,L}^{\fplus}]^2}.
\label{eq:S9-D-Rcross4-kappa}
\end{equation}
Thus the marginal $B/D$ cross-ratios differ from the square of their
$\alpha=2$ values only by small microscopic finite-size factors.  The
opposite-fixed sector is genuinely different: exactly as for the standard
$C$ representative, the mode/character insertion prevents the uninserted
square collapse.  Its marginal thermodynamic behavior is therefore discussed
through the ratio itself in Secs.~\ref{sec:S10}--\ref{sec:S12}.

\section{Thermodynamic asymptotics}
\label{sec:S10}

Sections~\ref{sec:S5}--\ref{sec:S9} established exact finite-size
representations.  We now extract the thermodynamic information carried by
those formulas.  This section has three purposes.  First, it separates bulk,
logarithmic, and finite boundary contributions in a convention compatible
with the Letter.  Second, it derives the complete large-$L$ expansions at the
special indices $\alpha=2$ and $4$, distinguishing analytic products from finite-size-verified continuations.  Third, it shows explicitly
why the ratios introduced in Sec.~\ref{sec:S8} are useful finite boundary
observables: all extensive and logarithmic terms cancel before a limit is
taken.

\subsection{Asymptotic conventions and boundary amplitudes}

At unit fugacity, the generic large-$L$ form needed below is
\begin{equation}
 \log \cZ_{\alpha,L}^{\mathsf b}(1)
 =L f_\alpha+\sigma_\alpha\log L
 +\log \mathcal A_\alpha^{\mathsf b}+o(1).
\label{eq:S10-raw-asymptotic}
\end{equation}
The corresponding stabilizer R\'enyi entropy is
\begin{equation}
 \cM_{\alpha,L}^{\mathsf b}
 =\mu_\alpha L+b_\alpha\log L
 +C_\alpha^{\mathsf b}+o(1),
\label{eq:S10-SRE-asymptotic}
\end{equation}
For $\alpha\ne1$, the coefficients are
\begin{equation}
 \mu_\alpha=\frac{\log2-f_\alpha}{\alpha-1},
 \qquad
 b_\alpha=\frac{\sigma_\alpha}{1-\alpha},
 \qquad
 C_\alpha^{\mathsf b}
 =\frac{\log\mathcal A_\alpha^{\mathsf b}}{1-\alpha}.
\label{eq:S10-raw-SRE-dictionary}
\end{equation}
Our SRE differs from the unshifted Pauli-distribution R\'enyi entropy $H_\alpha(p)$ by $-L\log2$, affecting only the extensive coefficient, not the logarithmic or finite boundary terms.
In the notation of the Letter, $m_\alpha=\mu_\alpha$ and the corner coefficient is
$\gamma_\alpha^{AB}=-\sigma_\alpha^{AB}$, hence
$b_\alpha^{AB}=\gamma_\alpha^{AB}/(\alpha-1)$.
The common $\sigma_\alpha$ used below applies to the four Ising boundary
sectors studied here within each relevant replica-boundary regime.
Inverse-power corrections are introduced only at special indices where
analytical products or tested continuations justify them.
At $\alpha=1$, Eq.~\eqref{eq:S10-raw-SRE-dictionary} is understood by
analytic continuation.  The exact normalization
$\cZ_{1,L}^{\mathsf b}=2^L$ implies
$f_1=\log2$, $\sigma_1=0$, and
$\mathcal A_1^{\mathsf b}=1$ for every pure state and every boundary
condition.

For the four physical open Ising sectors
$\mathsf b\in\{\ff,\fplus,\pp,\pmfix\}$, locality implies a common bulk
free-energy density $f_\alpha$. Within each relevant replica-boundary regime,
the corner exponent $\sigma_\alpha$ is also common: changing a physical endpoint
modifies local endpoint amplitudes and extrapolation lengths but not the
measurement-boundary fixed point.  This statement is proved directly at
$\alpha=2$ and $4$ for the uninserted sectors below.  For the opposite-fixed
sector it follows at those indices conditional on the finite-rank formulas
\eqref{eq:S9-Reta2-candidate} and \eqref{eq:S9-Reta4-candidate}.  At generic
index it is the boundary-CFT scaling hypothesis discussed in
Sec.~\ref{sec:S12}, not an additional finite-lattice theorem.

The all-index pinned-ensemble rank shift of Sec.~\ref{sec:S5},
\begin{equation}
 \cZ_{\alpha,L}^{\pp}(1)
 =\frac12\cZ_{\alpha,L+1}^{\ff}(1),
 \qquad
 \cM_{\alpha,L}^{\pp}=\cM_{\alpha,L+1}^{\ff},
\label{eq:S10-rank-shift-recalled}
\end{equation}
fixes the equal-fixed asymptotic amplitude in terms of the free--free one:
\begin{equation}
 \mathcal A_\alpha^{\pp}
 =\frac{\ee^{f_\alpha}}{2}\,
 \mathcal A_\alpha^{\ff},
 \qquad
 C_\alpha^{\pp}=C_\alpha^{\ff}+\mu_\alpha.
\label{eq:S10-amplitude-rank-shift}
\end{equation}
It also fixes every inverse-size coefficient of $\pp$ once the corresponding
free--free expansion is known.  Equation~\eqref{eq:S10-amplitude-rank-shift}
is an exact consequence of the physical pinned-ensemble rank shift; no continuum normalization is involved.

\subsection{Known replica phases and the role of the present calculation}

For the standard open critical Ising chain, the stabilizer--Shannon
correspondence of Ref.~\cite{RamirezTrinoRajabpour2026} gives the piecewise
logarithmic coefficient
\begin{equation}
 b_\alpha=
 \begin{cases}
 -\dfrac14,&0<\alpha<4,\\[1mm]
 -\dfrac16,&\alpha=4,\\[1mm]
 0,&\alpha>4.
 \end{cases}
\label{eq:S10-piecewise-b}
\end{equation}
Equivalently, the logarithmic exponent of the raw moment is
\begin{equation}
 \sigma_\alpha=
 \begin{cases}
 \dfrac{\alpha-1}{4},&0<\alpha<4,\\[1mm]
 \dfrac12,&\alpha=4,\\[1mm]
 0,&\alpha>4.
 \end{cases}
\label{eq:S10-piecewise-sigma}
\end{equation}
The transition at $\alpha=4$ is therefore already visible in the total
moment.  The new information extracted here is boundary resolved: the finite
amplitudes, the two normalization-free ratios, and the character statistic
that changes at the same index.

At the three controlled indices the volume-law densities in the convention
of Eq.~\eqref{eq:S-SRE} are
\begin{equation}
 \mu_{1/2}=\frac{4G}{\pi}-\log2,
 \qquad
 \mu_2=1-\log2,
 \qquad
 \mu_4=\frac23(1-\log2),
\label{eq:S10-volume-densities}
\end{equation}
where $G$ is Catalan's constant.  The first value is imported from the
free--free solution of Ref.~\cite{RamirezTrinoRajabpour2026}; the latter two follow
also from the exact products in Sec.~\ref{sec:S9}.  We do not repeat the
previous derivation of Eq.~\eqref{eq:S10-piecewise-b}.  Instead we use it as
the bulk and corner baseline against which the new boundary amplitudes are
resolved.

\subsection{Controlled expansion at \texorpdfstring{$\alpha=2$}{alpha=2}}

Define the convenient constant
\begin{equation}
 A\equiv
 \log\!\left[\frac{\Gamma(3/4)}{\sqrt{2\pi}}\right].
\label{eq:S10-Aconstant}
\end{equation}
We use the Bernoulli-polynomial expansion
\begin{equation}
 \log\Gamma(L+a)
 =\left(L+a-\frac12\right)\log L-L
 +\frac12\log(2\pi)
 +\frac{B_2(a)}{2L}
 -\frac{B_3(a)}{6L^2}
 +O(L^{-3}).
\label{eq:S10-shifted-Stirling}
\end{equation}
Substitution of the exact products
\eqref{eq:S9-ff2-u1}--\eqref{eq:S9-pp2-u1} gives
\begin{align}
 \cM_{2,L}^{\ff}
 ={}&(1-\log2)L-\frac14\log L
 +\frac12+A
 -\frac{11}{96L}
 +\frac{13}{384L^2}
 +O(L^{-3}),
\label{eq:S10-M2-ff}\\
 \cM_{2,L}^{\fplus}
 ={}&(1-\log2)L-\frac14\log L
 +1+A
 -\frac{47}{96L}
 +\frac{125}{384L^2}
 +O(L^{-3}),
\label{eq:S10-M2-fplus}\\
 \cM_{2,L}^{\pp}
 ={}&(1-\log2)L-\frac14\log L
 +\frac32-\log2+A
 -\frac{35}{96L}
 +\frac{35}{128L^2}
 +O(L^{-3}).
\label{eq:S10-M2-pp}
\end{align}
Equation~\eqref{eq:S10-M2-pp} can be obtained either from the pinned-ensemble
product or by expanding the rank shift
$\cM_{2,L}^{\pp}=\cM_{2,L+1}^{\ff}$.  Algebraically this is exact for the
pinned ensemble; its interpretation as the physical equal-fixed spin-chain
entropy follows from the S5 dual-parity column quotient.  The agreement is a
useful check on all subleading coefficients.

Conditional on the reconstructed opposite-fixed ratio
\eqref{eq:S9-Reta2-candidate}, its large-$L$ expansion is
\begin{equation}
 \cR_{2,L}^{(\eta)}
 =\frac34+\frac{5}{8L}-\frac{23}{64L^2}
 +\frac{19}{64L^3}+O(L^{-4}),
\label{eq:S10-Reta2-expansion}
\end{equation}
so that
\begin{align}
 \cM_{2,L}^{\pmfix}
 ={}&(1-\log2)L-\frac14\log L
 +\frac32-\log2+A+\log\frac43
 \nonumber\\
 &-\frac{115}{96L}
 +\frac{1267}{1152L^2}
 +O(L^{-3}).
\label{eq:S10-M2-pm}
\end{align}
The constant and the displayed inverse-size coefficients in
Eq.~\eqref{eq:S10-M2-pm} are therefore conjectural in precisely the same sense as Eq.~\eqref{eq:S9-Reta2-candidate}.  The free--free and free--fixed expansions follow from the analytic products in S9, and the pinned expansion follows exactly from the physical rank shift.

\begin{table}[H]
\caption{Boundary constants at $\alpha=2$ in
$\mathcal M_{2,L}^{\mathsf b}=(1-\log2)L-\frac14\log L+C_2^{\mathsf b}+o(1)$.}
\label{tab:S10-alpha2-constants}
\centering
\small
\begin{tabular}{@{}l l@{}}
\toprule
Boundary & $C_2^{\mathsf b}$ \\
\midrule
$\ff$ & $\frac12+A$ \\
$\fplus$ & $1+A$ \\
$\pp$ & $\frac32-\log2+A$ \\
$\pmfix$ & $\frac32-\log2+A+\log(4/3)$ \\
\bottomrule
\end{tabular}
\end{table}
The free--fixed entry uses the all-rank product proved in S9, the equal-fixed
entry follows from the exact rank shift, and the opposite-fixed entry also
uses the reconstructed ratio Eq.~\eqref{eq:S9-Reta2-candidate}.

The exact free--fixed product gives the second invariant in
Eq.~\eqref{eq:S9-Rcross2-finite}; expanding it gives
\begin{equation}
 \cR_{2,L}^{(\times)}
 =2-\frac1L+\frac{15}{16L^2}
 -\frac{19}{24L^3}+O(L^{-4}).
\label{eq:S10-Rcross2-expansion}
\end{equation}
Combining this exact result with the separately reconstructed opposite-fixed ratio gives
\begin{equation}
 \Delta_2^{(\eta)}=\log\frac43,
 \qquad
 \Delta_2^{(\times)}=-\log2,
 \qquad
 \Delta_2^{(\eta\times)}=-\log\frac32,
\label{eq:S10-alpha2-deltas}
\end{equation}
where
$\Delta_\alpha^{(\eta)}$ and $\Delta_\alpha^{(\times)}$ are the entropy
combinations in Eqs.~\eqref{eq:S8-Delta-eta} and
\eqref{eq:S8-Delta-cross}.  Here $\Delta_2^{(\eta)}$ inherits the opposite-fixed reconstruction,
$\Delta_2^{(\times)}=-\log2$ is exact, and
$\Delta_2^{(\eta\times)}$ inherits the opposite-fixed input.

\subsection{Controlled expansion at \texorpdfstring{$\alpha=4$}{alpha=4}}

For $\mathsf b\in\{\ff,\fplus,\pp\}$, continuation of the verified
square-collapse identity Eq.~\eqref{eq:S9-square-collapse} to all ranks implies
\begin{equation}
 \cM_{4,L}^{\mathsf b}=\frac23\cM_{2,L}^{\mathsf b}.
\label{eq:S10-M4-M2}
\end{equation}
Consequently,
\begin{align}
 \cM_{4,L}^{\ff}
 ={}&\frac23(1-\log2)L-\frac16\log L
 +\frac13+\frac23A
 -\frac{11}{144L}
 +\frac{13}{576L^2}
 +O(L^{-3}),
\label{eq:S10-M4-ff}\\
 \cM_{4,L}^{\fplus}
 ={}&\frac23(1-\log2)L-\frac16\log L
 +\frac23+\frac23A
 -\frac{47}{144L}
 +\frac{125}{576L^2}
 +O(L^{-3}),
\label{eq:S10-M4-fplus}\\
 \cM_{4,L}^{\pp}
 ={}&\frac23(1-\log2)L-\frac16\log L
 +1-\frac23\log2+\frac23A
 -\frac{35}{144L}
 +\frac{35}{192L^2}
 +O(L^{-3}).
\label{eq:S10-M4-pp}
\end{align}
The change from $-1/4$ to $-1/6$ in the SRE logarithm is therefore visible
directly in the controlled lattice closed forms, conditional on the all-rank
square-collapse continuation.

The reconstructed marginal ratio has the expansion
\begin{equation}
 \cR_{4,L}^{(\eta)}
 =\frac{143}{240}+\frac{67}{80L}
 -\frac{11}{128L^2}
 +\frac{107}{1280L^3}
 +O(L^{-4}).
\label{eq:S10-Reta4-expansion}
\end{equation}
Conditional on Eq.~\eqref{eq:S9-Reta4-candidate},
\begin{align}
 \cM_{4,L}^{\pmfix}
 ={}&\frac23(1-\log2)L-\frac16\log L
 +1-\frac23\log2+\frac23A
 +\frac13\log\frac{240}{143}
 \nonumber\\
 &-\frac{14653}{20592L}
 +\frac{2197307}{3926208L^2}
 +O(L^{-3}).
\label{eq:S10-M4-pm}
\end{align}
The unusually large $1/L$ correction in Eq.~\eqref{eq:S10-M4-pm} explains
why naive low-rank extrapolations of the marginal ratio are unreliable.

\begin{table}[H]
\caption{Boundary constants at $\alpha=4$ in
$\mathcal M_{4,L}^{\mathsf b}=\frac23(1-\log2)L-\frac16\log L+C_4^{\mathsf b}+o(1)$.}
\label{tab:S10-alpha4-constants}
\centering
\small
\begin{tabular}{@{}l l@{}}
\toprule
Boundary & $C_4^{\mathsf b}$ \\
\midrule
$\ff$ & $\frac13+\frac23A$ \\
$\fplus$ & $\frac23+\frac23A$ \\
$\pp$ & $1-\frac23\log2+\frac23A$ \\
$\pmfix$ & $1-\frac23\log2+\frac23A+\frac13\log(240/143)$ \\
\bottomrule
\end{tabular}
\end{table}
These constants are conditional on the all-rank square-collapse continuation;
the opposite-fixed value additionally uses the reconstructed marginal ratio.

Conditional on the all-rank square collapse for $\ff$, $\fplus$, and
$\pp$,
\begin{equation}
 \cR_{4,L}^{(\times)}
 =\left[\cR_{2,L}^{(\times)}\right]^2
 =4-\frac4L+\frac{19}{4L^2}+O(L^{-3}),
\label{eq:S10-Rcross4-expansion}
\end{equation}
and
\begin{equation}
 \Delta_4^{(\eta)}=\frac13\log\frac{240}{143},
 \qquad
 \Delta_4^{(\times)}=-\frac13\log4,
 \qquad
 \Delta_4^{(\eta\times)}=-\frac13\log\frac{143}{60}.
\label{eq:S10-alpha4-deltas}
\end{equation}
All three displayed $\alpha=4$ combinations are conditional: those containing $\eta$ inherit the marginal finite-rank reconstruction, while the cross-ratio contribution inherits the all-rank square-collapse continuation.

\subsection{Why the two ratios have finite limits}

Suppose the four open moments share $f_\alpha$ and $\sigma_\alpha$ as in
Eq.~\eqref{eq:S10-raw-asymptotic}.  Then
\begin{align}
 \log\cR_{\alpha,L}^{(\eta)}
 & =\log\mathcal A_\alpha^{\pmfix}
   -\log\mathcal A_\alpha^{\pp}+o(1),
\label{eq:S10-Reta-cancel}\\
 \log\cR_{\alpha,L}^{(\times)}
 & =\log\mathcal A_\alpha^{\ff}
  +\log\mathcal A_\alpha^{\pp}
  -2\log\mathcal A_\alpha^{\fplus}+o(1).
\label{eq:S10-Rcross-cancel}
\end{align}
The extensive free energy and the complete logarithmic corner term cancel
identically.  The endpoint-normalization theorem of Sec.~\ref{sec:S8} then
removes the remaining local scales.  This is the precise lattice reason the
ratios can carry universal junction information even though the individual
constants $C_\alpha^{\mathsf b}$ need not be universal.

A useful form of the second ratio follows from the rank shift:
\begin{equation}
 \cR_\alpha^{(\times)}
 =\frac{\ee^{f_\alpha}}{2}
 \left(\frac{\mathcal A_\alpha^{\ff}}
 {\mathcal A_\alpha^{\fplus}}\right)^2.
\label{eq:S10-Rcross-rankshift}
\end{equation}
Thus the all-index relation between $\pp$ and $\ff$ does not determine
$\cR_\alpha^{(\times)}$ by itself: the free--fixed amplitude remains an
independent boundary datum.  This explains why two invariants, rather than
one, are required.

\subsection{Locked phase and the order of limits}

The locking transition for the measurement boundary occurs for $\alpha>4$,
but it does not justify replacing the fixed-$\alpha$ thermodynamic sum by its
$\alpha\to\infty$ maximizers.  At fixed finite $L$, taking
$\alpha\to\infty$ indeed counts the unit-magnitude Pauli strings: the
parity-even free--free state has the identity and global parity string,
whereas a generic parity-breaking fixed-boundary state retains only the
identity.  The opposite order of limits is different.

For example, Eq.~\eqref{eq:S4-image-cosecant} in the $C_L$ free--free chain
gives $\abs{\avg{Z_j}}\to2/\pi$ for a positive fraction of bulk sites as
$L\to\infty$.  Hence for every fixed finite $\alpha$,
\begin{equation}
 \cZ_{\alpha,L}^{\ff}(1)
 \ge\sum_{j=1}^{L}\abs{\avg{Z_j}}^{2\alpha}
 \gtrsim L\left(\frac2\pi\right)^{2\alpha}.
\label{eq:S10-locked-lower-bound}
\end{equation}
Thus $\cZ_{\alpha,L}^{\ff}(1)$ cannot tend to the finite value two at fixed
$\alpha>4$.  Dominant-configuration counting describes the
$\alpha\to\infty$ limit at fixed size, not the fixed-$\alpha$ thermodynamic
limit.  The finite ratios $\cR_\alpha^{(\eta)}$ and
$\cR_\alpha^{(\times)}$ for $\alpha>4$ therefore remain to be determined by
a genuine large-$L$ calculation.

\begin{table}[H]
\caption{Thermodynamic organization of the open-boundary problem.}
\label{tab:S10-phase-summary}
\centering
\small
\begin{tabular}{@{}c c p{0.55\textwidth}@{}}
\toprule
Regime & $b_\alpha$ & Dominant boundary mechanism \\
\midrule
$0<\alpha<4$ & $-1/4$ & unlocked measurement boundary and collective log-gas fluctuations \\
$\alpha=4$ & $-1/6$ & marginal alias sector and the candidate square-collapse structure \\
$\alpha>4$ & $0$ & locked measurement-boundary phase; finite ratios require a fixed-$\alpha$ thermodynamic calculation \\
\bottomrule
\end{tabular}
\end{table}

The asymptotic analysis has therefore isolated the finite boundary-dependent quantities.
The common volume and corner terms reproduce the known replica phases; the
finite ratios retain the boundary-channel information.  Section~\ref{sec:S11}
now addresses the mechanism governing the first of these ratios below
locking.

\section{Gaussian-character limit and its scope}
\label{sec:S11}

The exact algebraic character identity of S8 converts the opposite/equal-fixed ratio into a single random variable in the parent $C_N$ ensemble; the S5 dual-parity quotient establishes its identification with the physical spin-chain ratio.  This section develops
the thermodynamic proposal behind the Letter: below locking, the fundamental
character behaves as the first Fourier mode of an unlocked log gas, while at
the marginal index its moments cease to follow the Gaussian continuation.
For the fixed-$\alpha$ locked regime the finite ratios are left open, as
explained in S10.
We separate exact finite-size statements from the probabilistic assumptions
needed for the limit.

\subsection{The character as a linear statistic}

Let
\begin{equation}
 X_{\alpha,N}(U)
 \equiv\chi_{\omega_1}^{(C_N)}(U)
 =2\sum_{\theta\in U}\cos(2\theta),
\label{eq:S11-Xdef}
\end{equation}
where $U$ is distributed with the normalized full parent measure
\begin{equation}
 \mathbb P_{\alpha,N}(U)
 =\frac{\mathfrak V_C(U)^{2\alpha}}
 {\displaystyle\sum_{|V|=N}\mathfrak V_C(V)^{2\alpha}}.
\label{eq:S11-parent-probability}
\end{equation}
the complement identity Eq.~\eqref{eq:S8-pinned-full} allows the full measure to be used in place
of the pinned one.  For the projected fixed-boundary construction one has exactly
\begin{equation}
 \cR_{\alpha,L}^{(\eta)}
 =\mathbb E_{\alpha,N}\abs{X_{\alpha,N}}^{2\alpha},
 \qquad N=L+1.
\label{eq:S11-Reta-moment}
\end{equation}

Complementation is an exact involution of the finite grid.  Since
\begin{equation}
 \mathfrak V_C(U^c)=\mathfrak V_C(U),
 \qquad
 X_{\alpha,N}(U^c)=-X_{\alpha,N}(U),
\label{eq:S11-complementation}
\end{equation}
all odd moments vanish at every rank:
\begin{equation}
 \mathbb E_{\alpha,N}[X_{\alpha,N}^{2m+1}]=0.
\label{eq:S11-odd-moments-zero}
\end{equation}
Thus centering is exact.  The only nontrivial questions are the limiting
variance, the higher cumulants, and the tails.

After unfolding the reflected interval, the continuous analogue of
Eq.~\eqref{eq:S11-parent-probability} is a $C_N$ log gas with effective Dyson
index
\begin{equation}
 \beta=2\alpha.
\label{eq:S11-beta}
\end{equation}
For the continuous reflected $C_N$ (Jacobi) $\beta$ ensemble, the asymptotic
variance of a smooth linear statistic $X_f=\sum_j f(x_j)$ is expressed in
terms of Fourier--Chebyshev coefficients $\widehat f_k$ as~\cite{Johansson1998,BorotGuionnet2013}
\begin{equation}
 \operatorname{Var}(X_f)
 \longrightarrow\frac{2}{\beta}
 \sum_{k=1}^{\infty}k\abs{\widehat f_k}^2.
\label{eq:S11-linear-stat-variance}
\end{equation}
For $f(x)=2x$, only $\widehat f_1=1$ contributes, and
Eq.~\eqref{eq:S11-linear-stat-variance} gives
\begin{equation}
 \operatorname{Var}(X_{\alpha,N})
 \longrightarrow\frac{1}{\alpha}.
\label{eq:S11-variance-prediction}
\end{equation}
This stiffness is the origin of the variance quoted in the Letter.

Equation~\eqref{eq:S11-linear-stat-variance} is not by itself a proof for the
present model.  The particle number and the number of allowed lattice sites
grow together, so the gas is a discrete $\beta$ ensemble at fixed filling.
Discrete-ensemble central-limit theorems require that the equilibrium measure
remain away from the upper-density constraint~\cite{BorodinGorinGuionnet2016}.
That requirement is precisely what is expected to fail at the locking
transition.  The variance argument is therefore a controlled guide to the
unlocked phase, not a theorem valid through $\alpha=4$.

\subsection{A precise conditional statement}

The assumptions needed to pass from a character central limit theorem to the
physical moment are stronger than weak convergence.  We formulate them
explicitly.

For every fixed $0<\alpha<4$,
\begin{equation}
 X_{\alpha,N}\xrightarrow{\mathrm d}
 X_\alpha,
 \qquad
 X_\alpha\sim\mathcal N\!\left(0,\frac1\alpha\right),
\label{eq:S11-CLT}
\end{equation}
and there exists $\varepsilon_\alpha>0$ such that
\begin{equation}
 \sup_N\mathbb E_{\alpha,N}
 \abs{X_{\alpha,N}}^{2\alpha+\varepsilon_\alpha}<\infty.
\label{eq:S11-uniform-integrability}
\end{equation}

The second condition is a convenient sufficient form of uniform
integrability.  It rules out a vanishing set of configurations whose
character is large enough to change the $2\alpha$-th moment while remaining
invisible to convergence in distribution.

Under these assumptions,
\begin{equation}
 \cR_\alpha^{(\eta)}
 =\lim_{L\to\infty}\cR_{\alpha,L}^{(\eta)}
 =\frac{2^\alpha\Gamma(\alpha+\frac12)}
 {\sqrt\pi\,\alpha^\alpha},
 \qquad 0<\alpha<4.
\label{eq:S11-Gaussian-Reta}
\end{equation}

\noindent\emph{Derivation.} For $X\sim\mathcal N(0,\sigma^2)$,
\begin{equation}
 \mathbb E\abs X^p
 =\sigma^p2^{p/2}
 \frac{\Gamma((p+1)/2)}{\sqrt\pi}.
\label{eq:S11-Gaussian-absolute-moment}
\end{equation}
Set $p=2\alpha$ and $\sigma^2=1/\alpha$.  Convergence in distribution
together with Eq.~\eqref{eq:S11-uniform-integrability} implies convergence
of the $2\alpha$-th absolute moment, yielding
Eq.~\eqref{eq:S11-Gaussian-Reta}.

For positive integer $\alpha$, Eq.~\eqref{eq:S11-Gaussian-Reta} reduces to
\begin{equation}
 \cR_\alpha^{(\eta)}
 =\frac{(2\alpha-1)!!}{\alpha^\alpha}.
\label{eq:S11-Gaussian-integer}
\end{equation}
Several values and the associated entropy differences are collected in
Table~\ref{tab:S11-Gaussian-values}.

\begin{table}[H]
\caption{Values implied by the conditional Gaussian-character formula.}
\label{tab:S11-Gaussian-values}
\centering
\small
\begin{tabular}{@{}c c c@{}}
\toprule
$\alpha$ & $\mathcal R_\alpha^{(\eta)}$ &
$\Delta_\alpha^{(\eta)}$ \\
\midrule
$\frac12$ & $2/\sqrt\pi$ & $\log(4/\pi)$ \\
$1$ & $1$ & $\gamma_{\rm E}+\log2-1$ \\
$\frac32$ & $8/[3\sqrt{3\pi}]$ & Gaussian formula \\
$2$ & $3/4$ & $\log(4/3)$ \\
$3$ & $5/9$ & $\frac12\log(9/5)$ \\
$4$ & $105/256$ & Gaussian continuation only \\
\bottomrule
\end{tabular}
\end{table}
Only the ratio at $\alpha=1$ is fixed independently by finite-state
normalization.  The $\alpha=2$ value also agrees with the reconstructed
finite-rank sequence; the $\alpha=4$ Gaussian continuation is not the marginal
candidate.

\subsection{The Shannon junction derivative}

At $\alpha=1$, every finite-size ratio equals one.  In differentiating the finite Pauli sums we use the standard convention $0\log0=0$ for exact zero weights, so the nontrivial information is in the derivative.  Define
\begin{equation}
 \mathscr S_\eta^{\rm junction}
 \equiv-\left.\frac{\partial}{\partial\alpha}
 \log\cR_\alpha^{(\eta)}\right|_{\alpha=1}.
\label{eq:S11-Shannon-junction-def}
\end{equation}
The Gaussian law gives
\begin{equation}
 \mathscr S_\eta^{\rm junction}
 =\gamma_{\rm E}+\log2-1
 =0.2703628455\ldots.
\label{eq:S11-Shannon-junction}
\end{equation}
This is exactly the $\alpha\to1$ limit of
$\Delta_\alpha^{(\eta)}$.

The same quantity can be computed directly without numerical differentiation.
Let
\begin{equation}
 a_P^{\mathsf b}=\abs{\avg{P}_{\mathsf b}}^2.
\label{eq:S11-aP}
\end{equation}
Since $\sum_Pa_P^{\mathsf b}=2^L$,
\begin{equation}
 \Delta_1^{(\eta)}(L)
 =-\frac1{2^L}\left[
 \sum_Pa_P^{\pmfix}\log a_P^{\pmfix}
 -\sum_Pa_P^{\pp}\log a_P^{\pp}
 \right].
\label{eq:S11-Shannon-direct}
\end{equation}
Equation~\eqref{eq:S11-Shannon-direct} is useful for a future large-size test
of Eq.~\eqref{eq:S11-Shannon-junction} because it avoids fitting moments at
nearby R\'enyi indices.

\subsection{A test away from the solvable indices}

At $\alpha=3$ the numerator and denominator use the same aliased measure, so
this index provides a useful test of the Gaussian-character
proposal.  Direct finite-rank character sums decrease toward the Gaussian
prediction $\mathcal R_3^{(\eta)}=5/9$, but the accessible ranks have sizable
corrections and do not yet discriminate the thermodynamic limit.  We therefore
use $5/9$ only as a conjectural prediction, not as numerical evidence for the
central-limit law.

\subsection{The second invariant}

The character CLT controls only $\cR_\alpha^{(\eta)}$.  For the second
normalization-free ratio, Pauli normalization gives
\begin{equation}
 \cR_1^{(\times)}=1,
\label{eq:S11-Rcross-a1}
\end{equation}
and the all-rank $\alpha=2$ free--fixed product proved in S9 gives
\begin{equation}
 \cR_2^{(\times)}=2.
\label{eq:S11-Rcross-a2}
\end{equation}
At $\alpha=4$, $\cR_4^{(\times)}=4$ remains conditional on the all-rank
rectangular square-collapse relation.

The exact and numerical data suggest a simple unlocked-phase law.  We
conjecture
\begin{equation}
 \cR_\alpha^{(\times)}=2^{\alpha-1},
 \qquad 0<\alpha<4.
\label{eq:S11-Rcross-conjecture}
\end{equation}
This conjecture is consistent with the exact support limit
$\cR_{0}^{(\times)}=1/2$, the normalization point
$\cR_{1}^{(\times)}=1$, and the proved value
$\cR_{2}^{(\times)}=2$.  At the exactly evaluable $\alpha=1/2$ Pfaffian
point, finite-size calculations through $L=60$ extrapolate to
\begin{equation}
 \cR_{1/2}^{(\times)}\simeq\frac1{\sqrt2},
\label{eq:S11-Rcross-half-numerical}
\end{equation}
in agreement with Eq.~\eqref{eq:S11-Rcross-conjecture}.  An independent
derivative calculation at the exact normalization point gives
\begin{equation}
 \left.\partial_\alpha\log\cR_\alpha^{(\times)}\right|_{\alpha=1}
 \longrightarrow\log2,
\label{eq:S11-Rcross-Shannon-slope}
\end{equation}
again exactly as predicted by Eq.~\eqref{eq:S11-Rcross-conjecture}.  Exact
finite-size data at $\alpha=3/2$ through $L=14$ approach
$\sqrt2=2^{1/2}$ with a clean leading $1/L$ correction.  At
$\alpha=5/2$, exact data through $L=14$ remain compatible with
$2^{3/2}$ but exhibit substantially slower finite-size corrections.  The
approach to the marginal point is therefore not controlled by the currently
accessible ranks, and Eq.~\eqref{eq:S11-Rcross-conjecture} should be regarded
as a thermodynamic conjecture for the unlocked phase rather than as an
all-rank finite-size identity.

If Eq.~\eqref{eq:S11-Rcross-conjecture} holds, the associated entropy
combination is independent of the R\'enyi index throughout the unlocked
phase:
\begin{equation}
 \Delta_\alpha^{(\times)}
 =\frac{\log\cR_\alpha^{(\times)}}{1-\alpha}
 =-\log2,
 \qquad 0<\alpha<4.
\label{eq:S11-Delta-cross-conjecture}
\end{equation}

At fixed $L$, support counting remains exact:
\begin{equation}
 \cZ_{0,L}^{\ff}=\binom{2L}{L},
 \qquad
 \cZ_{0,L}^{\fplus}=\binom{2L+1}{L},
 \qquad
 \cZ_{0,L}^{\pp}=\frac12\binom{2L+2}{L+1},
\label{eq:S11-support-counts}
\end{equation}
so
\begin{equation}
 \cR_{0,L}^{(\times)}=\frac{L+1}{2L+1}
 \longrightarrow\frac12.
\label{eq:S11-Rcross-alpha0}
\end{equation}
The conjecture Eq.~\eqref{eq:S11-Rcross-conjecture} is continuous with this
exact support limit, since $2^{\alpha-1}\to1/2$ as $\alpha\to0^+$.  Support
counting alone does not prove that the $L\to\infty$ and $\alpha\to0^+$
limits commute, but it provides no evidence for a mismatch between them.

\subsection{The marginal anomaly at \texorpdfstring{$\alpha=4$}{alpha=4}}

The Gaussian continuation gives
\begin{equation}
 \cR_{4,\mathrm G}^{(\eta)}=\frac{105}{256}
 =0.41015625,
\label{eq:S11-R4-Gauss}
\end{equation}
whereas the reconstructed finite-rank sequence gives
\begin{equation}
 \cR_4^{(\eta)}=\frac{143}{240}
 =0.59583333\ldots.
\label{eq:S11-R4-marginal}
\end{equation}
The absolute discrepancy and enhancement factor are
\begin{equation}
 \frac{143}{240}-\frac{105}{256}=\frac{713}{3840},
 \qquad
 \frac{\cR_4^{(\eta)}}{\cR_{4,\mathrm G}^{(\eta)}}
 =\frac{2288}{1575}=1.452698\ldots.
\label{eq:S11-R4-anomaly}
\end{equation}
This is much too large to be confused with the known $1/L$ correction in
Eq.~\eqref{eq:S10-Reta4-expansion}.  Additional exact enumeration through
$N=15$ indicates that the marginal anomaly is already visible in lower
moments.  In particular, the variance agrees at every tested rank with
\begin{equation}
 \mathbb E_{4,N}[X^2]=\frac{2N+1}{4N-1},
\label{eq:S11-marginal-variance-candidate}
\end{equation}
which would tend to $1/2$, rather than the Gaussian-continuation value
$1/4$.  This finite-rank rational pattern is strongly constrained but is not
proved here; it therefore provides evidence for a changed marginal law rather
than an additional theorem.

Three independent structures change at the same index:
\begin{enumerate}[label=(\roman*)]
\item the open SRE logarithm changes from $-1/4$ to $-1/6$;
\item the character eighth moment departs from the Gaussian continuation;
\item the character numerator acquires the extremal $\pm4D$ aliases absent
      from its denominator, as shown in Table~\ref{tab:S8-alias-kernels}.
\end{enumerate}
The third statement is an exact finite-size algebraic fact.  It suggests why
weak convergence of the central part of the distribution, even if present,
need not control the eighth moment: the new alias sector can encode rare
near-locked configurations or nonuniform tails.

At $\alpha=4$, the reconstructed eighth character moment approaches
$143/240$ rather than the Gaussian continuation $105/256$.  Together with
Eq.~\eqref{eq:S11-marginal-variance-candidate}, the finite-rank data favor a
marginal law whose low moments already differ from the unlocked Gaussian
continuation.  A Gaussian core with nonuniform tails is not excluded, because
weak convergence need not control even the variance without uniform
integrability, but the anomaly is not confined to the eighth moment.

\subsection{Thermodynamic status}

The finite character identity is exact, while the thermodynamic statements
have different levels of control.  For $0<\alpha<4$,
Eq.~\eqref{eq:S11-Gaussian-Reta} is a conditional consequence of the stated
CLT and uniform-integrability assumptions.  At $\alpha=4$, the value
$143/240$ remains conditional on the reconstructed finite-rank rational
form, with additional lower-moment data indicating a non-Gaussian marginal
regime.  For $\alpha>4$, the fixed-$\alpha$ thermodynamic ratios are not
determined by the $\alpha\to\infty$ maximizer count and are left open.

For the second invariant, the normalization point $\cR_1^{(\times)}=1$ and
the value $\cR_2^{(\times)}=2$ are exact, while the marginal value four is
conditional on the rectangular square collapse.  The combined support,
Pfaffian, Shannon-derivative, and finite-rank evidence motivates the
unlocked-phase conjecture Eq.~\eqref{eq:S11-Rcross-conjecture}; its behavior
arbitrarily close to $\alpha=4$ remains to be established analytically.

\section{Boundary interpretation and microscopic robustness tests}
\label{sec:S12}

This section separates two statements.  The $B_L,C_L,D_L$ Pauli geometries
are exact finite-lattice properties, whereas the identification of the
normalization-free thermodynamic ratios with replica-boundary junction
amplitudes is a continuum interpretation.  We then test the stability of the
ratios directly by varying the physical boundary field away from the
exactly solvable lattice representative.

\subsection{Shared infrared boundary data and exact Pauli geometry}

The three parity-preserving open chains have
\begin{equation}
 \varepsilon_n^{(R)}
 =\frac{\pi(n+\frac12)}{L_{\rm eff}^{(R)}}+O(L^{-3}),
 \qquad
 L_{\rm eff}^{(C,B,D)}=L+\frac12,\ L,\ L-\frac12.
\label{eq:S12-common-tower}
\end{equation}
After the nonuniversal effective-length shift is removed, all three realize
the free--free Ising Neveu--Schwarz tower.  Their exact Pauli minors nevertheless
carry different trigonometric wall factors,
\begin{equation}
 \prod_i\sin\theta_i,
 \qquad
 \prod_i\sin2\theta_i,
 \qquad
 1,
\label{eq:S12-wall-factors}
\end{equation}
for $B_L,C_L,D_L$, respectively.  Thus the common Cardy boundary state does
not determine the exact finite-lattice Pauli zero structure.  The distinction
is one of root lengths and trigonometric wall factors; $B_L$ and $C_L$ share
the same finite linear reflection hyperplanes.

For comparison, on a strip with the same conformal boundary
condition $a$ at both ends, the leading R\'enyi entropy of a
boundary-adjacent interval is~\cite{CalabreseCardy2009,AffleckLudwig1991,AffleckReview2009}
\begin{equation}
 S_n^{(a)}(\ell)
 =\frac{c}{12}\left(1+\frac1n\right)
 \log\!\left[
 \frac{2L}{\pi a_0}\sin\!\left(\frac{\pi\ell}{L}\right)
 \right]
 +\log g_a+s_n^{\rm bulk}+\cdots .
\label{eq:S12-boundary-entanglement}
\end{equation}
The tuned $B,C,D$ chains share $c=1/2$ and the free Cardy state, while their
exact Pauli distributions retain the different wall structures above.
Spatial entanglement can of course distinguish mixed boundary sectors through
boundary-condition-changing data in suitable geometries~\cite{Estienne2025}; the present claim
is only that the Pauli measurement probes a different replica object.

\subsection{\texorpdfstring{$B/C/D$}{B/C/D} microscopic realizations and orientation-resolved ratios}
\label{sec:S12-folded-ratios}

The distinction above suggests a direct test.  The exact $B,C,D$ wall
geometries should remain different at finite lattice spacing, but a
normalization-free quantity that depends only on the infrared physical
boundary class should lose that distinction.  The endpoint-tuned matrices of
\hyperref[sec:S5-folded-fixed]{Sec.~S5.E} make this test well defined without replacing
one sector of a ratio by an unrelated microscopic convention.

For $R=B,C,D$ we therefore use the orientation-resolved invariants
Eqs.~\eqref{eq:S8-Reta-folded} and~\eqref{eq:S8-Rcross-folded}.  In particular,
for the asymmetric $B$ realization
\begin{equation}
 \cR_{\alpha,L}^{(\times)\mid B}
 =\frac{\cZ_{\alpha,L}^{\ff\mid B}\cZ_{\alpha,L}^{\pp\mid B}}
 {\cZ_{\alpha,L}^{\fplus\mid B}\cZ_{\alpha,L}^{+f\mid B}},
\label{eq:S12-B-oriented-cross}
\end{equation}
rather than replacing only the $\ff$ factor in the standard ratio.  This
point is essential: changing a single microscopic endpoint convention would
mix local normalization schemes and would not test an infrared invariant.
For $C$ and $D$, reflection symmetry reduces Eq.~\eqref{eq:S12-B-oriented-cross}
to the squared-denominator form.

The special-index results of \hyperref[sec:S9-folded-special]{Sec.~S9.G} then separate
two roles cleanly.  The free moments retain exact $B,C,D$ microscopic
structure, including different Pfaffians and products, whereas the ratios can
be compared after all local endpoint factors have cancelled.  Equality of
their thermodynamic limits is not a finite-lattice theorem; it is the
infrared universality statement tested below.

\subsection{Continuum interpretation of the two ratios}

Recent SRE field-theory constructions represent Pauli/Bell measurement as an
$\alpha$-dependent replicated boundary~\cite{HoshinoOshikawaAshida2026,HoshinoAshida2026}.
A continuum treatment of the present open chains should therefore reproduce
the exact lattice ratios
\begin{equation}
 \cR_{\alpha}^{(\eta)}
 =\lim_{L\to\infty}\frac{\cZ_{\alpha,L}^{\pmfix}}
 {\cZ_{\alpha,L}^{\pp}},
 \qquad
 \cR_{\alpha}^{(\times)}
 =\lim_{L\to\infty}
 \frac{\cZ_{\alpha,L}^{\ff}\cZ_{\alpha,L}^{\pp}}
 {(\cZ_{\alpha,L}^{\fplus})^2},
\label{eq:S12-lattice-ratios}
\end{equation}
through the corresponding junction amplitudes of the Pauli replica boundary.
Schematically, the expected identification is
\begin{equation}
 \cR_{\alpha}^{(\eta)}
 \ \longleftrightarrow\ 
 \frac{\mathcal J_\alpha(+,-\mid\Gamma_\alpha^{\rm P})}
      {\mathcal J_\alpha(+,+\mid\Gamma_\alpha^{\rm P})},
 \qquad
 \cR_{\alpha}^{(\times)}
 \ \longleftrightarrow\ 
 \frac{\mathcal J_\alpha(f,f\mid\Gamma_\alpha^{\rm P})
       \mathcal J_\alpha(+,+\mid\Gamma_\alpha^{\rm P})}
      {\mathcal J_\alpha(f,+\mid\Gamma_\alpha^{\rm P})^2}.
\label{eq:S12-junction-targets}
\end{equation}
Equation~\eqref{eq:S12-junction-targets} is a continuum target, not an
additional lattice identity.  What is exact on the lattice is the cancellation
of the two independent endpoint normalizations established in Sec.~\ref{sec:S8}
and, for $\cR^{(\eta)}$, the finite-rank character representation of
Sec.~\ref{sec:S8}.  If the junction interpretation depends only on the
infrared boundary classes, the orientation-resolved auxiliary ratios of
\hyperref[sec:S12-folded-ratios]{Sec.~S12.B} should approach the same limits as the
standard $C$ representative.  The calculations below test precisely this
additional expectation.

\subsection{Boundary-field flow and direct microscopic tests}
\label{sec:S12-flow}

We use two complementary checks.  The first varies the longitudinal boundary
field while keeping the standard $C$ microscopic termination.  The second
keeps the physical boundary condition fixed but changes the transverse
termination among the $B,C,D$ realizations.

For the first test take
\[
 H_{\rm bulk}=-\frac12\sum_{j=1}^{L-1}X_jX_{j+1}
 -\frac12\sum_{j=1}^{L}Z_j,
\]
and
\begin{equation}
\begin{aligned}
 H_{\pp}(\eta_b)&=H_{\rm bulk}-\frac{\eta_b}{2}(X_1+X_L),
 &\qquad
 H_{\pmfix}(\eta_b)&=H_{\rm bulk}-\frac{\eta_b}{2}(X_1-X_L),\\
 H_{\fplus}(\eta_b)&=H_{\rm bulk}-\frac{\eta_b}{2}X_L,
 &
 H_{\ff}&=H_{\rm bulk}.
\end{aligned}
\label{eq:S12-variable-boundary-fields}
\end{equation}
The boundary magnetic operator has scaling dimension $x_b=1/2$, so its RG
eigenvalue is $y_b=1/2$.  If $u_b(\eta_b)$ is the nonlinear boundary scaling
field,
\begin{equation}
 \xi_b\propto |u_b(\eta_b)|^{-2},\qquad
 \frac{L}{\xi_b}\propto L\,u_b(\eta_b)^2,\qquad
 u_b(\eta_b)=c_b\eta_b+O(\eta_b^3).
\label{eq:S12-boundary-scaling-variable}
\end{equation}
Thus $L\eta_b^2$ is the leading small-field proxy for the crossover variable;
it is not meant as an exact coordinate at arbitrary bare field.

\begin{figure}[H]
\centering
\subfloat[$\cR_{2,L}^{(\eta)}$.  The dashed line is $3/4$.\label{fig:S12-boundary-flow-eta}]{%
\includegraphics[width=0.47\textwidth]{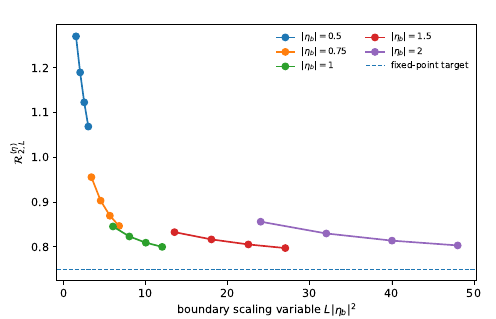}}
\hfill
\subfloat[$\cR_{2,L}^{(\times)}$.  The dashed line is $2$.\label{fig:S12-boundary-flow-cross}]{%
\includegraphics[width=0.47\textwidth]{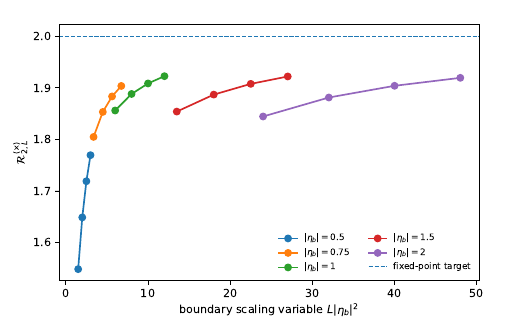}}
\caption{Direct-spin boundary-field flow at $\alpha=2$ for the standard $C$
representative.  The horizontal coordinate $L\eta_b^2$ is the leading
small-field scaling variable.  The data show the approach to the fixed-boundary
basin away from the exactly solvable value $\eta_b=1$.}
\label{fig:S12-boundary-flow}
\end{figure}

For $L=6,8,10,12$ we diagonalized the spin Hamiltonians directly and summed
the complete Pauli moments.  At $\eta_b=1$ the results reproduce the exact
minor representation.  Figure~\ref{fig:S12-boundary-flow} shows the evolution
away from that special lattice point.  The weakest field remains visibly in
the crossover regime, whereas stronger fields approach the same fixed-boundary
targets much more rapidly.

\begin{figure}[H]
\centering
\subfloat[$\alpha=2$: opposite/equal-fixed ratio.\label{fig:S12-BCD-a2-eta}]{%
\includegraphics[width=0.47\textwidth]{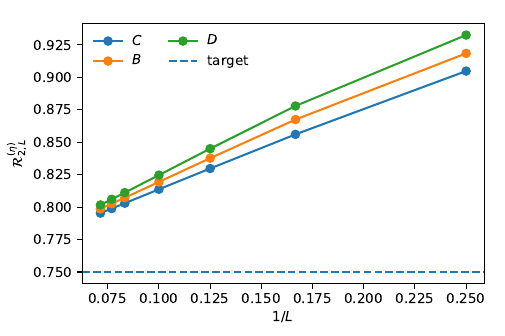}}
\hfill
\subfloat[$\alpha=2$: free/fixed cross-ratio.\label{fig:S12-BCD-a2-cross}]{%
\includegraphics[width=0.47\textwidth]{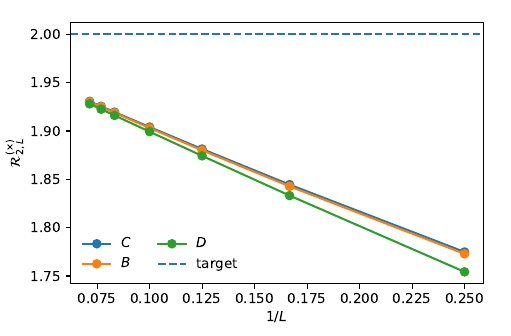}}\\[1mm]
\subfloat[$\alpha=4$: opposite/equal-fixed ratio.\label{fig:S12-BCD-a4-eta}]{%
\includegraphics[width=0.47\textwidth]{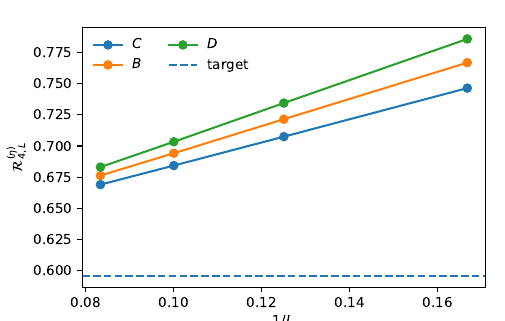}}
\hfill
\subfloat[$\alpha=4$: free/fixed cross-ratio.\label{fig:S12-BCD-a4-cross}]{%
\includegraphics[width=0.47\textwidth]{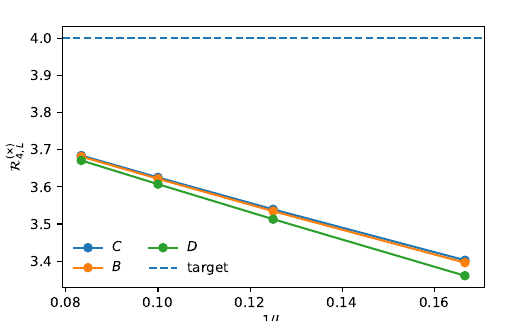}}
\caption{Convergence of the normalization-free ratios for the three microscopic
free-boundary realizations $C$, $B$, and $D$, using the stronger longitudinal
field $\eta_b=2$ to shorten the fixed-boundary crossover.  The dashed lines are
the thermodynamic targets $3/4$, $2$, $143/240$, and $4$, respectively.  The
finite-size sequences remain different, as expected from their distinct
microscopic terminations, while their separation decreases with increasing
$L$.}
\label{fig:S12-BCD-convergence}
\end{figure}

The second test is more directly tied to the root-system question.  We retain
the transverse endpoint tunings $\bm h_B$ or $\bm h_D$ while imposing the same
longitudinal free/fixed sectors.  For $B$ both mixed orientations are included
in the invariant according to Eq.~\eqref{eq:S8-Rcross-folded}.  The resulting
finite-size ratios are shown in Fig.~\ref{fig:S12-BCD-convergence}.  At
$\alpha=2$ the three microscopic realizations approach $3/4$ and $2$ from
different sides and with visibly different $1/L$ corrections.  At the
marginal point $\alpha=4$ the corrections are substantially larger, but the
same convergence pattern is recovered once the physical fixed boundary is
driven further into its RG basin.

For reference, Table~\ref{tab:S12-BCD-ratios} gives the largest directly
evaluated sizes.  At $\alpha=\tfrac12$ the table uses the solvable bare field
$\eta_b=1$; the reference values are $2/\sqrt\pi$ and $1/\sqrt2$, from the
unlocked-phase formulas of Eqs.~\eqref{eq:S11-Gaussian-Reta} and
\eqref{eq:S11-Rcross-conjecture}.  For $\alpha=2$ and $4$ the table uses
$\eta_b=2$.  The $\alpha=\tfrac12$ $\eta$ numerator still lacks the
single-Pfaffian compression of the uninserted sectors, whereas the
$\times$-ratio can be pushed much further with the Pfaffians of
\hyperref[sec:S9-folded-special]{Sec.~S9.G}.

\begin{table}[H]
\caption{Largest-size direct microscopic-realization data for the two
normalization-free ratios.  The targets at $\alpha=2$ and $4$ have the same
conditional status as the corresponding standard-$C$ formulas discussed in
Sec.~\ref{sec:S9}; the table is intended to display finite-size convergence,
not to assign statistical errors.}
\label{tab:S12-BCD-ratios}
\centering
\normalsize
\renewcommand{\arraystretch}{1.18}
\setlength{\tabcolsep}{5.5pt}
\begin{tabular}{@{}c c c c c c c c@{}}
\toprule
$\alpha$ & $\eta_b$ & $L$ & ratio & target & $C$ & $B$ & $D$\\
\midrule
$\tfrac12$ & $1$ & $12$ & $\cR^{(\eta)}$ & $1.128379$ & $1.095139$ & $1.075308$ & $1.059344$\\
$\tfrac12$ & $1$ & $12$ & $\cR^{(\times)}$ & $0.707107$ & $0.718331$ & $0.720764$ & $0.724052$\\
\addlinespace[2pt]
$2$ & $2$ & $14$ & $\cR^{(\eta)}$ & $0.750000$ & $0.795371$ & $0.798625$ & $0.801670$\\
$2$ & $2$ & $14$ & $\cR^{(\times)}$ & $2.000000$ & $1.930650$ & $1.930079$ & $1.928048$\\
\addlinespace[2pt]
$4$ & $2$ & $12$ & $\cR^{(\eta)}$ & $0.595833$ & $0.669001$ & $0.676274$ & $0.683079$\\
$4$ & $2$ & $12$ & $\cR^{(\times)}$ & $4.000000$ & $3.684449$ & $3.681643$ & $3.671216$\\
\bottomrule
\end{tabular}
\end{table}

At $\alpha=2$ and $\eta_b=2$, the $B/C/D$ spread has already fallen to
$6.3\times10^{-3}$ for $\cR^{(\eta)}$ and $2.6\times10^{-3}$ for
$\cR^{(\times)}$ at $L=14$.  At $\alpha=4$ the marginal corrections are
larger, but the three sequences again move coherently toward the same targets.
This behavior is consistent with different extrapolation lengths and
irrelevant boundary couplings multiplying a common infrared amplitude.

A useful analytic anchor is obtained in the strong-boundary limit.  When
$|\eta_b|\to\infty$, a fixed endpoint spin is polarized along $X$.  The local
transverse term at that site, including the $\sqrt2$ tuning that distinguishes
the $B$ and $D$ microscopic realizations, no longer affects the low-energy
state.  In the opposite/equal-fixed ratio the polarized endpoint factors
cancel.  Removing the two endpoint spins leaves the standard problem on
$L-2$ sites, and therefore
\begin{equation}
 \cR_{\alpha,L}^{(\eta)\mid B}(\infty)
 =\cR_{\alpha,L}^{(\eta)\mid C}(\infty)
 =\cR_{\alpha,L}^{(\eta)\mid D}(\infty)
 =\cR_{\alpha,L-2}^{(\eta)\mid C}(\eta_b=1).
\label{eq:S12-strong-field-Reta}
\end{equation}
At $\alpha=2$, Eq.~\eqref{eq:S9-Reta2-candidate} then gives
\begin{equation}
 \cR_{2,L}^{(\eta)\mid R}(\infty)
 =\frac{(2L-1)(6L-13)}{(4L-9)(4L-5)}
 \longrightarrow\frac34,
 \qquad R=B,C,D,
\label{eq:S12-strong-field-Reta2}
\end{equation}
with the same finite-rank qualification as the parent $C$ formula.

Within the reconstructed $\alpha=2$ products of
\hyperref[sec:S9-folded-special]{Sec.~S9.G}, the strong-field cross-ratios take the simple forms
\begin{equation}
 \cR_{2,L}^{(\times)\mid B}(\infty)
 =\frac{2L-1}{L},
 \qquad
 \cR_{2,L}^{(\times)\mid D}(\infty)
 =\frac{32L(L-1)+9}{(4L-1)^2},
\label{eq:S12-strong-field-Rcross2}
\end{equation}
so both tend directly to $2$.  At $\alpha=4$ the identity
\eqref{eq:S12-strong-field-Reta} reduces the $B/C/D$ comparison to the same
standard-$C$ marginal sequence and hence, with the finite-rank reconstruction
of Sec.~\ref{sec:S9}, to the common limiting value $143/240$.

The two numerical tests and the strong-boundary reduction therefore tell a
consistent story.  The complete finite-lattice Pauli spectrum retains the
microscopic $B_L$, $C_L$, and $D_L$ termination exactly, while the
normalization-free ratios progressively lose that distinction as the
physical boundaries enter the same infrared basin.  This supports the use of
the ratios as continuum junction targets, without turning the microscopic
root labels themselves into additional BCFT boundary labels.

\clearpage


\begin{thebibliography}{99}

\bibitem{AffleckLudwig1991} I.~Affleck and A.~W.~W. Ludwig,
Universal noninteger ``ground-state degeneracy'' in critical quantum systems, \href{https://doi.org/10.1103/PhysRevLett.67.161}{Phys. Rev. Lett. \textbf{67}, 161 (1991)}.

\bibitem{AffleckReview2009} I.~Affleck, N.~Laflorencie, and E.~S. S\o rensen,
Entanglement entropy in quantum impurity systems and systems with boundaries, \href{https://doi.org/10.1088/1751-8113/42/50/504009}{J. Phys. A \textbf{42}, 504009 (2009)}.

\bibitem{Cardy1984} J.~L. Cardy,
Conformal invariance and surface critical behavior, \href{https://doi.org/10.1016/0550-3213(84)90241-4}{Nucl. Phys. B \textbf{240}, 514 (1984)}.

\bibitem{Cardy1989} J.~L. Cardy,
Boundary conditions, fusion rules and the Verlinde formula, \href{https://doi.org/10.1016/0550-3213(89)90521-X}{Nucl. Phys. B \textbf{324}, 581 (1989)}.

\bibitem{Holzhey1994} C.~Holzhey, F.~Larsen, and F.~Wilczek,
Geometric and renormalized entropy in conformal field theory, \href{https://doi.org/10.1016/0550-3213(94)90402-2}{Nucl. Phys. B \textbf{424}, 443 (1994)}.

\bibitem{Vidal2003} G.~Vidal, J.~I. Latorre, E.~Rico, and A.~Kitaev,
Entanglement in quantum critical phenomena, \href{https://doi.org/10.1103/PhysRevLett.90.227902}{Phys. Rev. Lett. \textbf{90}, 227902 (2003)}.

\bibitem{CalabreseCardy2004} P.~Calabrese and J.~Cardy,
Entanglement entropy and quantum field theory, \href{https://doi.org/10.1088/1742-5468/2004/06/P06002}{J. Stat. Mech. (2004) P06002}.

\bibitem{CalabreseCardy2009} P.~Calabrese and J.~Cardy,
Entanglement entropy and conformal field theory, \href{https://doi.org/10.1088/1751-8113/42/50/504005}{J. Phys. A \textbf{42}, 504005 (2009)}.

\bibitem{Laflorencie2006} N.~Laflorencie, E.~S. S\o rensen, M.-S.~Chang, and I.~Affleck,
Boundary effects in the critical scaling of entanglement entropy in 1D systems, \href{https://doi.org/10.1103/PhysRevLett.96.100603}{Phys. Rev. Lett. \textbf{96}, 100603 (2006)}.

\bibitem{CornfeldSela2017} E.~Cornfeld and E.~Sela,
Entanglement entropy and boundary renormalization group flow: Exact results in the Ising universality class, \href{https://doi.org/10.1103/PhysRevB.96.075153}{Phys. Rev. B \textbf{96}, 075153 (2017)}.

\bibitem{XavierRajabpour2020} J.~C. Xavier and M.~A. Rajabpour,
Entanglement and boundary entropy in quantum spin chains with arbitrary direction of the boundary magnetic fields, \href{https://doi.org/10.1103/PhysRevB.101.235127}{Phys. Rev. B \textbf{101}, 235127 (2020)}.

\bibitem{Estienne2025} B.~Estienne, Y.~Ikhlef, and A.~Rotaru,
R\'enyi entropies for one-dimensional quantum systems with mixed boundary conditions, \href{https://doi.org/10.21468/SciPostPhys.19.5.119}{SciPost Phys. \textbf{19}, 119 (2025)}.

\bibitem{SakaiSatoh2008} K.~Sakai and Y.~Satoh,
Entanglement through conformal interfaces, \href{https://doi.org/10.1088/1126-6708/2008/12/001}{J. High Energy Phys. \textbf{12} (2008) 001}.

\bibitem{BrehmBrunner2015} E.~M. Brehm and I.~Brunner,
Entanglement entropy through conformal interfaces in the 2D Ising model, \href{https://doi.org/10.1007/JHEP09(2015)080}{J. High Energy Phys. \textbf{09} (2015) 080}.

\bibitem{GutperleMiller2017} M.~Gutperle and J.~D. Miller,
Entanglement entropy at CFT junctions, \href{https://doi.org/10.1103/PhysRevD.95.106008}{Phys. Rev. D \textbf{95}, 106008 (2017)}.

\bibitem{StephanEtAl2009} J.-M. St\'ephan, S.~Furukawa, G.~Misguich, and V.~Pasquier,
Shannon and entanglement entropies of one- and two-dimensional critical wave functions, \href{https://doi.org/10.1103/PhysRevB.80.184421}{Phys. Rev. B \textbf{80}, 184421 (2009)}.

\bibitem{StephanMisguichPasquier2010} J.-M. St\'ephan, G.~Misguich, and V.~Pasquier,
R\'enyi entropy of a line in two-dimensional Ising models, \href{https://doi.org/10.1103/PhysRevB.82.125455}{Phys. Rev. B \textbf{82}, 125455 (2010)}.

\bibitem{LuitzAletLaflorencie2014} D.~J. Luitz, F.~Alet, and N.~Laflorencie,
Universal behavior beyond multifractality in quantum many-body systems, \href{https://doi.org/10.1103/PhysRevLett.112.057203}{Phys. Rev. Lett. \textbf{112}, 057203 (2014)}.

\bibitem{AlcarazRajabpour2013} F.~C. Alcaraz and M.~A. Rajabpour,
Universal behavior of the Shannon mutual information of critical quantum chains, \href{https://doi.org/10.1103/PhysRevLett.111.017201}{Phys. Rev. Lett. \textbf{111}, 017201 (2013)}.

\bibitem{Stephan2014} J.-M. St\'ephan,
Shannon and R\'enyi mutual information in quantum critical spin chains, \href{https://doi.org/10.1103/PhysRevB.90.045424}{Phys. Rev. B \textbf{90}, 045424 (2014)}.

\bibitem{AlcarazRajabpour2014} F.~C. Alcaraz and M.~A. Rajabpour,
Universal behavior of the Shannon and R\'enyi mutual information of quantum critical chains, \href{https://doi.org/10.1103/PhysRevB.90.075132}{Phys. Rev. B \textbf{90}, 075132 (2014)}.

\bibitem{TarighiEtAl2022} B.~Tarighi, R.~Khasseh, M.~N. Najafi, and M.~A. Rajabpour,
Universal logarithmic correction to R\'enyi (Shannon) entropy in generic systems of critical quadratic fermions, \href{https://doi.org/10.1103/PhysRevB.105.245109}{Phys. Rev. B \textbf{105}, 245109 (2022)}.

\bibitem{OlivieroLeoneHamma2022} L.~Leone, S.~F.~E. Oliviero, and A.~Hamma,
Stabilizer R\'enyi entropy, \href{https://doi.org/10.1103/PhysRevLett.128.050402}{Phys. Rev. Lett. \textbf{128}, 050402 (2022)}.

\bibitem{Gottesman1997} D.~Gottesman,
\emph{Stabilizer Codes and Quantum Error Correction}, Ph.D. thesis, California Institute of Technology (1997), \href{https://doi.org/10.7907/rzr7-dt72}{doi:10.7907/rzr7-dt72}.

\bibitem{BravyiKitaev2005} S.~Bravyi and A.~Kitaev,
Universal quantum computation with ideal Clifford gates and noisy ancillas, \href{https://doi.org/10.1103/PhysRevA.71.022316}{Phys. Rev. A \textbf{71}, 022316 (2005)}.

\bibitem{Veitch2012} V.~Veitch, C.~Ferrie, D.~Gross, and J.~Emerson,
Negative quasi-probability as a resource for quantum computation, \href{https://doi.org/10.1088/1367-2630/14/11/113011}{New J. Phys. \textbf{14}, 113011 (2012)}.

\bibitem{Howard2014} M.~Howard, J.~Wallman, V.~Veitch, and J.~Emerson,
Contextuality supplies the ``magic'' for quantum computation, \href{https://doi.org/10.1038/nature13460}{Nature \textbf{510}, 351 (2014)}.

\bibitem{Veitch2014} V.~Veitch, S.~A.~H. Mousavian, D.~Gottesman, and J.~Emerson,
The resource theory of stabilizer quantum computation, \href{https://doi.org/10.1088/1367-2630/16/1/013009}{New J. Phys. \textbf{16}, 013009 (2014)}.

\bibitem{HaugPiroliMonotones2023} T.~Haug and L.~Piroli,
Stabilizer entropies and nonstabilizerness monotones, \href{https://doi.org/10.22331/q-2023-08-28-1092}{Quantum \textbf{7}, 1092 (2023)}.

\bibitem{LeoneBittel2024} L.~Leone and L.~Bittel,
Stabilizer entropies are monotones for magic-state resource theory, \href{https://doi.org/10.1103/PhysRevA.110.L040403}{Phys. Rev. A \textbf{110}, L040403 (2024)}.

\bibitem{LamiCollura2024} G.~Lami and M.~Collura,
Unveiling the stabilizer group of a matrix product state, \href{https://doi.org/10.1103/PhysRevLett.133.010602}{Phys. Rev. Lett. \textbf{133}, 010602 (2024)}.

\bibitem{ColluraEtAl2026} M.~Collura, J.~De Nardis, V.~Alba, and G.~Lami,
The non-stabilizerness of fermionic Gaussian states, \href{https://doi.org/10.22331/q-2026-03-23-2036}{Quantum \textbf{10}, 2036 (2026)}.

\bibitem{Sarkar2020} S.~Sarkar, C.~Mukhopadhyay, and A.~Bayat,
Characterization of an operational quantum resource in a critical many-body system, \href{https://doi.org/10.1088/1367-2630/aba919}{New J. Phys. \textbf{22}, 083077 (2020)}.

\bibitem{White2021} C.~D. White, C.~Cao, and B.~Swingle,
Conformal field theories are magical, \href{https://doi.org/10.1103/PhysRevB.103.075145}{Phys. Rev. B \textbf{103}, 075145 (2021)}.

\bibitem{LeoneIsing2022} S.~F.~E. Oliviero, L.~Leone, and A.~Hamma,
Magic-state resource theory for the ground state of the transverse-field Ising model, \href{https://doi.org/10.1103/PhysRevA.106.042426}{Phys. Rev. A \textbf{106}, 042426 (2022)}.

\bibitem{TarabungaCritical2024} P.~S. Tarabunga,
Critical behaviors of non-stabilizerness in quantum spin chains, \href{https://doi.org/10.22331/q-2024-07-17-1413}{Quantum \textbf{8}, 1413 (2024)}.

\bibitem{LammaBauerDalmonteCollura2026} E.~Lamma, T.~Bauer, M.~Dalmonte, and M.~Collura,
Magic of Kitaev spin liquids, \href{https://arxiv.org/abs/2608.25929}{arXiv:2608.25929 (2026)}.

\bibitem{FanEtAl2025} C.~Fan, X.~Qian, H.-C. Zhang, R.-Z. Huang, M.~Qin, and T.~Xiang,
Disentangling critical quantum spin chains with Clifford circuits, \href{https://doi.org/10.1103/PhysRevB.111.085121}{Phys. Rev. B \textbf{111}, 085121 (2025)}.

\bibitem{FrauEtAl2025} M.~Frau, P.~S. Tarabunga, M.~Collura, E.~Tirrito, and M.~Dalmonte,
Stabilizer disentangling of conformal field theories, \href{https://doi.org/10.21468/SciPostPhys.18.5.165}{SciPost Phys. \textbf{18}, 165 (2025)}.

\bibitem{FuxEtAl2024} G.~E. Fux, B.~B\'eri, R.~Fazio, and E.~Tirrito,
Disentangling magic states with classically simulable quantum circuits, \href{https://doi.org/10.1103/ggp1-byj1}{Phys. Rev. Lett. \textbf{135}, 260605 (2025)}.

\bibitem{Korbany2025} D.~A. Korbany, M.~J. Gullans, and L.~Piroli,
Long-range nonstabilizerness and phases of matter, \href{https://doi.org/10.1103/1hlj-h6t9}{Phys. Rev. Lett. \textbf{135}, 160404 (2025)}.

\bibitem{Turkeshi2025} X.~Turkeshi, A.~Dymarsky, and P.~Sierant,
Pauli spectrum and nonstabilizerness of typical quantum many-body states, \href{https://doi.org/10.1103/PhysRevB.111.054301}{Phys. Rev. B \textbf{111}, 054301 (2025)}.

\bibitem{TurkeshiSierantTarabunga2026} X.~Turkeshi, P.~Sierant, and P.~S. Tarabunga,
Pauli spectrum and nonstabilizerness of random fermionic Gaussian states, \href{https://arxiv.org/abs/2609.15739}{arXiv:2609.15739 (2026)}.

\bibitem{Hallam2026} A.~Hallam, R.~Smith, and Z.~Papi\'c,
Spectral signatures of nonstabilizerness and criticality in infinite matrix product states, \href{https://doi.org/10.1103/77wh-qv29}{Phys. Rev. B \textbf{113}, 245113 (2026)}.

\bibitem{OlivieroProcessor2022} S.~F.~E. Oliviero, L.~Leone, A.~Hamma, and S.~Lloyd,
Measuring magic on a quantum processor, \href{https://doi.org/10.1038/s41534-022-00666-5}{npj Quantum Inf. \textbf{8}, 148 (2022)}.

\bibitem{HaugLeeKim2024} T.~Haug, S.~Lee, and M.~S. Kim,
Efficient quantum algorithms for stabilizer entropies, \href{https://doi.org/10.1103/PhysRevLett.132.240602}{Phys. Rev. Lett. \textbf{132}, 240602 (2024)}.

\bibitem{HaugPiroli2023} T.~Haug and L.~Piroli,
Quantifying nonstabilizerness of matrix product states, \href{https://doi.org/10.1103/PhysRevB.107.035148}{Phys. Rev. B \textbf{107}, 035148 (2023)}.

\bibitem{TarabungaMPS2024} P.~S. Tarabunga, E.~Tirrito, M.~C. Ba\~nuls, and M.~Dalmonte,
Nonstabilizerness via matrix product states in the Pauli basis, \href{https://doi.org/10.1103/PhysRevLett.133.010601}{Phys. Rev. Lett. \textbf{133}, 010601 (2024)}.

\bibitem{LamiCollura2023} G.~Lami and M.~Collura,
Nonstabilizerness via perfect Pauli sampling of matrix product states, \href{https://doi.org/10.1103/PhysRevLett.131.180401}{Phys. Rev. Lett. \textbf{131}, 180401 (2023)}.

\bibitem{DingWangYan2025} Y.-M. Ding, Z.~Wang, and Z.~Yan,
Evaluating many-body stabilizer R\'enyi entropy by sampling reduced Pauli strings: Singularities, volume law, and nonlocal magic, \href{https://doi.org/10.1103/pyzr-jmvw}{PRX Quantum \textbf{6}, 030328 (2025)}.

\bibitem{LiChang2026} Y.-L.~Li and P.-Y.~Chang,
Pauli spectrum and stabilizer R\'enyi entropy in gapless symmetry-protected topological phases, \href{https://arxiv.org/abs/2607.03762}{arXiv:2607.03762}.

\bibitem{HoshinoOshikawaAshida2026} M.~Hoshino, M.~Oshikawa, and Y.~Ashida,
Stabilizer R\'enyi entropy and conformal field theory, \href{https://doi.org/10.1103/ylsz-dm3y}{Phys. Rev. X \textbf{16}, 011037 (2026)}.

\bibitem{MatsudaHoshinoAshida2026} R.~Matsuda, M.~Hoshino, and Y.~Ashida,
Quantum computational resources and conformal field theory: Unifying spins, bosons, and fermions, \href{https://arxiv.org/abs/2607.05343}{arXiv:2607.05343}.

\bibitem{HoshinoAshida2026} M.~Hoshino and Y.~Ashida,
Stabilizer R\'enyi entropy encodes fusion rules of topological defects and boundaries, \href{https://doi.org/10.1103/1tyr-rlbb}{Phys. Rev. Lett. \textbf{136}, 080402 (2026)}.

\bibitem{RamirezTrinoRajabpour2026} E.~A. Ramirez Trino and M.~A. Rajabpour,
Equivalence of stabilizer and Shannon R\'enyi entropies: Exact results for quantum critical chains, \href{https://doi.org/10.1103/2frt-tdg9}{Phys. Rev. Lett. \textbf{137}, 090403 (2026)}.

\bibitem{KhassehRajabpour2026} R.~Khasseh and M.~A. Rajabpour,
Hidden conformal boundary data in finite-temperature stabilizer entropy, \href{https://arxiv.org/abs/2606.08606}{arXiv:2606.08606}.

\bibitem{KhassehRamirezTrinoRajabpour2026} R.~Khasseh, E.~A. Ramirez Trino, and M.~A. Rajabpour,
Universal crossovers of stabilizer entropy beyond criticality, \href{https://arxiv.org/abs/2606.13810}{arXiv:2606.13810}.

\bibitem{ZhangZhouSun2026} P.~Zhang, S.~Zhou, and N.~Sun,
Stabilizer R\'enyi entropy and its transition in the coupled Sachdev--Ye--Kitaev model, \href{https://doi.org/10.1103/5c15-4g5n}{Phys. Rev. Lett. \textbf{136}, 080201 (2026)}.

\bibitem{RajabpourBoundary2026} M.~A. Rajabpour,
Quantized stabilizer-R\'enyi boundary response across fermionic SPT transitions, \href{https://arxiv.org/abs/2608.09749}{arXiv:2608.09749}.

\bibitem{Companion} R.~Khasseh and M.~A. Rajabpour,
Fugacity-resolved stabilizer entropy in critical quantum chains: Discrete Selberg sums and exactly solvable R\'enyi indices, \href{https://arxiv.org/abs/2608.06995}{arXiv:2608.06995}.

\bibitem{Pfeuty1970} P.~Pfeuty,
The one-dimensional Ising model with a transverse field, \href{https://doi.org/10.1016/0003-4916(70)90270-8}{Ann. Phys. \textbf{57}, 79 (1970)}.

\bibitem{Campostrini2015} M.~Campostrini, A.~Pelissetto, and E.~Vicari,
Quantum Ising chains with boundary fields, \href{https://doi.org/10.1088/1742-5468/2015/11/P11015}{J. Stat. Mech. (2015) P11015}.

\bibitem{JordanWigner1928} P.~Jordan and E.~Wigner,
\"Uber das Paulische \"Aquivalenzverbot, \href{https://doi.org/10.1007/BF01331938}{Z. Phys. \textbf{47}, 631 (1928)}.

\bibitem{LiebSchultzMattis1961} E.~Lieb, T.~Schultz, and D.~Mattis,
Two soluble models of an antiferromagnetic chain, \href{https://doi.org/10.1016/0003-4916(61)90115-4}{Ann. Phys. \textbf{16}, 407 (1961)}.

\bibitem{Peschel2003} I.~Peschel,
Calculation of reduced density matrices from correlation functions, \href{https://doi.org/10.1088/0305-4470/36/14/101}{J. Phys. A \textbf{36}, L205 (2003)}.

\bibitem{PeschelEisler2009} I.~Peschel and V.~Eisler,
Reduced density matrices and entanglement entropy in free lattice models, \href{https://doi.org/10.1088/1751-8113/42/50/504003}{J. Phys. A \textbf{42}, 504003 (2009)}.

\bibitem{Sutherland1971} B.~Sutherland,
Quantum many-body problem in one dimension: Ground state, \href{https://doi.org/10.1063/1.1665584}{J. Math. Phys. \textbf{12}, 246 (1971)}.

\bibitem{Haldane1988} F.~D.~M. Haldane,
Exact Jastrow--Gutzwiller resonating-valence-bond ground state of the spin-$1/2$ antiferromagnetic Heisenberg chain with $1/r^2$ exchange, \href{https://doi.org/10.1103/PhysRevLett.60.635}{Phys. Rev. Lett. \textbf{60}, 635 (1988)}.

\bibitem{Shastry1988} B.~S. Shastry,
Exact solution of an $S=1/2$ Heisenberg antiferromagnetic chain with long-ranged interactions, \href{https://doi.org/10.1103/PhysRevLett.60.639}{Phys. Rev. Lett. \textbf{60}, 639 (1988)}.

\bibitem{Gaudin1973} M.~Gaudin,
Gaz coulombien discret \`a une dimension, \href{https://doi.org/10.1051/jphys:01973003407051100}{J. Phys. France \textbf{34}, 511--522 (1973)}.

\bibitem{Mehta1975} M.~L. Mehta and G.~C. Mehta,
Discrete Coulomb gas in one dimension: Correlation functions, \href{https://doi.org/10.1063/1.522665}{J. Math. Phys. \textbf{16}, 1256 (1975)}.

\bibitem{Tsukerman2017} E.~Tsukerman,
Inverse participation ratios in the XX spin chain, \href{https://doi.org/10.1103/PhysRevB.95.115121}{Phys. Rev. B \textbf{95}, 115121 (2017)}.

\bibitem{StephanPollmann2017} J.-M. St\'ephan and F.~Pollmann,
Full counting statistics in the Haldane--Shastry chain, \href{https://doi.org/10.1103/PhysRevB.95.035119}{Phys. Rev. B \textbf{95}, 035119 (2017)}.

\bibitem{Selberg1944} A.~Selberg,
Remarks on a multiple integral, \href{https://cir.nii.ac.jp/crid/1370026059417066004}{Norsk Mat. Tidsskr. \textbf{26}, 71--78 (1944)}.

\bibitem{ForresterWarnaar2008} P.~J. Forrester and S.~O. Warnaar,
The importance of the Selberg integral, \href{https://doi.org/10.1090/S0273-0979-08-01221-4}{Bull. Am. Math. Soc. \textbf{45}, 489 (2008)}.

\bibitem{deBruijn1955} N.~G. de Bruijn,
On some multiple integrals involving determinants, \href{https://research.tue.nl/en/publications/on-some-multiple-integrals-involving-determinants/}{J. Indian Math. Soc. \textbf{19}, 133 (1955)}.

\bibitem{Dyson1962} F.~J. Dyson,
Statistical theory of the energy levels of complex systems. I, \href{https://doi.org/10.1063/1.1703773}{J. Math. Phys. \textbf{3}, 140 (1962)}.

\bibitem{Morris1982} W.~G. Morris,
\emph{Constant Term Identities for Finite and Affine Root Systems: Conjectures and Theorems}, Ph.D. thesis, University of Wisconsin--Madison (1982).

\bibitem{Macdonald1982} I.~G. Macdonald,
Some conjectures for root systems, \href{https://doi.org/10.1137/0513070}{SIAM J. Math. Anal. \textbf{13}, 988 (1982)}.

\bibitem{Stanley1989} R.~P. Stanley,
Some combinatorial properties of Jack symmetric functions, \href{https://doi.org/10.1016/0001-8708(89)90015-7}{Adv. Math. \textbf{77}, 76 (1989)}.

\bibitem{Kadell1994} K.~W.~J. Kadell,
A proof of the $q$-Macdonald--Morris conjecture for $BC_n$, \href{https://bookstore.ams.org/memo-108-516}{Mem. Am. Math. Soc. \textbf{108}, No.~516 (1994)}.

\bibitem{BalianBrezin1969} R.~Balian and E.~Br\'ezin,
Nonunitary Bogoliubov transformations and extension of Wick's theorem, \href{https://doi.org/10.1007/BF02710281}{Nuovo Cimento B \textbf{64}, 37 (1969)}.

\bibitem{KramersWannier1941} H.~A. Kramers and G.~H. Wannier,
Statistics of the two-dimensional ferromagnet. Part I, \href{https://doi.org/10.1103/PhysRev.60.252}{Phys. Rev. \textbf{60}, 252 (1941)}.

\bibitem{IshikawaWakayama1995} M.~Ishikawa and M.~Wakayama,
Minor summation formula of Pfaffians, \href{https://doi.org/10.1080/03081089508818403}{Linear Multilinear Algebra \textbf{39}, 285 (1995)}.

\bibitem{Johansson1998} K.~Johansson,
On fluctuations of eigenvalues of random Hermitian matrices, \href{https://doi.org/10.1215/S0012-7094-98-09108-6}{Duke Math. J. \textbf{91}, 151 (1998)}.

\bibitem{BorotGuionnet2013} G.~Borot and A.~Guionnet,
Asymptotic expansion of $\beta$ matrix models in the one-cut regime, \href{https://doi.org/10.1007/s00220-012-1619-4}{Commun. Math. Phys. \textbf{317}, 447 (2013)}.

\bibitem{BorodinGorinGuionnet2016} A.~Borodin, V.~Gorin, and A.~Guionnet,
Gaussian asymptotics of discrete $\beta$-ensembles, \href{https://doi.org/10.1007/s10240-016-0085-5}{Publ. Math. Inst. Hautes \'Etudes Sci. \textbf{125}, 1 (2017)}.

\end{thebibliography}
\end{document}